\documentclass[a4paper,fleqn]{cas-sc}

\usepackage{graphicx}
\usepackage{tikz}
\usetikzlibrary{arrows.meta,calc}
\usepackage[numbers]{natbib}

\definecolor{manuscriptrevision}{RGB}{0,0,0}
\newenvironment{revisiontext}{\color{manuscriptrevision}}{}

\definecolor{manuscriptworkprecision}{RGB}{0,0,0}
\newcommand{\workprecisionrevision}[1]{{\color{manuscriptworkprecision}#1}}
\newenvironment{workprecisionrevisiontext}{\color{manuscriptworkprecision}}{}

\definecolor{manuscriptaccuracyrevision}{RGB}{0,0,0}
\newcommand{\accuracyrevision}[1]{{\color{manuscriptaccuracyrevision}#1}}
\newenvironment{accuracyrevisiontext}{\color{manuscriptaccuracyrevision}}{}

\definecolor{manuscriptintrorevision}{RGB}{0,0,0}

\definecolor{manuscripttheory}{RGB}{0,0,0}
\newcounter{remark}
\newenvironment{remarkblock}[2][manuscripttheory]
{%
    \par\medskip
    \refstepcounter{remark}%
    \color{#1}%
    \noindent\textbf{Remark \theremark\space (#2).}\ \ignorespaces
}
{\par\medskip}

\newcommand{\manuscriptdisplayspacing}{%
    \setlength{\abovedisplayskip}{6pt plus 1pt minus 1pt}%
    \setlength{\belowdisplayskip}{8pt plus 1pt minus 1pt}%
    \setlength{\abovedisplayshortskip}{6pt plus 1pt minus 1pt}%
    \setlength{\belowdisplayshortskip}{8pt plus 1pt minus 1pt}%
}
\AddToHook{env/equation/begin}{\manuscriptdisplayspacing}
\AddToHook{env/equation*/begin}{\manuscriptdisplayspacing}
\AddToHook{env/align/begin}{\manuscriptdisplayspacing}
\AddToHook{env/align*/begin}{\manuscriptdisplayspacing}
\allowdisplaybreaks[4]

\newcounter{algorithm}
\renewcommand{\thealgorithm}{\arabic{algorithm}}
\newenvironment{algorithmblock}[1]
{%
    \refstepcounter{algorithm}
    \par\medskip
    \begingroup
    \small
    \hrule\nobreak\vspace{5pt}\nobreak
    \noindent\textbf{Algorithm \thealgorithm:} #1
    \par\nobreak\vspace{5pt}\nobreak\hrule\nobreak\vspace{5pt}\nobreak
}
{%
    \par\nobreak\vspace{3pt}\nobreak\hrule
    \endgroup
    \par\medskip
}
\newcommand{\algline}[2]{%
    \noindent\makebox[2em][r]{#1}\hspace{0.7em}%
    \parbox[t]{\dimexpr\linewidth-2.7em\relax}{#2}\par
}

\begin{document}

\colorlet{frontmatterlinkcolor}{black}
\colorlet{hscolor}{black}

\let\WriteBookmarks\relax
\hypersetup{hypertexnames=false}
\let\printorcid\relax
\ExplSyntaxOn
\keys_set:nn { stm / mktitle } { nologo = true }
\ExplSyntaxOff

\shorttitle{Local ADER central differences for GPU DNS}
\shortauthors{H. Zhang et al.}

\title[mode=title]{A GPU-efficient local ADER finite-difference method for compressible-flow DNS}

\author[1]{Hong Zhang}
\author[1]{Xian Wang}
\author[2,3]{Xinliang Li}
\author[4]{Yibing Chen}
\author[1]{Xing Ji}
\cormark[1]
\ead{jixing@xjtu.edu.cn}

\affiliation[1]{
    organization={School of Aerospace Engineering, Xi'an Jiaotong University},
    city={Xi'an},
    state={Shaanxi},
    country={China}
}

\affiliation[2]{
    organization={State Key Laboratory of High Temperature Gas Dynamics, Institute of Mechanics, Chinese Academy of Sciences},
    city={Beijing},
    postcode={100190},
    country={China}
}

\affiliation[3]{
    organization={School of Engineering Science, University of Chinese Academy of Sciences},
    city={Beijing},
    postcode={100049},
    country={China}
}

\affiliation[4]{
    organization={National Key Laboratory of Computational Physics, Institute of Applied Physics and Computational Mathematics},
    city={Beijing},
    postcode={100088},
    country={China}
}

\cortext[1]{Corresponding author}

\begin{abstract}
Algorithm design tailored to graphics processing units (GPUs) can
substantially improve the efficiency of computational fluid dynamics
(CFD) simulations. Efforts to adapt CFD solvers to GPUs often focus on
spatial discretization, while time advancement commonly relies on
multistage Runge--Kutta (RK) methods. However, these RK methods require
repeated residual evaluations and full-domain operations, increasing the
cost of compressible-flow direct numerical simulation (DNS). To reduce
this overhead, we develop a local arbitrary high-order derivatives (ADER)
finite-difference method that couples a compact ten-point quadratic
predictor to a conservative central corrector through operator matching. With two
quadrature nodes, one Picard correction, and selective dissipation, smooth
convective tests recover third-order convergence under fixed-CFL refinement
and near-seventh-order spatial behavior when background spatial error dominates.
Temporal states remain kernel-local, and halo communication overlaps
interior computation. Tests of smooth advection, periodic turbulence, and
wall-bounded flows assess accuracy and flow statistics. On a $302^3$-node
viscous Taylor--Green benchmark without heat conduction, throughput
reaches 230.87 million node updates per second on one V100 GPU in double
precision, 1.944 times that of spatially matched three-stage Runge--Kutta
integration. Weak-scaling efficiency reaches 99.44\% from 4 to 32 V100
GPUs at 1.812 billion nodes.
\end{abstract}

\begin{highlights}
\item Local ADER couples compact prediction to one high-order central-difference update.
\item Local ADER achieves 1.944 times the complete-step throughput of matched RK3-F.
\item Kernel-local prediction avoids full-domain storage of temporal states.
\end{highlights}


\let\frontmattertextcolor\textcolor
\renewcommand{\textcolor}[3][]{\frontmattertextcolor{black}{#3}}
\maketitle
\let\textcolor\frontmattertextcolor

\section{Introduction}

Direct numerical simulation (DNS) resolves the dynamically relevant scales
of turbulent flow and places stringent demands on numerical resolution
and computational resources \citep{MoinMahesh1998}. In compressible DNS,
broadband vortical fluctuations coexist with localized compression and
steep thermodynamic gradients. The discretization must therefore preserve
small-scale flow content while controlling nonphysical oscillations near
under-resolved gradients \citep{JohnsenEtAl2010}. Meeting these requirements
over physically relevant time intervals makes computational efficiency an
integral part of numerical-method design. For large GPU simulations, this
efficiency depends on the spatial and temporal discretizations together
with the data movement and communication required to execute them.

High-order central differences, nonlinear reconstructions, and
element-based formulations provide different balances between resolution
and nonlinear stabilization
\citep{Lele1992,JiangShu1996,FuHuAdams2016,CockburnShu1998}. Central
differences combined with selective dissipation retain a low-dissipation
approximation in smooth regions while introducing additional damping near
localized steep gradients \citep{SciacovelliEtAl2021}. These spatial
operators are commonly advanced through the method of lines using explicit
multistage Runge--Kutta (RK) schemes. The separation of spatial and
temporal discretizations allows reconstruction, molecular transport, and
boundary treatment to enter a common residual evaluation, making the time
integrator reusable across different spatial formulations. Low-storage RK
methods reduce the number of solution-sized work arrays
\citep{KennedyCarpenterLewis2000}, while retaining the stagewise
evaluation of that residual. An $s$-stage method evaluates the residual
$s$ times, and each intermediate state generally requires boundary
treatment and halo completion. On distributed GPUs, these dependencies
connect local residual evaluation to repeated access to global solution
fields and exchanges of updated halo data.

High-order one-step methods use the governing equations to construct the
temporal evolution within each step. Lax--Wendroff-type methods replace
temporal derivatives with spatial derivatives, including finite-difference
WENO realizations \citep{LaxWendroff1960,QiuShu2003}. This approach links
temporal accuracy to the approximation of spatial derivatives of the
physical fluxes. Its direct high-order extension, however,
produces increasingly elaborate, equation-dependent expressions. Shu and
Osher \citep{ShuOsher1988} identified the implementation complexity of
Lax--Wendroff time discretization, particularly for multidimensional
problems with forcing terms, as a motivation for TVD RK formulations.
Subsequent approaches have sought to construct the required temporal
information with less explicit differentiation. In finite-volume and discontinuous Galerkin ADER
formulations, an element-local space--time predictor supplies
time-integrated fluxes to a conservative corrector
\citep{DumbserEtAl2008}. Local prediction constructs temporal information
before the global correction, thereby changing the dependencies between
temporal evolution and the exchange of neighboring solution data. Related
finite-difference approaches include Picard integral formulations based on
time-averaged physical fluxes and Jacobian- and Hessian-free constructions
\citep{ChristliebGucluSeal2015,LeeLee2021}. These approaches organize
high-order advancement around the construction of fluxes over a complete
time step. Efficiency studies of structured-mesh ADER schemes and
communication-avoiding transport formulations demonstrate the potential of
reorganizing temporal dependencies to reduce advancement costs within
their respective numerical settings \citep{BalsaraEtAl2013,Norman2014}.

Our experience with space--time coupled gas-kinetic schemes (GKS)
motivates constructing time-dependent fluxes from a local evolution
model to reduce repeated reconstruction and global stage operations.
Gao, Ji, and co-workers \citep{GaoEtAl2026} provided a theoretical basis
for an additional benefit: their Fourier analysis showed that the
second-order single-stage compact GKS considered admits a larger stable
CFL range than the RK--DG formulation of the same order. This result
supports formulations that combine fewer global stages with stable
advancement at comparatively large time steps.
Realizing these benefits in finite-difference DNS requires a suitable
local evolution model. GKS and its compact
variants have been developed primarily in finite-volume frameworks
\citep{WangWangPan2024}. Finite-difference alternatives have been proposed
\citep{XuanXu2018,ZhaoChenXu2026}, but their reported validation is
concentrated on one- and two-dimensional problems; broader assessment in
large-scale three-dimensional compressible turbulence remains important.
Kinetic flux evaluation concentrates substantial arithmetic locally,
offering the potential for high arithmetic intensity, but the
distribution-function derivatives and velocity-moment calculations
required by high-order extensions can also entail considerable
floating-point work and algebraic complexity \citep{LiChenJiang2020}.
This tradeoff motivates a formulation that preserves local space--time
coupling while controlling the cost of its evolution model.

GPU solvers have also enabled large-scale simulations of compressible
flows. STREAmS-2 and OpenCFD-SCU support high-resolution simulations of
compressible turbulence \citep{BernardiniEtAl2023,DangEtAl2022}, while HTR
addresses hypersonic aerothermodynamics through a task-based multi-GPU
framework \citep{DiRenzoEtAl2020}. MFC provides a portable and scalable
GPU approach for multiphase compressible flow
\citep{RadhakrishnanEtAl2024}. Performance studies of STREAmS-2 and MFC
identify memory reuse, kernel organization, and inter-GPU communication as
central performance considerations
\citep{SathyanarayanaEtAl2023,RadhakrishnanEtAl2024}. Roofline analysis of
STREAmS-2 further shows that the same kernel can exhibit different
arithmetic intensities across GPU architectures
\citep{SathyanarayanaEtAl2023}. These findings connect application
performance to the numerical workload, the realized data traffic, and the
target architecture.

Alongside these architectural advances, computer algebra
and artificial-intelligence approaches to symbolic integration and
differential equations \citep{LampleCharton2020} offer support for the
analytical work required by high-order temporal formulations. These tools
address formulation effort, while GPU parallelism and local data reuse
alter the cost of evaluating the resulting expressions. Their combined
development motivates renewed investigation of single-stage schemes that
exchange additional local arithmetic for fewer full-domain memory
accesses and synchronization points.

For central-difference DNS, the resulting objective is to
construct time-integrated Navier--Stokes fluxes directly from local
macroscopic states and use them in one global conservative update. Its
efficiency depends on both the predictor--corrector coupling and the
organization of intermediate data. A compact local predictor must be coupled consistently
to a wide-stencil conservative corrector and retain the
molecular-transport terms of the compressible Navier--Stokes equations,
including velocity and temperature gradients. The predictor introduces
arithmetic and temporal states; storing these states as full-domain arrays
would also introduce global storage and repeated memory accesses. Reducing
global stage dependencies therefore requires an assessment of the
additional local prediction work and the data exchanged with the
corrector. The resulting problem is to preserve spatial accuracy and
conservation while organizing temporal evolution locally, and to establish
whether this organization yields a GPU performance advantage over a
matched multistage method.

This work develops a conservative local ADER finite-difference method for
compressible-flow DNS on GPUs. Operator matching couples a compact local
space--time predictor to a high-order central corrector. The complete Euler
consistency analysis retains the mixed errors introduced by the compact
representation and gives third-order consistency under fixed-CFL refinement.
Smooth convective tests recover this convergence and near-seventh-order
behavior in the regime dominated by background spatial error. The GPU realization keeps temporal-node states
within the predictor kernel, contracts their contributions into compact
corrector data, and performs one global conservative update. Smooth
advection tests examine spatial and temporal error behavior, while
periodic turbulence and wall-bounded compressible flows assess flow
statistics, molecular transport, and boundary treatment. A controlled
comparison with a three-stage full-residual RK reference (RK3-F) evaluates
complete-step efficiency with the spatial discretization, dissipation,
transport model, precision, and time-step selector held fixed. Comparisons
with established GPU solvers and distributed scaling tests provide
complementary performance evidence.

Section~\ref{sec:numerical_method_rewrite} presents the numerical
formulation. Section~\ref{sec:gpu_performance}
describes the GPU realization and performance assessment.
Section~\ref{sec:numerical_results} reports the accuracy studies and
flow-validation results, and Section~\ref{sec:conclusions} summarizes the
conclusions.

\colorlet{hscolor}{frontmatterlinkcolor}

\section{Local ADER finite-difference method}
\label{sec:numerical_method_rewrite}

\begingroup
\definecolor{sectiontworevision}{RGB}{0,0,0}
\newenvironment{s2revision}{\color{sectiontworevision}\colorlet{hscolor}{sectiontworevision}}{}

\subsection{Governing equations}
\label{sec:governing_equations_rewrite}

We consider the nondimensional three-dimensional compressible Navier--Stokes
equations for a calorically perfect gas. In Cartesian coordinates
$\boldsymbol{x}$, the conservation laws are
\begin{equation}
    \frac{\partial\boldsymbol{W}}{\partial t}
    +\sum_{j=1}^{3}\frac{\partial\boldsymbol{F}_j}{\partial x_j}
    =\boldsymbol{S},
    \qquad
    \boldsymbol{F}_j
    =\boldsymbol{F}^{\mathrm{c}}_j-\boldsymbol{F}^{\mathrm{v}}_j,
    \label{eq:compressible_ns_rewrite}
\end{equation}
where the conservative state and the convective and viscous fluxes are
\begin{equation*}
\begin{aligned}
    \boldsymbol{W}
    &=\left(\rho,\,\rho u_i,\,\rho E\right)^{\mathsf T},
    \qquad
    \boldsymbol{F}^{\mathrm{c}}_j
    =\left(\rho u_j,\,\rho u_i u_j+p\delta_{ij},\,
             (\rho E+p)u_j\right)^{\mathsf T},\\
    \boldsymbol{F}^{\mathrm{v}}_j
    &=\left(0,\,\tau_{ij},\,u_i\tau_{ij}-q_j\right)^{\mathsf T}.
\end{aligned}
\end{equation*}
Here, $\rho$, $u_i$, $p$, $T$, and $E$ denote density, velocity,
pressure, temperature, and specific total energy, respectively;
$\delta_{ij}$ is the Kronecker delta. Cartesian component indices take
the values $1,2,3$, and repeated indices imply summation unless an explicit
sum is written. The perfect-gas closure is
\begin{equation*}
    p=(\gamma-1)\left(\rho E-\frac{1}{2}\rho u_i u_i\right),
    \qquad T=\frac{p}{\rho},
\end{equation*}
where $\gamma$ is the ratio of specific heats and the nondimensional gas
constant is unity. The speed of sound is $c=(\gamma p/\rho)^{1/2}$.

The viscous stress and conductive heat flux follow the Newtonian
constitutive relation with Stokes' hypothesis and Fourier's law,
\begin{equation*}
    \tau_{ij}
    =\mu\left(
        \frac{\partial u_i}{\partial x_j}
        +\frac{\partial u_j}{\partial x_i}
        -\frac{2}{3}\delta_{ij}\frac{\partial u_k}{\partial x_k}
      \right),
    \qquad
    q_j=-\kappa\frac{\partial T}{\partial x_j}.
\end{equation*}
With density, velocity, and length scaled by $\rho_{\mathrm{ref}}$,
$U_{\mathrm{ref}}$, and $L_{\mathrm{ref}}$, respectively, the dimensionless
viscosity and thermal conductivity are
\begin{equation*}
    \mu(T)=\frac{1}{\mathrm{Re}}
           \left(\frac{T}{T_{\mathrm{ref}}}\right)^{s_\mu},
    \qquad
    \kappa(T)=\frac{\gamma}{(\gamma-1)\mathrm{Pr}}\mu(T).
\end{equation*}
Here, $\mathrm{Re}$ and $\mathrm{Pr}$ are the Reynolds and Prandtl
numbers, and $T_{\mathrm{ref}}$ is the nondimensional reference
temperature. The exponent $s_\mu=0$ gives constant viscosity, whereas a
nonzero exponent specifies temperature-dependent transport. Heat
conduction can be disabled by setting $\kappa=0$ while retaining the
viscous stress; setting $\mu=\kappa=0$ recovers the Euler equations.
The physical source $\boldsymbol{S}$ represents prescribed forcing when
required and vanishes in unforced configurations.

\begingroup
\color{black}
\colorlet{hscolor}{black}

\subsection{Single-stage conservative update framework}
\label{sec:step_initialization_rewrite}

\begin{s2revision}
Each physical step combines a local space--time prediction with one global conservative correction. The predictor supplies the temporal data needed to integrate the physical fluxes, and the corrector advances the nodal conservative solution from $t^n$ to $t^{n+1}$ once \citep{DumbserEtAl2008,BalsaraEtAl2013}. We first specify this update and the data that the predictor must provide. The continuous physical evolution operator is denoted by $\mathcal{L}$:
\end{s2revision}
\begin{equation*}
\begin{aligned}
    \mathcal{L}[\boldsymbol{W}](\boldsymbol{x},t)
    &:=-\sum_{j=1}^{3}
       \frac{\partial\boldsymbol{F}_j}{\partial x_j}(\boldsymbol{x},t)
       +\boldsymbol{S}(\boldsymbol{x},t),\\
    \frac{\partial\boldsymbol{W}}{\partial t}(\boldsymbol{x},t)
    &=\mathcal{L}[\boldsymbol{W}](\boldsymbol{x},t).
\end{aligned}
\end{equation*}
The operator acts on the spatial state field; its fluxes and source
are evaluated from that field and the prescribed physical data at
position $\boldsymbol{x}$ and time $t$.
To identify the temporal data required by the method, introduce
$\tau=(t-t^n)/\Delta t$, with $\Delta t=t^{n+1}-t^n$.
Integration of the evolution equation at a fixed physical position gives
\begin{equation}
    \boldsymbol{W}(\boldsymbol{x},t^n+\tau\Delta t)
    =\boldsymbol{W}(\boldsymbol{x},t^n)
     +\Delta t\int_0^\tau
       \mathcal{L}[\boldsymbol{W}]
       (\boldsymbol{x},t^n+\tau'\Delta t)\,\mathrm{d}\tau'.
\label{eq:time_integral_problem_rewrite}
\end{equation}

For a smooth solution, spatial differentiation commutes with time
integration. At $\tau=1$, the flux and source integrals therefore
provide the data for the complete step. The numerical method
approximates these averages and applies the conservative update
\begin{equation}
\begin{aligned}
    \overline{\mathcal{L}}_{\boldsymbol{m}}^n
    &=-\sum_{j=1}^{3}
       \bigl(\mathcal{D}_j\overline{\boldsymbol{F}}_j^n\bigr)_{\boldsymbol{m}}
      +\overline{\boldsymbol{S}}_{\boldsymbol{m}}^n
      +\overline{\mathcal{A}}_{\boldsymbol{m}}^n,\\
    \boldsymbol{W}_{\boldsymbol{m}}^{n+1}
    &=\boldsymbol{W}_{\boldsymbol{m}}^n
      +\Delta t\,\overline{\mathcal{L}}_{\boldsymbol{m}}^n.
\end{aligned}
\label{eq:complete_update_rewrite}
\end{equation}

\begin{s2revision}
Here, $\overline{\boldsymbol{F}}_j^n$ and $\overline{\boldsymbol{S}}^n$ approximate the physical flux and source averaged over $[t^n,t^{n+1}]$. The radius-four derivative $\mathcal{D}_j$ is the eighth-order central difference defined in Eq.~\eqref{eq:eighth_order_central_rewrite}. Molecular transport is included in $\boldsymbol{F}_j=\boldsymbol{F}^{\mathrm{c}}_j-\boldsymbol{F}^{\mathrm{v}}_j$, whereas $\overline{\mathcal{A}}^n$ is the selective artificial-dissipation contribution defined in Section~\ref{sec:global_corrector_rewrite}. Thus, $\overline{\mathcal{L}}_{\boldsymbol{m}}^n$ is the assembled discrete right-hand side of the accepted update. The local predictor uses a compact approximation of the physical evolution law to construct its temporal data; the wide derivative acts on the resulting nodal flux averages. This division allows the intermediate prediction to remain local while the final update couples neighboring nodes conservatively.
\end{s2revision}

The principal unknown in this update is the physical flux throughout
the time interval. We approximate its integral with the two-point
Gauss--Legendre rule,
\begin{equation}
\begin{gathered}
    \int_0^1\boldsymbol{F}_j
       (\boldsymbol{x}_{\boldsymbol{m}},t^n+\tau\Delta t)
       \,\mathrm{d}\tau
    \approx\sum_{\alpha=1}^{2}w_\alpha
       \boldsymbol{F}_j
       (\boldsymbol{x}_{\boldsymbol{m}},t^n+c_\alpha\Delta t),\\[4pt]
    c_{1,2}=\frac{1}{2}\left(1\mp\frac{1}{\sqrt{3}}\right),
    \qquad w_1=w_2=\frac{1}{2}.
\end{gathered}
\label{eq:temporal_quadrature_target_rewrite}
\end{equation}

\begin{s2revision}
The source uses the same nodes and weights. The Gauss rule integrates temporal polynomials of degree at most three exactly, so the remaining task is to supply sufficiently accurate states at those nodes. Both states are constructed from beginning-of-step data before the accepted solution is changed. The two temporal evaluations therefore belong to one conservative step. Section~\ref{sec:accuracy_consistency_rewrite} analyzes their accuracy together with the spatial representation used during prediction.
\end{s2revision}

\begin{s2revision}
Let $\widetilde{\boldsymbol{W}}$ denote the local predicted field and $\widetilde{\boldsymbol{W}}_{\alpha,\boldsymbol{m}}$ its center state at $t^n+c_\alpha\Delta t$. Equation~\eqref{eq:temporal_quadrature_target_rewrite} requires these two center states for convection and their velocity and temperature gradients for molecular transport. Constructing them requires the local flux divergence and its spatial variation, which are obtained from the known initial neighborhood in the next subsection.
\end{s2revision}

Each realization of this construction starts with the known nodal
field $\boldsymbol{W}_{\boldsymbol{m}}^n$ and a time interval selected
from convective and viscous rate estimates:
\begin{equation}
\begin{aligned}
    \sigma_{\mathrm{c},\boldsymbol{m}}^n
    &=\sum_{j=1}^{3}
      \frac{|u_{j,\boldsymbol{m}}^n|}{h_{j,\boldsymbol{m}}}
      +c_{\boldsymbol{m}}^n
       \left(\sum_{j=1}^{3}\frac{1}{h_{j,\boldsymbol{m}}^2}\right)^{1/2},
    \qquad
    \sigma_{\mathrm{v},\boldsymbol{m}}^n
    =2\frac{\mu_{\boldsymbol{m}}^n}{\rho_{\boldsymbol{m}}^n}
      \max\left(\frac{4}{3},\frac{\gamma}{\mathrm{Pr}}\right)
      \sum_{j=1}^{3}\frac{1}{h_{j,\boldsymbol{m}}^2},\\
    \Delta t
    &=\min\Bigl(
        t_{\mathrm{end}}-t^n,\,
        \frac{\mathrm{CFL}}{
          \max\nolimits_{\text{all nodes and ranks}}
          \bigl(\sigma_{\mathrm{c},\boldsymbol{m}}^n
                +\sigma_{\mathrm{v},\boldsymbol{m}}^n\bigr)}
      \Bigr).
\end{aligned}
\label{eq:cfl_time_step_rewrite}
\end{equation}

Here, $h_{j,\boldsymbol{m}}$ is the local spacing in direction $j$,
and $\mu_{\boldsymbol{m}}^n=\mu(T_{\boldsymbol{m}}^n)$.
The advective contributions are summed over the coordinate directions,
whereas the acoustic contribution uses the Euclidean norm of the inverse
grid spacings. The convective and viscous rates are added locally before
the maximum is taken over all nodes and MPI ranks. For the Euler equations,
the viscous rate vanishes; without heat conduction, the factor
$\max(4/3,\gamma/\mathrm{Pr})$ reduces to $4/3$.
Physical boundary conditions and halo exchange supply the neighboring
initial states required by the spatial stencils.

\begin{s2revision}
With the time interval fixed, Section~\ref{sec:local_spatial_operator_rewrite} constructs the local spatial data and their center matching to the wide derivative. Section~\ref{sec:local_ader_predictor_rewrite} uses these data to predict and correct the temporal-node states. Section~\ref{sec:time_averaged_fluxes_rewrite} evaluates the physical fluxes and forms their averages; Section~\ref{sec:global_corrector_rewrite} adds selective dissipation and completes Eq.~\eqref{eq:complete_update_rewrite}. The dependency order is therefore initialization, local prediction, flux quadrature, and one conservative correction.
\end{s2revision}

\subsection{Initial spatial representation and operator matching}
\label{sec:local_spatial_operator_rewrite}

\begin{s2revision}
The temporal-node states required by Eq.~\eqref{eq:temporal_quadrature_target_rewrite} depend on the evolution of the local neighborhood. The center flux divergence determines the center-state variation, and its spatial variation determines the changes in neighboring predicted states and in the gradients needed by molecular transport. We therefore construct both the center divergence and its directional linear modes from the beginning-of-step nodal field. These quantities provide the spatial input to the temporal predictor.
\end{s2revision}

\begin{s2revision}
A complete quadratic spatial representation is chosen because its pure and mixed curvature coefficients supply the second spatial derivatives entering the divergence gradients. Differentiating a quadratic convective-flux polynomial yields constant and linear divergence modes exactly within this representation. Quadratic velocity and temperature polynomials supply the derivatives entering the viscous construction. The resulting compact operator predicts within-step physical evolution; the accepted update in Eq.~\eqref{eq:complete_update_rewrite} retains its separate wide-stencil derivative.
\end{s2revision}

\begin{s2revision}
All local fields below are centered at a target node $\boldsymbol{m}$. Their initial samples are neighboring values of $\boldsymbol{W}^n$. During temporal prediction, the same sampling positions receive values from this target node's own provisional field. The spatial construction can therefore be reused for operator reevaluation without requiring predicted fields centered at neighboring nodes.
\end{s2revision}

On the Cartesian stencil centered at $\boldsymbol{m}$, define
\begin{equation*}
    \xi_r=\frac{x_r-x_{r,\boldsymbol{m}}}{h_r},
    \qquad
    \frac{\partial}{\partial x_r}
    =\frac{1}{h_r}\frac{\partial}{\partial\xi_r},
    \qquad r=1,2,3.
\end{equation*}
All local polynomials and their derivatives below use
$\boldsymbol{\xi}$; $h_r$ converts their derivatives to physical
units. A complete quadratic representation contains a constant
term, three linear terms, three pure-quadratic terms, and three
mixed terms. For a sampled scalar or vector quantity
$\boldsymbol{a}$, we use the common form
\begin{equation}
    \boldsymbol{a}(\boldsymbol{\xi})
    =\boldsymbol{a}_0
     +\sum_{r=1}^{3}\boldsymbol{a}_r\xi_r
     +\sum_{r=1}^{3}\boldsymbol{a}_{rr}\xi_r^2
     +\sum_{1\leq r<s\leq3}\boldsymbol{a}_{rs}\xi_r\xi_s.
    \label{eq:local_p2_polynomial_rewrite}
\end{equation}
This basis is used for the initial conservative field and for the
sampled fluxes and primitive variables entering operator evaluations.
To determine its ten coefficients directly from pointwise data, we
select the following one-layer sampling geometry:
\begin{equation*}
\begin{aligned}
    \mathcal{S}_{\boldsymbol{m}}=\{\,
    &\boldsymbol{m},\;
    \boldsymbol{m}\mathbin{\pm}\boldsymbol{e}_1,\;
    \boldsymbol{m}\mathbin{\pm}\boldsymbol{e}_2,\;
    \boldsymbol{m}\mathbin{\pm}\boldsymbol{e}_3,\\
    &\boldsymbol{m}+\boldsymbol{e}_1+\boldsymbol{e}_2,\;
    \boldsymbol{m}+\boldsymbol{e}_1+\boldsymbol{e}_3,\;
    \boldsymbol{m}+\boldsymbol{e}_2+\boldsymbol{e}_3\,\},
\end{aligned}
\end{equation*}
Here, $\boldsymbol{e}_r$ is a unit grid-index vector, and the
corresponding local coordinates are the index offsets
$\boldsymbol{s}\in\mathcal{S}_{\boldsymbol{m}}-\boldsymbol{m}$.
The ten coefficients follow directly from the ten sampled values:
\begin{equation*}
\begin{aligned}
    \boldsymbol{a}_0&=\boldsymbol{a}_{\boldsymbol{m}},\qquad
    \boldsymbol{a}_r
    =\frac{\boldsymbol{a}_{\boldsymbol{m}+\boldsymbol{e}_r}
           -\boldsymbol{a}_{\boldsymbol{m}-\boldsymbol{e}_r}}{2},\\
    \boldsymbol{a}_{rr}
    &=\frac{\boldsymbol{a}_{\boldsymbol{m}+\boldsymbol{e}_r}
            +\boldsymbol{a}_{\boldsymbol{m}-\boldsymbol{e}_r}}{2}
      -\boldsymbol{a}_{\boldsymbol{m}},\\
    \boldsymbol{a}_{rs}
    &=\boldsymbol{a}_{\boldsymbol{m}+\boldsymbol{e}_r+\boldsymbol{e}_s}
      -\boldsymbol{a}_{\boldsymbol{m}+\boldsymbol{e}_r}
      -\boldsymbol{a}_{\boldsymbol{m}+\boldsymbol{e}_s}
      +\boldsymbol{a}_{\boldsymbol{m}},\qquad r<s.
\end{aligned}
\end{equation*}
The center and six axial neighbors determine the constant, three
linear, and three pure-quadratic coefficients, giving seven
independent interpolation conditions. Each mixed-direction neighbor
fixes one of the remaining cross terms. These ten independent
conditions make the interpolation matrix nonsingular, so arbitrary
sampled values determine a unique polynomial, and every polynomial
of total degree at most two is reproduced exactly. The ten-point
stencil is therefore a minimal unisolvent set for a complete
three-dimensional quadratic representation from pointwise data.
Its geometry permits the explicit coefficient formulas above.

\begin{s2revision}
Analytic differentiation gives the center first, pure-second, and mixed-second derivatives as $\boldsymbol{a}_r$, $2\boldsymbol{a}_{rr}$, and $\boldsymbol{a}_{rs}$, respectively. The three off-axis samples therefore supply the cross-direction information needed by the divergence gradients. The explicit coefficient formulas determine the complete polynomial without a local matrix solve or an overdetermined fit. Limiting the sampling set to ten points also limits the pointwise flux and primitive-variable evaluations repeated during the local prediction.
\end{s2revision}

Applying these coefficient relations to
$\boldsymbol{a}_{\boldsymbol{m}+\boldsymbol{s}}
=\boldsymbol{W}_{\boldsymbol{m}+\boldsymbol{s}}^n$
gives the initial spatial field
\begin{equation*}
    \boldsymbol{W}^n(\boldsymbol{\xi})
    =\boldsymbol{W}_0^n
     +\sum_{r=1}^{3}\boldsymbol{W}_r^n\xi_r
     +\sum_{r=1}^{3}\boldsymbol{W}_{rr}^n\xi_r^2
     +\sum_{1\leq r<s\leq3}\boldsymbol{W}_{rs}^n\xi_r\xi_s.
\end{equation*}
\begin{s2revision}
This polynomial interpolates the accepted nodal data, $\boldsymbol{W}^n(\boldsymbol{s})=\boldsymbol{W}_{\boldsymbol{m}+\boldsymbol{s}}^n$, at every stencil point. Its ten known coefficients prescribe the initial spatial slice. The next step is to construct the flux-divergence information that determines its temporal prediction.
\end{s2revision}

To obtain the local operator, physical quantities are first
evaluated at the input states on this stencil. The same procedure
applies to the initial samples and, later, to provisional samples
from the target node's evolving local field. For convection, the
input values are the physical fluxes
$\boldsymbol{F}^{\mathrm{c}}_j
(\boldsymbol{W}_{\boldsymbol{m}+\boldsymbol{s}})$.
Applying Eq.~\eqref{eq:local_p2_polynomial_rewrite} to these values
gives their quadratic interpolant,
\begin{equation}
    \boldsymbol{F}^{\mathrm{c}}_j(\boldsymbol{\xi})
    =\boldsymbol{F}^{\mathrm{c}}_{j,0}
     +\sum_{r=1}^{3}\boldsymbol{F}^{\mathrm{c}}_{j,r}\xi_r
     +\sum_{r=1}^{3}\boldsymbol{F}^{\mathrm{c}}_{j,rr}\xi_r^2
     +\sum_{1\leq r<s\leq3}
       \boldsymbol{F}^{\mathrm{c}}_{j,rs}\xi_r\xi_s.
    \label{eq:local_convective_flux_polynomial_rewrite}
\end{equation}
\begin{s2revision}
The coefficient relations following Eq.~\eqref{eq:local_p2_polynomial_rewrite} are applied with $\boldsymbol{a}_{\boldsymbol{m}+\boldsymbol{s}}=\boldsymbol{F}^{\mathrm{c}}_j(\boldsymbol{W}_{\boldsymbol{m}+\boldsymbol{s}})$. Thus, physical-flux evaluation precedes interpolation. At a provisional time, the flux samples and all flux-polynomial coefficients are recomputed. This construction preserves the distinction between the conservative-state polynomial and the interpolant of nonlinear flux samples; the latter supplies the changing divergence used by the temporal correction.
\end{s2revision}

Molecular transport requires velocity and temperature gradients as
well as the state. We apply the same coefficient relations to the
sampled primitive values
$a_{\boldsymbol{m}+\boldsymbol{s}}
=u_i(\boldsymbol{W}_{\boldsymbol{m}+\boldsymbol{s}})$ and
$a_{\boldsymbol{m}+\boldsymbol{s}}
=T(\boldsymbol{W}_{\boldsymbol{m}+\boldsymbol{s}})$, obtaining
\begin{equation}
\begin{aligned}
    u_i(\boldsymbol{\xi})
    &=u_{i,0}
      +\sum_{r=1}^{3}u_{i,r}\xi_r
      +\sum_{r=1}^{3}u_{i,rr}\xi_r^2
      +\sum_{1\leq r<s\leq3}u_{i,rs}\xi_r\xi_s,\\
    T(\boldsymbol{\xi})
    &=T_0
      +\sum_{r=1}^{3}T_r\xi_r
      +\sum_{r=1}^{3}T_{rr}\xi_r^2
      +\sum_{1\leq r<s\leq3}T_{rs}\xi_r\xi_s.
\end{aligned}
\label{eq:local_primitive_polynomials_rewrite}
\end{equation}
\begin{s2revision}
The coefficient subscripts have the same meaning as in Eq.~\eqref{eq:local_p2_polynomial_rewrite}. Their analytic derivatives provide the velocity and temperature gradients and curvature entering stress and heat transport. These primitive polynomials are reconstructed whenever the input samples change.
\end{s2revision}

Before differentiating the viscous flux, the transport coefficient
fields must also be specified. The beginning-of-step center value
$\mu_{\boldsymbol{m}}^n=\mu(T_{\boldsymbol{m}}^n)$ is evaluated
from the constitutive law. At a later local evaluation, let
$T_\star$ denote the current center temperature and define
$\eta_\star=(T_\star-T_{\boldsymbol{m}}^n)/T_{\boldsymbol{m}}^n$.
The corresponding center transport values are
\begin{equation}
\begin{aligned}
    \widetilde{\mu}_{\boldsymbol{m}}^n(T_\star)
    &=\mu_{\boldsymbol{m}}^n
      \left[1+s_\mu\eta_\star
          +\frac{s_\mu(s_\mu-1)}{2}\eta_\star^2\right],\\
    \widetilde{\kappa}_{\boldsymbol{m}}^n(T_\star)
    &=\frac{\gamma}{(\gamma-1)\mathrm{Pr}}
      \widetilde{\mu}_{\boldsymbol{m}}^n(T_\star).
\end{aligned}
\label{eq:temporal_transport_rewrite}
\end{equation}
This temporal expansion is exact for constant viscosity and agrees
with the constitutive law through second order in $\eta_\star$.
For a smooth step, $\eta_\star=O(\Delta t)$, so the omitted value
is $O(\Delta t^3)$ and contributes $O(\Delta t^4)$ after time
integration. At the beginning of the step, $\eta_\star=0$.
The same evaluation is used for the corrected flux-quadrature
states constructed later.

The center value
$\mu_\star=\widetilde{\mu}_{\boldsymbol{m}}^n(T_\star)$ supplies
the reference for the local spatial property law,
\begin{equation*}
    \mu(\boldsymbol{\xi})
    =\mu_\star\left[\frac{T(\boldsymbol{\xi})}{T_\star}\right]^{s_\mu},
    \qquad
    \kappa(\boldsymbol{\xi})
    =\frac{\gamma}{(\gamma-1)\mathrm{Pr}}\mu(\boldsymbol{\xi}).
\end{equation*}
The spatial variation of the transport coefficients must therefore
be included when the viscous flux is differentiated. The required
center derivatives are
\begin{equation*}
    \mu_T=\frac{s_\mu\mu_\star}{T_\star},\qquad
    \mu_{TT}=\frac{s_\mu(s_\mu-1)\mu_\star}{T_\star^2},\qquad
    (\kappa_T,\kappa_{TT})
    =\frac{\gamma}{(\gamma-1)\mathrm{Pr}}(\mu_T,\mu_{TT}).
\end{equation*}
With the primitive polynomials and property fields specified,
analytic differentiation gives the stress, heat flux, and viscous
flux:
\begin{equation}
\begin{aligned}
    \tau_{ij}(\boldsymbol{\xi})
    &=\mu(\boldsymbol{\xi})\left[
       \frac{1}{h_j}\frac{\partial u_i(\boldsymbol{\xi})}{\partial\xi_j}
      +\frac{1}{h_i}\frac{\partial u_j(\boldsymbol{\xi})}{\partial\xi_i}
      -\frac{2}{3}\delta_{ij}\sum_{k=1}^{3}
       \frac{1}{h_k}\frac{\partial u_k(\boldsymbol{\xi})}{\partial\xi_k}
      \right],\\
    q_j(\boldsymbol{\xi})
    &=-\frac{\kappa(\boldsymbol{\xi})}{h_j}
       \frac{\partial T(\boldsymbol{\xi})}{\partial\xi_j},\\
    \boldsymbol{F}^{\mathrm{v}}_j(\boldsymbol{\xi})
    &=\left(
        0,\;\tau_{ij}(\boldsymbol{\xi}),\;
        \sum_{i=1}^{3}u_i(\boldsymbol{\xi})\tau_{ij}(\boldsymbol{\xi})
        -q_j(\boldsymbol{\xi})
      \right)^{\mathsf T}.
\end{aligned}
\label{eq:local_viscous_flux_construction_rewrite}
\end{equation}
\begin{s2revision}
The stress relation has free indices $i$ and $j$ and an explicit sum over $k$. Products in the constitutive laws make the viscous flux generally nonquadratic. Differentiating Eq.~\eqref{eq:local_viscous_flux_construction_rewrite}, including the displayed property derivatives, supplies its center divergence and spatial variation. The retained modes therefore include the velocity, temperature, and transport-coefficient variations required by the product rule, using the same local sampling geometry.
\end{s2revision}

The preceding constructions now provide the total physical flux
needed by the local evolution law. For either initial or provisional
input, we retain the constant and linear terms of its divergence:
\begin{equation}
\begin{aligned}
    \boldsymbol{R}_{\mathrm{loc}}[\boldsymbol{W}](\boldsymbol{\xi})
    &=\boldsymbol{R}_0+\sum_{r=1}^{3}\boldsymbol{R}_r\xi_r,\\
    \boldsymbol{R}_0
    &=\left.\sum_{j=1}^{3}\frac{1}{h_j}
      \frac{\partial}{\partial\xi_j}
      \left[\boldsymbol{F}^{\mathrm{c}}_j(\boldsymbol{\xi})
            -\boldsymbol{F}^{\mathrm{v}}_j(\boldsymbol{\xi})\right]
      \right|_{\boldsymbol{\xi}=\boldsymbol{0}},\\
    \boldsymbol{R}_r
    &=\left.\sum_{j=1}^{3}\frac{1}{h_j}
      \frac{\partial^2}{\partial\xi_r\,\partial\xi_j}
      \left[\boldsymbol{F}^{\mathrm{c}}_j(\boldsymbol{\xi})
            -\boldsymbol{F}^{\mathrm{v}}_j(\boldsymbol{\xi})\right]
      \right|_{\boldsymbol{\xi}=\boldsymbol{0}},\qquad r=1,2,3.
\end{aligned}
\label{eq:local_operator_modes_rewrite}
\end{equation}
\begin{s2revision}
The input field in square brackets determines the fluxes. The center mode $\boldsymbol{R}_0$ controls the center-state change, while $\boldsymbol{R}_r$ supplies the divergence variation in direction $\xi_r$. Its definition includes mixed flux derivatives when the differentiation directions differ. These terms represent multidirectional coupling in the local prediction. Because $\boldsymbol{R}_r$ multiplies a dimensionless coordinate, the divergence factor $h_j^{-1}$ requires no additional $h_r^{-1}$. The convective divergence is exactly linear for the quadratic interpolant; the viscous part retains its center and linear Taylor terms.
\end{s2revision}

Including the source gives the local evolution operator,
\begin{equation}
    \boldsymbol{H}[\boldsymbol{W}](\boldsymbol{\xi})
    =\boldsymbol{R}_{\mathrm{loc}}[\boldsymbol{W}](\boldsymbol{\xi})
     -\boldsymbol{S}[\boldsymbol{W}](\boldsymbol{\xi}).
    \label{eq:local_evolution_operator_rewrite}
\end{equation}
\begin{s2revision}
We use $\boldsymbol{H}$ exclusively for this compact physical operator in the predictor. For any specified initial or provisional input field, $-\boldsymbol{H}$ approximates the continuous evolution operator $\mathcal{L}$ locally. It is distinct from the assembled corrector $\overline{\mathcal{L}}^n$ in Eq.~\eqref{eq:complete_update_rewrite}, which uses averaged nodal fluxes, the wide derivative, and artificial dissipation. Replacing the input field changes the evaluated fluxes and divergence modes but leaves the definition of $\boldsymbol{H}$ unchanged.
\end{s2revision}

\begin{s2revision}
The compact predictor and wide corrector must now be connected at the beginning-of-step center. Differentiating the quadratic convective-flux interpolant there gives
\end{s2revision}
\begin{equation}
    \boldsymbol{R}_{\mathrm{loc}}^{\mathrm{c}}
       [\boldsymbol{W}^n](\boldsymbol{0})
    =\sum_{j=1}^{3}
      \frac{\boldsymbol{F}^{\mathrm{c}}_j
                (\boldsymbol{W}_{\boldsymbol{m}+\boldsymbol{e}_j}^n)
           -\boldsymbol{F}^{\mathrm{c}}_j
                (\boldsymbol{W}_{\boldsymbol{m}-\boldsymbol{e}_j}^n)}
           {2h_j}.
    \label{eq:initial_compact_divergence_rewrite}
\end{equation}
\begin{s2revision}
This center value is a sum of second-order directional central differences. Its spatial variation is supplied by the pure and mixed quadratic flux coefficients. The final corrector uses the radius-four derivative in Eq.~\eqref{eq:complete_update_rewrite}, defined by
\end{s2revision}
\begin{equation}
\begin{aligned}
    (\mathcal{D}_j\boldsymbol{a})_{\boldsymbol{m}}
    &=\frac{1}{h_j}\bigg[
       \frac{4}{5}\left(
          \boldsymbol{a}_{\boldsymbol{m}+\boldsymbol{e}_j}
         -\boldsymbol{a}_{\boldsymbol{m}-\boldsymbol{e}_j}\right)
      -\frac{1}{5}\left(
          \boldsymbol{a}_{\boldsymbol{m}+2\boldsymbol{e}_j}
         -\boldsymbol{a}_{\boldsymbol{m}-2\boldsymbol{e}_j}\right)\\
    &\qquad
      +\frac{4}{105}\left(
          \boldsymbol{a}_{\boldsymbol{m}+3\boldsymbol{e}_j}
         -\boldsymbol{a}_{\boldsymbol{m}-3\boldsymbol{e}_j}\right)
      -\frac{1}{280}\left(
          \boldsymbol{a}_{\boldsymbol{m}+4\boldsymbol{e}_j}
         -\boldsymbol{a}_{\boldsymbol{m}-4\boldsymbol{e}_j}\right)
      \bigg].
\end{aligned}
\label{eq:eighth_order_central_rewrite}
\end{equation}
Here, $\boldsymbol{a}$ is a generic nodal field. On a uniform grid,
we denote the one-dimensional compact and wide derivatives by $D_2$
and $D_8$, respectively. The index $j$ in $\mathcal{D}_j$ identifies
the coordinate direction, while the subscripts in $D_2$ and $D_8$
identify accuracy order. For a sufficiently smooth field, their
initial derivative values differ by $O(h^2)$. The Taylor estimates
and their contribution to the complete update are discussed in
Section~\ref{sec:accuracy_consistency_rewrite}.

To remove this initial center difference from the convective part
of the temporal construction, let
$\boldsymbol{R}_{\mathrm{loc}}
=\boldsymbol{R}_{\mathrm{loc}}^{\mathrm{c}}
-\boldsymbol{R}_{\mathrm{loc}}^{\mathrm{v}}$
separate the convective and viscous divergences and define
\begin{equation}
    \boldsymbol{C}_{\boldsymbol{m}}^n
    =\sum_{j=1}^{3}
       \bigl(\mathcal{D}_j\boldsymbol{F}_j^{\mathrm{c},n}\bigr)_{\boldsymbol{m}}
     -\boldsymbol{R}_{\mathrm{loc}}^{\mathrm{c}}
       [\boldsymbol{W}^n](\boldsymbol{0}),
    \label{eq:operator_matching_correction_rewrite}
\end{equation}
where
$\boldsymbol{F}_{j,\boldsymbol{m}}^{\mathrm{c},n}
=\boldsymbol{F}^{\mathrm{c}}_j(\boldsymbol{W}_{\boldsymbol{m}}^n)$.
Both terms use the known beginning-of-step state, so
$\boldsymbol{C}_{\boldsymbol{m}}^n$ is computed once per physical
step and held fixed during prediction. The matching condition
specifies the operator value at $\boldsymbol{\xi}=\boldsymbol{0}$.
Only its constant mode contributes at that location. A spatially
constant addition enforces this center-value condition while
preserving the linear coefficients:
\begin{equation*}
    \boldsymbol{R}_{\mathrm{loc}}[\boldsymbol{W}^n](\boldsymbol{\xi})
    +\boldsymbol{C}_{\boldsymbol{m}}^n
    =\left(\boldsymbol{R}_0+\boldsymbol{C}_{\boldsymbol{m}}^n\right)
     +\sum_{r=1}^{3}\boldsymbol{R}_r\xi_r.
\end{equation*}
\begin{s2revision}
The modes $\boldsymbol{R}_0$ and $\boldsymbol{R}_r$ are those of Eq.~\eqref{eq:local_operator_modes_rewrite}. Adding a spatially constant correction changes the center value and retains the directional modes that predict neighboring states and conservative gradients. The correction therefore enforces the specified center matching with no additional spatial coefficients. At the beginning-of-step center,
\end{s2revision}
\begin{equation*}
    \boldsymbol{H}[\boldsymbol{W}^n](\boldsymbol{0})
    +\boldsymbol{C}_{\boldsymbol{m}}^n
    =\sum_{j=1}^{3}
       \bigl(\mathcal{D}_j\boldsymbol{F}_j^{\mathrm{c},n}\bigr)_{\boldsymbol{m}}
     -\boldsymbol{R}_{\mathrm{loc}}^{\mathrm{v}}
       [\boldsymbol{W}^n](\boldsymbol{0})
     -\boldsymbol{S}_{\boldsymbol{m}}^n.
\end{equation*}
For the Euler equations, this matches the full physical center
operator. With molecular transport, the correction targets the
convective contribution. Its wide-stencil reference is available
directly from the initial conservative states. The viscous flux
also depends on velocity and temperature gradients, so its discrete
approximation is determined by the complete gradient--flux--divergence
construction. The order of the outer derivative
$\mathcal{D}_j$ alone does not determine that approximation's
accuracy.

\begin{s2revision}
For molecular transport, the analytic compact viscous divergence remains in the predictor, and the corrected states and gradients determine the final viscous fluxes. The frozen convective matching does not alter the ten-point provisional sampling geometry. In the temporal construction below, the compact physical operator first generates the starting prediction; the correction is then added when the reevaluated operators are assembled for the Picard update.
\end{s2revision}

\begin{s2revision}
The temporal predictor therefore receives three known ingredients: the initial spatial polynomial $\boldsymbol{W}^n(\boldsymbol{\xi})$, the compact physical operator $\boldsymbol{H}$, and the frozen center correction $\boldsymbol{C}_{\boldsymbol{m}}^n$. The following subsection specifies when each is used and constructs the corrected states and gradients required by the flux quadrature. Section~\ref{sec:accuracy_consistency_rewrite} assesses their combined contribution to the accepted update.
\end{s2revision}

\endgroup

\begin{revisiontext}
\colorlet{hscolor}{black}

\subsection{Taylor-initialized local space--time prediction}
\label{sec:local_ader_predictor_rewrite}

\begin{s2revision}
The flux quadrature in Eq.~\eqref{eq:temporal_quadrature_target_rewrite} needs two within-step center states and, for molecular transport, their velocity and temperature gradients. The spatial construction provides the initial field and its evolution information, but these temporal-node data are still unknown. We obtain them by forming a compact-operator Taylor starting iterate, reevaluating the physical operator on its samples, and integrating the matched operator response once. The resulting corrected polynomial supplies the temporal data consumed by the flux quadrature.

At the beginning of the step, $\boldsymbol{H}[\boldsymbol{W}^n]$ is known from Section~\ref{sec:local_spatial_operator_rewrite}. Its compact physical slope generates an explicit initial prediction. We reserve the frozen center correction for the subsequent Picard update and define the starting iterate and its Gauss-node fields by

\begin{equation}
\begin{aligned}
    \widetilde{\boldsymbol{W}}^{[0]}(\boldsymbol{\xi},\tau)
    &=\boldsymbol{W}^n(\boldsymbol{\xi})
      -\tau\Delta t\,
       \boldsymbol{H}[\boldsymbol{W}^n](\boldsymbol{\xi}),\\
    \widetilde{\boldsymbol{W}}_\alpha^{[0]}(\boldsymbol{\xi})
    &=\widetilde{\boldsymbol{W}}^{[0]}(\boldsymbol{\xi},c_\alpha),
      \qquad \alpha=1,2.
\end{aligned}
\label{eq:taylor_seed_rewrite}
\end{equation}

The index $[0]$ denotes this starting approximation, and $\tau\Delta t$ is the elapsed physical time. The polynomial reproduces the initial spatial field at $\tau=0$ and supplies provisional states at $\tau=c_\alpha$. Its first-order temporal dependence provides the input needed to reevaluate the changing compact physical operator. Center matching is enforced when that response is integrated by the correction, rather than by modifying the sampling law at this initialization stage. The matched integral is introduced below, and Section~\ref{sec:accuracy_consistency_rewrite} includes the contribution of this initialization choice.

The initial prediction must now supply a spatial sampling set at each temporal node, because reevaluating the evolution law requires flux divergence. Since $\boldsymbol{W}^n(\boldsymbol{s})=\boldsymbol{W}_{\boldsymbol{m}+\boldsymbol{s}}^n$ on the interpolation stencil, evaluating Eq.~\eqref{eq:taylor_seed_rewrite} at $\boldsymbol{\xi}=\boldsymbol{s}$ and $\tau=c_\beta$ gives

\begin{equation}
\begin{aligned}
    \widetilde{\boldsymbol{W}}_\beta^{[0]}(\boldsymbol{s})
    &=\boldsymbol{W}_{\boldsymbol{m}+\boldsymbol{s}}^n
      -c_\beta\Delta t
       \left(\boldsymbol{R}_0+\sum_{r=1}^{3}\boldsymbol{R}_r s_r
             -\boldsymbol{S}[\boldsymbol{W}^n](\boldsymbol{s})\right),\\
    &\hspace{24mm}
      \boldsymbol{s}\in\mathcal{S}_{\boldsymbol{m}}-\boldsymbol{m},
      \qquad \beta=1,2.
\end{aligned}
\label{eq:provisional_samples_rewrite}
\end{equation}

Here, $\boldsymbol{R}_0$ and $\boldsymbol{R}_r$ are the beginning-of-step modes centered at $\boldsymbol{m}$. They are shared by all samples in that target-node construction; the offsets determine the local spatial variation. The values in Eq.~\eqref{eq:provisional_samples_rewrite} come from this node's own polynomial, rather than a separate temporal update at each neighboring node. Each Gauss node therefore provides a complete local input set for reevaluating the physical operator.

Applying the spatial construction of Section~\ref{sec:local_spatial_operator_rewrite} to these provisional states updates the flux samples and divergence modes. For example, the center convective divergence becomes

\begin{equation}
\boldsymbol{R}_{\mathrm{loc}}^{\mathrm{c}}
       [\widetilde{\boldsymbol{W}}_\beta^{[0]}](\boldsymbol{0})
    =\sum_{j=1}^{3}
      \frac{\boldsymbol{F}^{\mathrm{c}}_j
              (\widetilde{\boldsymbol{W}}_\beta^{[0]}(\boldsymbol{e}_j))
           -\boldsymbol{F}^{\mathrm{c}}_j
              (\widetilde{\boldsymbol{W}}_\beta^{[0]}(-\boldsymbol{e}_j))}
           {2h_j}.
\label{eq:provisional_compact_divergence_rewrite}
\end{equation}

The surrounding fluxes determine the center mode, and the mixed-direction samples supply the linear divergence modes. For molecular transport, the provisional states also provide new velocity and temperature polynomials, property values, and analytic derivatives. The physical source is evaluated from the same input field. The frozen center matching is then added to each reevaluated operator:

\begin{equation}
\widehat{\boldsymbol{H}}_\beta^{[0]}(\boldsymbol{\xi})
    =\boldsymbol{H}[\widetilde{\boldsymbol{W}}_\beta^{[0]}]
       (\boldsymbol{\xi})+\boldsymbol{C}_{\boldsymbol{m}}^n,
    \qquad \beta=1,2.
\label{eq:matched_operator_rewrite}
\end{equation}

Both temporal evaluations use the same $\boldsymbol{C}_{\boldsymbol{m}}^n$, computed from the initial field. Their physical parts are recomputed from different provisional states. Thus, the constant matching supplies the initial center connection to the wide derivative, while the changing compact operator supplies the response within the step.

With these matched samples available, one Picard correction approximates the local integral relation

\begin{equation}
\widetilde{\boldsymbol{W}}(\boldsymbol{\xi},\tau)
    =\boldsymbol{W}^n(\boldsymbol{\xi})
     -\Delta t\int_0^\tau
       \left(
       \boldsymbol{H}[\widetilde{\boldsymbol{W}}(\,\cdot\,,\tau')]
          (\boldsymbol{\xi})
       +\boldsymbol{C}_{\boldsymbol{m}}^n
       \right)\,\mathrm{d}\tau'.
\label{eq:local_spacetime_integral_rewrite}
\end{equation}

The dot denotes the local spatial field at a fixed time. We evaluate the integrand using the starting prediction and represent its temporal dependence by the linear interpolant through the two values in Eq.~\eqref{eq:matched_operator_rewrite}. The Lagrange basis and its partial-step integration coefficients are

\begin{equation*}
\ell_1(\tau)=\frac{c_2-\tau}{c_2-c_1},
    \qquad
    \ell_2(\tau)=\frac{\tau-c_1}{c_2-c_1},
    \qquad
    A_{\alpha\beta}
       =\int_0^{c_\alpha}\ell_\beta(\tau)\,\mathrm{d}\tau.
\end{equation*}

The operator samples are values, rather than time-integrated quantities. Their interpolation gives the continuous approximation needed inside the integral:

\begin{equation*}
\boldsymbol{H}[\widetilde{\boldsymbol{W}}^{[0]}(\,\cdot\,,\tau)]
       (\boldsymbol{\xi})+\boldsymbol{C}_{\boldsymbol{m}}^n
    \approx\sum_{\beta=1}^{2}\ell_\beta(\tau)
       \widehat{\boldsymbol{H}}_\beta^{[0]}(\boldsymbol{\xi}).
\end{equation*}

Integrating this linear-in-time operator from the prescribed initial field produces a quadratic-in-time state polynomial. The explicit Picard correction is \citep{DumbserEtAl2008,BalsaraEtAl2013}

\begin{equation}
\widetilde{\boldsymbol{W}}^{[1]}(\boldsymbol{\xi},\tau)
    =\boldsymbol{W}^n(\boldsymbol{\xi})
      -\Delta t\sum_{\beta=1}^{2}
       \left[\int_0^\tau\ell_\beta(\tau')\,\mathrm{d}\tau'\right]
       \widehat{\boldsymbol{H}}_\beta^{[0]}(\boldsymbol{\xi}).
\label{eq:corrected_spacetime_polynomial_rewrite}
\end{equation}

The temporal increment vanishes at $\tau=0$, preserving the initial spatial field. Only its values at the two Gauss nodes are needed for flux quadrature. Evaluating Eq.~\eqref{eq:corrected_spacetime_polynomial_rewrite} at those nodes gives

\begin{equation}
\widetilde{\boldsymbol{W}}_\alpha^{[1]}(\boldsymbol{\xi})
    =\boldsymbol{W}^n(\boldsymbol{\xi})
     -\Delta t\sum_{\beta=1}^{2}A_{\alpha\beta}
      \widehat{\boldsymbol{H}}_\beta^{[0]}(\boldsymbol{\xi}),
    \qquad \alpha=1,2.
\label{eq:one_picard_predictor_rewrite}
\end{equation}

The coefficients are fixed by the Gauss-node geometry:

\begin{equation*}
\boldsymbol{A}
    =\begin{bmatrix}
        \dfrac{1}{4}&\dfrac{1}{4}-\dfrac{\sqrt{3}}{6}\\[4pt]
        \dfrac{1}{4}+\dfrac{\sqrt{3}}{6}&\dfrac{1}{4}
      \end{bmatrix}.
\end{equation*}

The two corrected fields are evaluations of one space--time polynomial. Since $\sum_\beta\ell_\beta(\tau)=1$, the fixed matching term contributes exactly $-\tau\Delta t\boldsymbol{C}_{\boldsymbol{m}}^n$ to that polynomial. The provisional operator reevaluations supply its remaining temporal response. The matrix moments and their contribution to the complete update are examined in Section~\ref{sec:accuracy_consistency_rewrite}.

For source-free flow, the corrected spatial coefficients make the output of this construction explicit:

\begin{equation}
\begin{aligned}
    \widetilde{\boldsymbol{W}}^{[1]}(\boldsymbol{\xi},\tau)
    &=\widetilde{\boldsymbol{W}}_0^{[1]}(\tau)
      +\sum_{r=1}^{3}\widetilde{\boldsymbol{W}}_r^{[1]}(\tau)\xi_r\\
    &\quad+\sum_{r=1}^{3}\boldsymbol{W}_{rr}^n\xi_r^2
      +\sum_{1\leq r<s\leq3}\boldsymbol{W}_{rs}^n\xi_r\xi_s.
\end{aligned}
\label{eq:corrected_spatial_modes_rewrite}
\end{equation}

The center and linear coefficients evolve quadratically in time because the retained divergence has constant and linear spatial modes. Its temporal integral therefore changes those state coefficients, while the quadratic spatial coefficients remain those of the initial field. This does not freeze the flux polynomials: their quadratic and mixed coefficients are recomputed from each provisional sampling set and enter the reevaluated divergence. Physical sources contribute through $\boldsymbol{H}$ to the modes required by the predictor.

At each temporal node, the corrected constant coefficient supplies the conservative center state; the corrected linear coefficients supply the conservative derivatives from which the primitive gradients are recovered. These are the inputs needed by Section~\ref{sec:time_averaged_fluxes_rewrite}. The local construction has thus supplied the physical-flux quadrature data for Eq.~\eqref{eq:complete_update_rewrite}, whose global spatial correction follows after those averages have been formed.
\end{s2revision}

\subsection{Time-averaged physical fluxes and corrector data}
\label{sec:time_averaged_fluxes_rewrite}

\begin{s2revision}
The predictor has now supplied the temporal-node fields. To form the flux averages required by Eq.~\eqref{eq:complete_update_rewrite}, we extract their corrected states at the target center:
\end{s2revision}
\begin{equation}
    \widetilde{\boldsymbol{W}}_{\alpha,\boldsymbol{m}}
    =\widetilde{\boldsymbol{W}}_\alpha^{[1]}(\boldsymbol{0}),
    \qquad \alpha=1,2.
    \label{eq:corrected_center_states_rewrite}
\end{equation}

\begin{s2revision}
These center states determine the Euler flux. Molecular transport additionally requires the velocity and temperature gradients at the same temporal nodes. They are obtained from the corrected conservative center and linear coefficients, so the final physical-flux evaluation uses the state and gradients produced by the same temporal correction.
\end{s2revision}

The conversion to primitive gradients follows the state relations in
Section~\ref{sec:governing_equations_rewrite}. Suppressing the
temporal-node index, the following identities apply to one corrected
local state and its conservative derivatives:
\begin{equation}
\begin{aligned}
    \frac{\partial u_i}{\partial\xi_j}
    &=\frac{1}{\rho}\left[
        \frac{\partial(\rho u_i)}{\partial\xi_j}
        -u_i\frac{\partial\rho}{\partial\xi_j}\right],\\
    \frac{\partial p}{\partial\xi_j}
    &=(\gamma-1)\left[
        \frac{\partial(\rho E)}{\partial\xi_j}
        -u_i\frac{\partial(\rho u_i)}{\partial\xi_j}
        +\frac{1}{2}u_i u_i\frac{\partial\rho}{\partial\xi_j}
      \right],\\
    \frac{\partial T}{\partial\xi_j}
    &=\frac{1}{\rho}\left[
        \frac{\partial p}{\partial\xi_j}
        -T\frac{\partial\rho}{\partial\xi_j}\right].
\end{aligned}
\label{eq:primitive_gradient_conversion_rewrite}
\end{equation}

\begin{s2revision}
The index $i$ is summed in the pressure relation. At the center, the conservative derivatives are the corrected linear coefficients; the conversion therefore requires only those coefficients and the corrected center state.
\end{s2revision}

The physical gradients entering molecular transport are
\begin{equation}
\begin{aligned}
    \bigl[(\nabla\boldsymbol{u})_{\alpha,\boldsymbol{m}}\bigr]_{ij}
    &=\left.\frac{1}{h_j}\frac{\partial}{\partial\xi_j}
       u_i\!\left(\widetilde{\boldsymbol{W}}_\alpha^{[1]}
                    (\boldsymbol{\xi})\right)
      \right|_{\boldsymbol{\xi}=\boldsymbol{0}},\\
    \bigl[(\nabla T)_{\alpha,\boldsymbol{m}}\bigr]_j
    &=\left.\frac{1}{h_j}\frac{\partial}{\partial\xi_j}
       T\!\left(\widetilde{\boldsymbol{W}}_\alpha^{[1]}
                    (\boldsymbol{\xi})\right)
      \right|_{\boldsymbol{\xi}=\boldsymbol{0}},
      \qquad \alpha=1,2.
\end{aligned}
\label{eq:predicted_physical_gradients_rewrite}
\end{equation}

\begin{s2revision}
Here $h_j^{-1}$ converts the local derivative to physical units, without summation over $j$. These gradients belong to the corrected conservative polynomial. The sampled primitive polynomials used in earlier operator evaluations serve a different purpose: they determine the provisional divergence. Using the corrected gradients in the final flux keeps its molecular contribution consistent with the corrected center state.
\end{s2revision}

At each temporal node, the total physical flux is formed from this
state, its gradients, and the corresponding transport coefficients:
\begin{equation*}
    \mu_{\alpha,\boldsymbol{m}}
    =\widetilde{\mu}_{\boldsymbol{m}}^n
       \!\left(T(\widetilde{\boldsymbol{W}}_{\alpha,\boldsymbol{m}})\right),
    \qquad
    \kappa_{\alpha,\boldsymbol{m}}
    =\frac{\gamma\mu_{\alpha,\boldsymbol{m}}}
           {(\gamma-1)\mathrm{Pr}}.
\end{equation*}

The viscosity evaluation is the one specified in
Eq.~\eqref{eq:temporal_transport_rewrite}; the thermal conductivity
uses the same constitutive relation. Both the convective and molecular
contributions are evaluated before temporal quadrature. In the Euler
specialization, the molecular contribution vanishes and these final
primitive-gradient and transport-coefficient evaluations are omitted.

An overbar in Eq.~\eqref{eq:complete_update_rewrite} denotes the
computed approximation to an average over the current physical step.
For a nodal quantity $\boldsymbol{a}$, the quadrature convention is
\begin{equation*}
    \frac{1}{\Delta t}\int_{t^n}^{t^{n+1}}
       \boldsymbol{a}_{\boldsymbol{m}}(t)\,\mathrm{d}t
    \;\approx\;
    \sum_{\alpha=1}^{2}w_\alpha
       \boldsymbol{a}_{\alpha,\boldsymbol{m}},
    \qquad w_1=w_2=\frac{1}{2}.
\end{equation*}

Applying this convention to the corrected states, the already evaluated
physical fluxes, and the physical source gives
\begin{align}
    \overline{\boldsymbol{W}}_{\boldsymbol{m}}^n
    &=\sum_{\alpha=1}^{2}w_\alpha
       \widetilde{\boldsymbol{W}}_{\alpha,\boldsymbol{m}},\notag\\
    \overline{\boldsymbol{F}}_{j,\boldsymbol{m}}^n
    &=\sum_{\alpha=1}^{2}w_\alpha\,
       \boldsymbol{F}_j\!\left(
          \widetilde{\boldsymbol{W}}_{\alpha,\boldsymbol{m}},
          (\nabla\boldsymbol{u})_{\alpha,\boldsymbol{m}},
          (\nabla T)_{\alpha,\boldsymbol{m}};
          \mu_{\alpha,\boldsymbol{m}}\right),\notag\\
    \overline{\boldsymbol{S}}_{\boldsymbol{m}}^n
    &=\sum_{\alpha=1}^{2}w_\alpha\,
       \boldsymbol{S}\!\left(
          \widetilde{\boldsymbol{W}}_{\alpha,\boldsymbol{m}},
          \boldsymbol{x}_{\boldsymbol{m}},t^n+c_\alpha\Delta t
       \right).
\label{eq:time_averaged_quantities_rewrite}
\end{align}

Here, $\boldsymbol{F}_j=\boldsymbol{F}^{\mathrm{c}}_j-
\boldsymbol{F}^{\mathrm{v}}_j$ denotes the total physical flux. Its
arguments specify the state, primitive gradients, and constitutive
values used at each temporal node. The source is evaluated at
$\boldsymbol{x}_{\boldsymbol{m}}$, corresponding to the same local
center, and at the indicated physical time.

\begin{s2revision}
For a nonlinear flux, each temporal-node flux is evaluated before averaging. In general, evaluating a flux from the averaged conservative state does not reproduce this quadrature. The averaged physical flux supplies the central derivative, the averaged state supplies the dissipative differences, and the averaged source enters the accepted update directly. For the implemented linear body force with a coefficient fixed within the step, source quadrature is equivalently evaluated from $\overline{\boldsymbol{W}}_{\boldsymbol{m}}^n$.
\end{s2revision}

The directional propagation speeds needed by the dissipative flux
are averaged with the same temporal weights:
\begin{equation}
    \overline{\lambda}_{j,\boldsymbol{m}}^n
    =\sum_{\alpha=1}^{2}w_\alpha
       \lambda_j(\widetilde{\boldsymbol{W}}_{\alpha,\boldsymbol{m}}),
    \qquad
    \lambda_j(\boldsymbol{W})=|u_j(\boldsymbol{W})|+c(\boldsymbol{W}).
    \label{eq:averaged_signal_radius_rewrite}
\end{equation}

Each direction retains its own normal velocity and sound-speed
contribution. These averages provide the directional scaling data for
the face construction in Section~\ref{sec:global_corrector_rewrite};
the physical-flux average, conservative-state average, and signal-radius
average retain their distinct roles in the accepted update.

\begin{s2revision}
The averaged state, physical flux, source, and directional signal radii now form the complete nodal input to the conservative corrector. Their compact storage is described in Section~\ref{sec:gpu_resident_execution}.
\end{s2revision}

\begin{s2revision}
Repeating the prediction at every node produces one physical-flux average associated with each grid location. The wide corrector at $\boldsymbol{m}$ reads the center averages produced by the predictors at its stencil neighbors. This shared nodal field is the spatial input to Eq.~\eqref{eq:complete_update_rewrite}; the provisional samples used internally by each predictor have already served their operator-reevaluation role. The next subsection constructs the remaining dissipative contribution from the averaged states, signal radii, and initial sensor.
\end{s2revision}

\subsection{Selective dissipation and conservative update}
\label{sec:global_corrector_rewrite}

\begin{s2revision}
The averages constructed above supply the physical-flux and source terms in Eq.~\eqref{eq:complete_update_rewrite}. The wide central derivative provides no built-in short-wave damping, so selective artificial dissipation \citep{Jameson1981} completes the update. A beginning-of-step sensor activates the stronger branch in compressive regions, while the averaged state and directional signal radii set the dissipative flux. This construction controls unresolved oscillations and retains the high-order background branch in smooth regions.
\end{s2revision}

\begin{s2revision}
The sensor is evaluated from the initial field and remains fixed within the physical step. Artificial dissipation is assembled from the averaged corrector data after the physical-flux quadrature; it does not enter the local physical-operator reevaluation. The activation factors are
\end{s2revision}
\begin{align}
    \nu_{\boldsymbol{m}}^n
    &=\chi_{p,\boldsymbol{m}}^n
      \chi_{D,\boldsymbol{m}}^n
      \chi_{c,\boldsymbol{m}}^n,\notag\\
    \chi_{p,\boldsymbol{m}}^n
    &=\max_{j=1,2,3}\left|
      \frac{
        p_{\boldsymbol{m}+\boldsymbol{e}_j}^n
        -2p_{\boldsymbol{m}}^n
        +p_{\boldsymbol{m}-\boldsymbol{e}_j}^n
      }{
        p_{\boldsymbol{m}+\boldsymbol{e}_j}^n
        +2p_{\boldsymbol{m}}^n
        +p_{\boldsymbol{m}-\boldsymbol{e}_j}^n
      }\right|,\notag\\
    \chi_{D,\boldsymbol{m}}^n
    &=\frac{(\theta_{\boldsymbol{m}}^n)^2}{
        (\theta_{\boldsymbol{m}}^n)^2
        +|\boldsymbol{\omega}_{\boldsymbol{m}}^n|^2+\varepsilon_s},\notag\\
    \chi_{c,\boldsymbol{m}}^n
    &=\frac{1}{2}\left[
       1-\tanh\left(2.5+10h_{\min}
             \frac{\theta_{\boldsymbol{m}}^n}{c_{\boldsymbol{m}}^n}
            \right)\right].
\label{eq:dissipation_sensor_rewrite}
\end{align}

Here, $\theta=\nabla\!\cdot\boldsymbol{u}$ and
$\boldsymbol{\omega}=\nabla\!\times\boldsymbol{u}$ are evaluated
from the initial velocity using $\mathcal{D}_j$,
$h_{\min}=\min_jh_j$, and $\varepsilon_s=10^{-16}$.
The pressure-curvature factor identifies rapid compressive variation.
The dilatation--vorticity factor suppresses activation in predominantly
vortical regions \citep{Ducros1999,BhagatwalaLele2009}, while the
compression switch distinguishes compression from expansion. The last
factor can be set to unity when this additional discrimination is not
used.

\begin{s2revision}
The frozen sensor and the time-averaged state and propagation speeds have complementary roles: the former selects the dissipative branch, and the latter determine the differences and scaling entering its face flux.
\end{s2revision}

At the positive algebraic face normal to direction $j$, the signal
radius and the activated and background coefficients are
\begin{equation}
\begin{aligned}
    \lambda_{j,\boldsymbol{m}+\frac{1}{2}\boldsymbol{e}_j}^n
    &=\max\left(
       \overline{\lambda}_{j,\boldsymbol{m}}^n,
       \overline{\lambda}_{j,\boldsymbol{m}+\boldsymbol{e}_j}^n
       \right),\\
    \epsilon_{2,\boldsymbol{m}+\frac{1}{2}\boldsymbol{e}_j}^n
    &=k_2\max\left(
       \nu_{\boldsymbol{m}}^n,
       \nu_{\boldsymbol{m}+\boldsymbol{e}_j}^n\right),\\
    \epsilon_{8,\boldsymbol{m}+\frac{1}{2}\boldsymbol{e}_j}^n
    &=\max\left(0,\,
       k_8-\frac{3}{70}
       \epsilon_{2,\boldsymbol{m}+\frac{1}{2}\boldsymbol{e}_j}^n
       \right).
\end{aligned}
\label{eq:dissipation_coefficients_rewrite}
\end{equation}

The parameters $k_2$ and $k_8$ control the two branches. Taking the
sensor maximum at the face makes activation respond to either adjacent
node. The coupling of the coefficients reduces the background branch
as the activated contribution increases. The signal radius remains
specific to the face-normal direction and uses the averages in
Eq.~\eqref{eq:averaged_signal_radius_rewrite}.

The face flux acts on the following undivided differences of the
averaged conservative state:
\begin{equation}
\begin{aligned}
    \delta_j^1\overline{\boldsymbol{W}}_{
          \boldsymbol{m}+\frac{1}{2}\boldsymbol{e}_j}^n
    &=\overline{\boldsymbol{W}}_{\boldsymbol{m}+\boldsymbol{e}_j}^n
      -\overline{\boldsymbol{W}}_{\boldsymbol{m}}^n,\\
    \delta_j^7\overline{\boldsymbol{W}}_{
          \boldsymbol{m}+\frac{1}{2}\boldsymbol{e}_j}^n
    &=-\overline{\boldsymbol{W}}_{\boldsymbol{m}-3\boldsymbol{e}_j}^n
      +7\overline{\boldsymbol{W}}_{\boldsymbol{m}-2\boldsymbol{e}_j}^n
      -21\overline{\boldsymbol{W}}_{\boldsymbol{m}-\boldsymbol{e}_j}^n
      +35\overline{\boldsymbol{W}}_{\boldsymbol{m}}^n\\
    &\quad
      -35\overline{\boldsymbol{W}}_{\boldsymbol{m}+\boldsymbol{e}_j}^n
      +21\overline{\boldsymbol{W}}_{\boldsymbol{m}+2\boldsymbol{e}_j}^n
      -7\overline{\boldsymbol{W}}_{\boldsymbol{m}+3\boldsymbol{e}_j}^n
      +\overline{\boldsymbol{W}}_{\boldsymbol{m}+4\boldsymbol{e}_j}^n.
\end{aligned}
\label{eq:undivided_differences_rewrite}
\end{equation}

Together, these differences define the dissipative face flux and its
nodal contribution:
\begin{equation}
\begin{aligned}
    \boldsymbol{d}_{j,\boldsymbol{m}+\frac{1}{2}\boldsymbol{e}_j}^n
    &=\lambda_{j,\boldsymbol{m}+\frac{1}{2}\boldsymbol{e}_j}^n
      \left[
        \epsilon_{2,\boldsymbol{m}+\frac{1}{2}\boldsymbol{e}_j}^n
        \delta_j^1\overline{\boldsymbol{W}}_{
             \boldsymbol{m}+\frac{1}{2}\boldsymbol{e}_j}^n
        -\epsilon_{8,\boldsymbol{m}+\frac{1}{2}\boldsymbol{e}_j}^n
        \delta_j^7\overline{\boldsymbol{W}}_{
             \boldsymbol{m}+\frac{1}{2}\boldsymbol{e}_j}^n
      \right],\\
    \overline{\mathcal{A}}_{\boldsymbol{m}}^n
    &=\sum_{j=1}^{3}\frac{
        \boldsymbol{d}_{j,\boldsymbol{m}+\frac{1}{2}\boldsymbol{e}_j}^n
        -\boldsymbol{d}_{j,\boldsymbol{m}-\frac{1}{2}\boldsymbol{e}_j}^n
       }{h_j}.
\end{aligned}
\label{eq:selective_dissipation_rewrite}
\end{equation}

\begin{s2revision}
The seventh-difference sign gives a dissipative eighth-difference contribution after adjacent faces are differenced. The overbar on $\overline{\mathcal{A}}_{\boldsymbol{m}}^n$ identifies a nodal term assembled from averaged states and speeds. It is not a separate temporal quadrature of artificial fluxes.
\end{s2revision}

Where the activated branch is negligible, the remaining background
dissipation is $O(h^7)$; in constant-pressure regions the
pressure-curvature sensor vanishes. In smooth compressive regions,
normalized pressure curvature is $O(h^2)$ and the activated
second-difference contribution is third-order consistent. Near a
shock, the sensor factors can become $O(1)$ and activate the stronger
branch. These conditions distinguish the background and activated
regimes without assigning the same spatial error to both.

\begin{s2revision}
Specifying $\overline{\mathcal{A}}^n$ completes the right-hand side of Eq.~\eqref{eq:complete_update_rewrite}. To establish conservation, we express its central and dissipative contributions as shared face-flux differences.
\end{s2revision}

The central derivative has an exact algebraic face-flux
representation:
\begin{equation*}
    \bigl(\mathcal{D}_j\overline{\boldsymbol{F}}_j^n\bigr)_{\boldsymbol{m}}
    =\frac{
       \widehat{\boldsymbol{F}}_{j,\boldsymbol{m}+\frac{1}{2}\boldsymbol{e}_j}^n
       -\widehat{\boldsymbol{F}}_{j,\boldsymbol{m}-\frac{1}{2}\boldsymbol{e}_j}^n
      }{h_j},
\end{equation*}

where
\begin{equation}
\begin{aligned}
    \widehat{\boldsymbol{F}}_{j,\boldsymbol{m}+\frac{1}{2}\boldsymbol{e}_j}^n
    &=\frac{1}{840}\bigg[
       533\left(
         \overline{\boldsymbol{F}}_{j,\boldsymbol{m}}^n
        +\overline{\boldsymbol{F}}_{j,\boldsymbol{m}+\boldsymbol{e}_j}^n\right)
      -139\left(
         \overline{\boldsymbol{F}}_{j,\boldsymbol{m}-\boldsymbol{e}_j}^n
        +\overline{\boldsymbol{F}}_{j,\boldsymbol{m}+2\boldsymbol{e}_j}^n\right)\\
    &\qquad
      +29\left(
         \overline{\boldsymbol{F}}_{j,\boldsymbol{m}-2\boldsymbol{e}_j}^n
        +\overline{\boldsymbol{F}}_{j,\boldsymbol{m}+3\boldsymbol{e}_j}^n\right)
      -3\left(
         \overline{\boldsymbol{F}}_{j,\boldsymbol{m}-3\boldsymbol{e}_j}^n
        +\overline{\boldsymbol{F}}_{j,\boldsymbol{m}+4\boldsymbol{e}_j}^n\right)
      \bigg].
\end{aligned}
\label{eq:central_face_flux_rewrite}
\end{equation}

The half-index denotes an algebraic flux associated with the nodal
central difference. The formula uses the existing averaged physical
fluxes and requires neither a Riemann problem nor a separate interface
reconstruction. Combining this representation with the dissipative
face flux gives the conservative form of the accepted update:
\begin{equation}
\begin{aligned}
    \boldsymbol{W}_{\boldsymbol{m}}^{n+1}
    &=\boldsymbol{W}_{\boldsymbol{m}}^n
      -\Delta t\sum_{j=1}^{3}\frac{1}{h_j}
       \bigg[
       \left(\widehat{\boldsymbol{F}}_j^n-\boldsymbol{d}_j^n\right)_{
              \boldsymbol{m}+\frac{1}{2}\boldsymbol{e}_j}
       -\left(\widehat{\boldsymbol{F}}_j^n-\boldsymbol{d}_j^n\right)_{
              \boldsymbol{m}-\frac{1}{2}\boldsymbol{e}_j}
       \bigg]
      +\Delta t\,\overline{\boldsymbol{S}}_{\boldsymbol{m}}^n.
\end{aligned}
\label{eq:conservative_face_update_rewrite}
\end{equation}

An internal face contributes with opposite signs to its two adjacent
nodes. For periodic boundaries and no physical source, these
contributions cancel in the domain sum, conserving every component of
$\boldsymbol{W}$ to floating-point roundoff. With physical boundaries,
the net change is determined by the prescribed boundary fluxes and
physical source. Center matching influences the predicted flux values;
the common face-flux difference is the source of conservation.

\begin{s2revision}
Algorithm~\ref{alg:complete_update_rewrite} summarizes the construction in dependency order: beginning-of-step spatial data, provisional operator evaluations, one Picard correction, temporal quadrature, and one global conservative update. The required stencil data are completed before their respective spatial operations.
\end{s2revision}

\begin{algorithmblock}{One local ADER prediction and conservative update}
\label{alg:complete_update_rewrite}
\algline{1}{Given $\boldsymbol{W}^n$, compute $\Delta t$ using
Eq.~\eqref{eq:cfl_time_step_rewrite} and complete the required
beginning-of-step stencil data.}
\algline{2}{\textbf{for each} owned node $\boldsymbol{m}$
\textbf{in parallel do}}
\algline{3}{Evaluate the nodal fluxes and primitive variables; construct
the four local flux-divergence modes and include the physical source
using Section~\ref{sec:local_spatial_operator_rewrite}.}
\algline{4}{Compute $\boldsymbol{C}_{\boldsymbol{m}}^n$ and
$\nu_{\boldsymbol{m}}^n$ from Eqs.~\eqref{eq:operator_matching_correction_rewrite}
and \eqref{eq:dissipation_sensor_rewrite}.}
\algline{5}{Form $\widetilde{\boldsymbol{W}}_\alpha^{[0]}$ at both
temporal nodes using Eq.~\eqref{eq:taylor_seed_rewrite}; reevaluate the
local operators and add the frozen matching correction.}
\algline{6}{Apply Eq.~\eqref{eq:one_picard_predictor_rewrite} to obtain
$\widetilde{\boldsymbol{W}}_\alpha^{[1]}$, then recover its center
state and, when needed, the primitive gradients.}
\algline{7}{Form the averaged state, physical flux, source contribution,
and directional signal radii using
Eqs.~\eqref{eq:time_averaged_quantities_rewrite} and
\eqref{eq:averaged_signal_radius_rewrite}. Release the local predictor
data. \textbf{end for}}
\algline{8}{Complete the neighboring averaged corrector data and sensor
values required by the final spatial stencil.}
\algline{9}{\textbf{for each} owned node $\boldsymbol{m}$
\textbf{in parallel do} Assemble the central flux derivative and
dissipative contribution, then update
$\boldsymbol{W}_{\boldsymbol{m}}^{n+1}$ using
Eq.~\eqref{eq:complete_update_rewrite}. \textbf{end for}}
\algline{10}{Set $t^{n+1}=t^n+\Delta t$ and apply the physical boundary
conditions for the accepted state.}
\end{algorithmblock}

\begin{s2revision}
The completed step changes the accepted solution once, using the averaged nodal fields supplied by local prediction. Section~\ref{sec:accuracy_consistency_rewrite} assesses the temporal moments, center matching, and compact spatial response together. The implementation of these data dependencies is described in Section~\ref{sec:gpu_performance}.
\end{s2revision}

\subsection{Accuracy and consistency}
\label{sec:accuracy_consistency_rewrite}

\begin{s2revision}
The accuracy statement for the implemented method concerns the complete update, including its compact prediction, center matching, wide physical-flux derivative, and selective dissipation. For smooth Euler flow on a uniform periodic grid, the estimate below establishes third-order coupled convergence under fixed-CFL refinement and the stated stability conditions. We explain this result through three contributions: the temporal moments of the Taylor--Picard construction, the integrated-flux term removed by center matching, and the spatial response retained in the actual predictor.
\end{s2revision}

\begin{s2revision}
Remark~\ref{rem:temporal_order_rewrite} first examines a common-operator reference to identify the temporal coefficients generated by the explicit correction and Gauss quadrature. The matching calculation then connects the compact and wide center derivatives. Finally, Remark~\ref{rem:complete_euler_consistency_rewrite} retains the implemented compact response and combines all contributions in the complete Euler estimate. The reference moment calculation and the complete-update estimate consequently have distinct, complementary roles.
\end{s2revision}

\begin{remarkblock}[manuscriptrevision]{Temporal moment conditions of the Taylor-initialized Picard construction}
\label{rem:temporal_order_rewrite}
This remark identifies the temporal moment conditions of the
Taylor-initialized Picard construction using the smooth autonomous
reference problem
$\mathrm{d}\boldsymbol{W}/\mathrm{d}t=\mathcal{L}_{\mathrm{ref}}[\boldsymbol{W}]$.
The initial prediction, Picard evaluation, and final quadrature
use a common evolution operator for this temporal analysis.
For the compact local reference, $\mathcal{L}_{\mathrm{ref}}=-\boldsymbol{H}$.
Write $\mathcal{L}_{\mathrm{ref},n}=\mathcal{L}_{\mathrm{ref}}[\boldsymbol{W}^n]$, and denote its
first and second Fr\'echet derivatives at $\boldsymbol{W}^n$ by
$\mathcal{L}'_{\mathrm{ref},n}$ and $\mathcal{L}''_{\mathrm{ref},n}$, respectively. The exact
time derivatives are
\begin{equation*}
\begin{aligned}
    \boldsymbol{W}_t^n&=\mathcal{L}_{\mathrm{ref},n},\qquad
    \boldsymbol{W}_{tt}^n=\mathcal{L}'_{\mathrm{ref},n}\mathcal{L}_{\mathrm{ref},n},\\
    \boldsymbol{W}_{ttt}^n
    &=(\mathcal{L}'_{\mathrm{ref},n})^2\mathcal{L}_{\mathrm{ref},n}
      +\mathcal{L}''_{\mathrm{ref},n}[\mathcal{L}_{\mathrm{ref},n},\mathcal{L}_{\mathrm{ref},n}].
\end{aligned}
\end{equation*}
The Taylor initialization and one Picard correction for this
evolution problem take the form
\begin{equation*}
\begin{aligned}
    \widetilde{\boldsymbol{W}}_\beta^{[0]}
    &=\boldsymbol{W}^n+c_\beta\Delta t\,\mathcal{L}_{\mathrm{ref},n},\\
    \widetilde{\boldsymbol{W}}_\alpha^{[1]}
    &=\boldsymbol{W}^n+\Delta t\sum_{\beta=1}^{2}
       A_{\alpha\beta}\,
       \mathcal{L}_{\mathrm{ref}}[\widetilde{\boldsymbol{W}}_\beta^{[0]}].
\end{aligned}
\end{equation*}
The Lagrange basis reproduces constants and linear functions,
$\sum_\beta\ell_\beta(\tau)=1$ and
$\sum_\beta c_\beta\ell_\beta(\tau)=\tau$. Consequently,
\begin{equation*}
    \sum_{\beta=1}^{2}A_{\alpha\beta}=c_\alpha,
    \qquad
    \sum_{\beta=1}^{2}A_{\alpha\beta}c_\beta
      =\frac{c_\alpha^2}{2}.
\end{equation*}
The first identity also shows that the Taylor initialization is
equivalent to one Picard application starting from a temporally
constant field. Thus, the stated single correction follows an
initial approximation that already contains the first temporal
derivative.

Expanding the operator about $\boldsymbol{W}^n$ gives
\begin{equation*}
\begin{aligned}
    \mathcal{L}_{\mathrm{ref}}[\widetilde{\boldsymbol{W}}_\beta^{[0]}]
    &=\mathcal{L}_{\mathrm{ref},n}+c_\beta\Delta t\,
      \mathcal{L}'_{\mathrm{ref},n}\mathcal{L}_{\mathrm{ref},n}+O(\Delta t^2),\\
    \widetilde{\boldsymbol{W}}_\alpha^{[1]}
    &=\boldsymbol{W}^n+c_\alpha\Delta t\,\mathcal{L}_{\mathrm{ref},n}
      +\frac{c_\alpha^2\Delta t^2}{2}
       \mathcal{L}'_{\mathrm{ref},n}\mathcal{L}_{\mathrm{ref},n}+O(\Delta t^3).
\end{aligned}
\end{equation*}
The corrected state therefore contains the exact quadratic
temporal coefficient. Evaluating the nonlinear operator at this
state retains both contributions required at the next order:
\begin{equation*}
\begin{aligned}
    \mathcal{L}_{\mathrm{ref}}[\widetilde{\boldsymbol{W}}_\alpha^{[1]}]
    &=\mathcal{L}_{\mathrm{ref},n}+c_\alpha\Delta t\,
      \mathcal{L}'_{\mathrm{ref},n}\mathcal{L}_{\mathrm{ref},n}\\
    &\quad+\frac{c_\alpha^2\Delta t^2}{2}
      \left((\mathcal{L}'_{\mathrm{ref},n})^2\mathcal{L}_{\mathrm{ref},n}
           +\mathcal{L}''_{\mathrm{ref},n}[\mathcal{L}_{\mathrm{ref},n},\mathcal{L}_{\mathrm{ref},n}]\right)
      +O(\Delta t^3).
\end{aligned}
\end{equation*}
The Gauss weights and the predictor matrix satisfy the four
third-order moment conditions
\begin{equation*}
    \sum_{\alpha=1}^{2}w_\alpha=1,
    \qquad
    \sum_{\alpha=1}^{2}w_\alpha c_\alpha=\frac12,
    \qquad
    \sum_{\alpha=1}^{2}w_\alpha c_\alpha^2=\frac13,
    \qquad
    \sum_{\alpha,\beta=1}^{2}
       w_\alpha A_{\alpha\beta}c_\beta=\frac16.
\end{equation*}
The final quadrature consequently yields
\begin{equation*}
\begin{aligned}
    \boldsymbol{W}^{n+1}
    &=\boldsymbol{W}^n+\Delta t\sum_{\alpha=1}^{2}
      w_\alpha\mathcal{L}_{\mathrm{ref}}
      [\widetilde{\boldsymbol{W}}_\alpha^{[1]}]\\
    &=\boldsymbol{W}^n+\Delta t\,\mathcal{L}_{\mathrm{ref},n}
      +\frac{\Delta t^2}{2}\mathcal{L}'_{\mathrm{ref},n}\mathcal{L}_{\mathrm{ref},n}\\
    &\quad+\frac{\Delta t^3}{6}
      \left((\mathcal{L}'_{\mathrm{ref},n})^2\mathcal{L}_{\mathrm{ref},n}
           +\mathcal{L}''_{\mathrm{ref},n}[\mathcal{L}_{\mathrm{ref},n},\mathcal{L}_{\mathrm{ref},n}]\right)
      +O(\Delta t^4).
\end{aligned}
\end{equation*}
\begin{s2revision}
The reference expansion reproduces the exact evolution through $\Delta t^3$, giving an $O(\Delta t^4)$ local temporal error and third-order global temporal accuracy under stable integration over a fixed interval. The derivatives $\mathcal{L}'_{\mathrm{ref},n}$ and $\mathcal{L}''_{\mathrm{ref},n}$ express coefficients generated by operator reevaluation; they are not evaluated explicitly in the algorithm. These moment conditions identify the temporal contribution of the Taylor starting iterate and one Picard correction. Remark~\ref{rem:complete_euler_consistency_rewrite} then evaluates the actual compact response, frozen center matching, and wide corrector together.
\end{s2revision}
\end{remarkblock}

\begin{s2revision}
Center matching addresses the difference between the compact and wide center derivatives. For its one-dimensional estimate, use the operators $D_2$ and $D_8$ introduced in Section~\ref{sec:local_spatial_operator_rewrite}, with $f_m=f(x_m)$ and $(D_2f)_m=(f_{m+1}-f_{m-1})/(2h)$. Taylor expansion of a sufficiently smooth scalar flux gives
\end{s2revision}
\begin{equation*}
\begin{aligned}
    (D_2f)_m
    &=f'(x_m)+\frac{h^2}{6}f^{(3)}(x_m)+O(h^4),\\
    (D_8f)_m
    &=f'(x_m)-\frac{h^8}{630}f^{(9)}(x_m)+O(h^{10}).
\end{aligned}
\end{equation*}
\begin{s2revision}
The notation $f^{(k)}$ denotes the $k$th physical-coordinate derivative. Antisymmetry and the wide-stencil weights give the displayed leading errors, so the compact and wide derivatives differ by $O(h^2)$. Matching removes this beginning-of-step center difference; the directional compact modes continue to describe the local spatial variation.
\end{s2revision}

The contribution of center matching to the complete convective
update can be isolated for smooth scalar advection
$\partial_t w+a\partial_x w=0$, with constant speed $a$, on a
uniform periodic grid without sources or artificial dissipation.
Use the one-dimensional operators $D_2$ and
$D_8$ introduced above, and let
$w^n=\{w_m^n\}$ be the scalar nodal field, with
$w_m^n=w(x_m,t^n)$. The scalar counterpart of
Eq.~\eqref{eq:operator_matching_correction_rewrite} is
$C_m^n=a\bigl[(D_8-D_2)w^n\bigr]_m$.
For the same provisional fields, adding $C_m^n$ to both Picard
operators changes the predicted state at node $m$ and time
$t^n+c_\alpha\Delta t$ by $-c_\alpha\Delta t C_m^n$. Since
$\sum_\alpha w_\alpha c_\alpha=1/2$, the time-averaged scalar
flux at that node changes by $-a\Delta t C_m^n/2$. Applying the final derivative
$D_8$ gives the exact difference between the two linear
updates:
\begin{equation*}
\begin{aligned}
    w_{m,\mathrm{unmatched}}^{n+1}-w_{m,\mathrm{matched}}^{n+1}
    &=\frac{a^2\Delta t^2}{2}\,
       \bigl[D_8(D_2-D_8)w^n\bigr]_m\\
    &=\frac{a^2\Delta t^2h^2}{12}\,
       \partial_x^4 w(x_m,t^n)+O(\Delta t^2h^4).
\end{aligned}
\end{equation*}
Thus, matching removes a specific $O(\Delta t^2h^2)$ single-step
contribution caused by the compact initial derivative. Equivalently,
the term quadratic in $\Delta t$ changes from
$a^2\Delta t^2(D_8D_2w^n)_m/2$ to
$a^2\Delta t^2(D_8^2w^n)_m/2$, consistent with the corresponding
Taylor coefficient of the semidiscrete equation
$\mathrm{d}w_m/\mathrm{d}t=-a(D_8w)_m$. This calculation identifies the
convective consistency condition enforced by the constant-mode
correction. The remaining cubic temporal coefficient depends on
the compact spatial representation, as quantified in
Remark~\ref{rem:complete_euler_consistency_rewrite}.
Molecular transport retains the gradient approximation specified in
Section~\ref{sec:local_spatial_operator_rewrite}.

Center matching is introduced in the Picard integral update, while
the Taylor initialization reuses the compact initial operator.
For a smooth field, $\boldsymbol{C}_{\boldsymbol{m}}^n=O(h^2)$
is the difference between compact and wide convective derivatives.
For Euler flow, this initialization choice changes the corrected
center state by $O(\Delta t^2\boldsymbol{C}_{\boldsymbol{m}}^n)$
and the accepted physical-flux update by $O(\Delta t^3h^2)$ for
a smooth truncation-error field. Under $\Delta t=O(h)$ these
contributions are $O(h^4)$ and $O(h^5)$, respectively, and enter
as higher-order initialization terms in the fixed-CFL consistency
estimate. This separates the role of the Taylor starting iterate
from the retained compact-mode contribution quantified below.

\begin{remarkblock}[manuscriptrevision]{Consistency of the complete Euler update}
\label{rem:complete_euler_consistency_rewrite}
\begin{s2revision}
We now assess the physical-flux part of the accepted update in Eq.~\eqref{eq:complete_update_rewrite}. Consider a smooth source-free Euler solution on a uniform periodic Cartesian grid, with positive density and pressure and inactive positivity limiting. Let $h_j\asymp h$. For this analysis, the instantaneous wide-stencil physical operator is the spatial reference, denoted by $\mathcal{L}_h$; its subscript identifies the mesh-dependent discretization. Define this operator and the flux Jacobians by
\end{s2revision}
\begin{equation*}
    \mathcal{L}_h[\boldsymbol{W}]
    =-\sum_{j=1}^{3}\mathcal{D}_j
       \boldsymbol{F}^{\mathrm{c}}_j(\boldsymbol{W}),
    \qquad
    \boldsymbol{J}_{j,\boldsymbol{m}}^n
    =(\boldsymbol{F}^{\mathrm{c}}_j)'
       (\boldsymbol{W}_{\boldsymbol{m}}^n).
\end{equation*}
Here $\mathcal{L}_h^n=\mathcal{L}_h[\boldsymbol{W}^n]$, and
$\mathcal{L}_h'$ and $\mathcal{L}_h''$ denote derivatives with
respect to the nodal state field. The actual Taylor samples use the
local slope
$\boldsymbol{V}_{\boldsymbol{m}}(\boldsymbol{s})
=-\boldsymbol{R}_0-\sum_r\boldsymbol{R}_rs_r$.
Linearizing the provisional compact center divergence in time gives
the nodal coefficient $\boldsymbol{B}^n=\{\boldsymbol{B}_{\boldsymbol{m}}^n\}$,
\begin{equation*}
    \boldsymbol{B}_{\boldsymbol{m}}^n
    =-\sum_{j=1}^{3}\frac{
      \boldsymbol{J}_{j,\boldsymbol{m}+\boldsymbol{e}_j}^n
        \boldsymbol{V}_{\boldsymbol{m}}(\boldsymbol{e}_j)
      -\boldsymbol{J}_{j,\boldsymbol{m}-\boldsymbol{e}_j}^n
        \boldsymbol{V}_{\boldsymbol{m}}(-\boldsymbol{e}_j)}
      {2h_j}.
\end{equation*}
This coefficient describes the implemented operator response and
introduces no additional evaluation. The frozen matching condition and
the identities $\sum_\beta A_{\alpha\beta}=c_\alpha$ and
$\sum_\beta A_{\alpha\beta}c_\beta=c_\alpha^2/2$ yield
\begin{equation*}
\begin{aligned}
    -\widehat{\boldsymbol{H}}_\beta^{[0]}(\boldsymbol{0})
    &=\mathcal{L}_{h,\boldsymbol{m}}^n
      +c_\beta\Delta t\,\boldsymbol{B}_{\boldsymbol{m}}^n
      +O(\Delta t^2),\\
    \widetilde{\boldsymbol{W}}_\alpha^{[1]}(\boldsymbol{0})
    &=\boldsymbol{W}_{\boldsymbol{m}}^n
      +c_\alpha\Delta t\,\mathcal{L}_{h,\boldsymbol{m}}^n
      +\frac{c_\alpha^2\Delta t^2}{2}
         \boldsymbol{B}_{\boldsymbol{m}}^n
      +O(\Delta t^3).
\end{aligned}
\end{equation*}
Expanding the nonlinear physical flux at these corrected states,
using $\sum_\alpha w_\alpha c_\alpha=1/2$ and
$\sum_\alpha w_\alpha c_\alpha^2=1/3$, and applying the final
spatial derivative gives the physical-flux part of the implemented update:
\begin{equation*}
\begin{aligned}
    \boldsymbol{W}^{n+1}
    &=\boldsymbol{W}^n+\Delta t\,\mathcal{L}_h^n
      +\frac{\Delta t^2}{2}(\mathcal{L}_h')^n\mathcal{L}_h^n\\
    &\quad+\frac{\Delta t^3}{6}
      \left((\mathcal{L}_h')^n\boldsymbol{B}^n
      +(\mathcal{L}_h'')^n[\mathcal{L}_h^n,\mathcal{L}_h^n]\right)
      +O(\Delta t^4).
\end{aligned}
\end{equation*}
The linear and quadratic temporal coefficients agree with the
wide-stencil semidiscrete equation. The cubic nonlinear-flux
contribution also agrees; the remaining difference from its exact
Taylor update is
\begin{equation*}
    \frac{\Delta t^3}{6}(\mathcal{L}_h')^n
       \left[\boldsymbol{B}^n
             -(\mathcal{L}_h')^n\mathcal{L}_h^n\right]
       +O(\Delta t^4).
\end{equation*}

The spatial size of this difference follows from the ten-point
coefficient formulas. The center derivative is second-order accurate.
For a smooth sampled flux component $a$, the mixed coefficient obeys
\begin{equation*}
\begin{aligned}
    \frac{a_{rs}}{h_rh_s}
    &=\frac{\partial^2a}{\partial x_r\,\partial x_s}
      +\frac{h_r}{2}
        \frac{\partial^3a}{\partial x_r^2\,\partial x_s}\\
    &\quad+\frac{h_s}{2}
        \frac{\partial^3a}{\partial x_r\,\partial x_s^2}
      +O(h^2),\qquad r<s,
\end{aligned}
\end{equation*}
with derivatives evaluated at the target node and
$x_r=x_{\boldsymbol{m},r}+h_r\xi_r$.
Consequently, $\boldsymbol{R}_r/h_r$ approximates the physical
gradient of the divergence with $O(h)$ error in general.
The provisional center response therefore satisfies
$\boldsymbol{B}^n-(\mathcal{L}_h')^n\mathcal{L}_h^n=O(h)$
on smooth fields. The physical-flux one-step consistency error
relative to the Euler solution is
$O(\Delta t^4)+O(\Delta t^3h)+O(\Delta t\,h^8)$.

Artificial dissipation adds $O(\Delta t\,h^7)$ when the background
branch dominates, or $O(\Delta t\,h^3)$ for the activated branch
in smooth compressive regions, as specified in
Section~\ref{sec:global_corrector_rewrite}.
For uniformly stable integration over a fixed interval and smooth
truncation-error fields, the accumulated consistency estimate is
\begin{equation*}
\begin{aligned}
    E(T)&\lesssim C_T
      \left[\Delta t_{\max}^3+\Delta t_{\max}^2h
            +h^8+\mathcal{E}_{\mathrm{diss}}(h)\right],\\
    \mathcal{E}_{\mathrm{diss}}(h)
    &=\begin{cases}
        O(h^7),&\text{background branch},\\
        O(h^3),&\text{smooth activated branch}.
      \end{cases}
\end{aligned}
\end{equation*}
For the smooth Euler solution on the uniform periodic grid,
under the stated stability assumption, fixed-CFL refinement gives
$\Delta t_{\max}\asymp h$ and the estimate establishes third-order
coupled convergence. In the tested finite smooth-regime interval
where the background dissipation branch applies, background spatial
error dominates and time-dependent error contributions are controlled
by a sufficiently small CFL; the observed slopes approach seven.

Retaining center matching suppresses the compact-initial-slope
contribution identified by the scalar calculation above: an
$O(\Delta t^2h^2)$ one-step term and its
$O(\Delta t_{\max}h^2)$ accumulated contribution.
This unmatched contribution scales as $h^3$ at fixed CFL and
decreases linearly with CFL. Center matching therefore suppresses
this integrated-flux error contribution while preserving the compact local
prediction, including in tests where both matched and unmatched
formulations exhibit a third-order mesh slope.

\begin{s2revision}
The coupled bound retains the actual compact response through its $O(\Delta t_{\max}^2h)$ contribution. Together with the temporal moments and the center-matching calculation, it connects the local representation to the accuracy of the complete conservative update. Fixed-CFL third-order convergence follows under the stated smooth-Euler, mesh, and stability conditions; the near-seventh-order slopes belong to the specified background-dissipation regime with sufficiently controlled temporal contributions.
\end{s2revision}
\end{remarkblock}

\end{revisiontext}

\endgroup

\section{GPU implementation and performance}
\label{sec:gpu_performance}

The local predictor in Section~\ref{sec:numerical_method_rewrite}
changes both the arithmetic and the global data dependencies of
high-order time integration. The resident fields and kernel organization
in Section~\ref{sec:gpu_resident_execution} establish the implementation
basis for evaluating these changes. Section~\ref{sec:single_gpu_performance}
first defines the common single-GPU benchmark and measurement protocol,
then quantifies the complete-step benefit relative to RK3-F within a
shared spatial-discretization framework. Independent kernel profiles
identify the associated computation and data movement.
Section~\ref{sec:external_solver_comparison} extends the comparison to
established GPU solvers with their native numerical and runtime choices.
Finally, Section~\ref{sec:multi_gpu_scalability} measures how the local-ADER
advancement rate scales when distributed communication enters the cost.

\subsection{GPU implementation of the local ADER method}
\label{sec:gpu_resident_execution}

Algorithm~\ref{alg:complete_update_rewrite} is organized around a fused local predictor and a fused global
corrector. At each node, the predictor reads $\boldsymbol{W}^{n}$, constructs
the compact quadratic representation and its local operator, evaluates the
two temporal nodes, and immediately combines their contributions by
Gauss--Legendre quadrature. Polynomial coefficients, Picard intermediates,
and temporal-node states remain kernel-local throughout this construction.
The globally materialized products are limited to the time-averaged state,
compact physical-flux data, three averaged directional signal radii, and the frozen
dissipation sensor. Figure~\ref{fig:gpu_resident_execution}(a) shows how
these products connect the predictor to one conservative correction.
This organization avoids full-domain storage and halo exchange of the
temporal-node states.

The implementation assigns one MPI rank to one GPU. The accepted solution
and all time-integration work fields remain device resident, with storage
allocated once during initialization and reused at every step. The regular
advancement path neither allocates solution-sized temporary arrays nor
transfers a complete field to the host; only scalar controls required for
time-step selection and MPI orchestration become host visible. Each
allocated block, including halos, uses a structure-of-arrays layout,
\[
N_{\mathrm a}=N_x^{\mathrm a}N_y^{\mathrm a}N_z^{\mathrm a},\qquad
\mathcal I(i,j,k)=i+N_x^{\mathrm a}(j+N_y^{\mathrm a}k),\qquad
\mathcal O(a,i,j,k)=aN_{\mathrm a}+\mathcal I(i,j,k).
\]
A component therefore occupies one contiguous segment, and consecutive
threads along the logical $x$ direction access consecutive locations.
Interior, boundary, and halo kernels share this indexing convention.

\begin{figure}[pos=!htbp]
    \centering
    \resizebox{\linewidth}{!}{%
        \definecolor{aderblue}{RGB}{32,91,164}
\definecolor{aderteal}{RGB}{27,112,110}
\definecolor{aderorange}{RGB}{221,92,19}
\definecolor{adergray}{RGB}{105,105,105}
\begin{tikzpicture}[
    x=1cm,
    y=1cm,
    >=Latex,
    line cap=round,
    line join=round,
    every node/.style={font=\rmfamily\scriptsize, text=black},
    flow/.style={draw=black, line width=0.42pt, -{Latex[length=2.0mm,width=1.25mm]}},
    connector/.style={draw=black, line width=0.42pt},
    process/.style={draw=black!70, line width=0.38pt, fill=white,
        minimum height=0.44cm, align=center, inner sep=1.0pt,
        font=\rmfamily\tiny},
    state/.style={draw=aderblue, line width=0.62pt, fill=aderblue!4,
        align=center, inner sep=2pt},
    compute/.style={draw=aderteal, line width=0.60pt, fill=aderteal!4,
        align=center, inner sep=2pt},
    comm/.style={draw=aderorange, line width=0.60pt, fill=aderorange!4,
        align=center, inner sep=2pt}
]
\path[use as bounding box] (0,0) rectangle (18,8.55);

\node[anchor=west, font=\rmfamily\bfseries\normalsize] at (0.05,8.30)
    {(a) GPU-resident dataflow};

\node[state, minimum width=1.85cm, minimum height=1.38cm] (state-n) at (1.05,6.63)
    {\footnotesize Accepted state\\[2pt]
     $\boldsymbol{W}^{n}$\\[2pt]
     \itshape device-resident SoA};

\draw[aderteal, line width=0.62pt, fill=aderteal!2]
    (2.62,5.18) rectangle (8.26,8.02);
\draw[aderteal, line width=0.50pt] (2.62,7.57) -- (8.26,7.57);
\node[font=\rmfamily\footnotesize] at (5.44,7.79) {Fused local predictor};

\draw[adergray, line width=0.40pt, dashed]
    (2.83,5.78) rectangle (8.06,7.42);
\node[font=\rmfamily\scriptsize] at (5.445,7.22)
    {kernel-local space--time construction};

\node[process, text width=1.02cm] (p2) at (3.45,6.62)
    {compact P2\\representation};
\node[process, text width=0.92cm] (gauss) at (4.72,6.62)
    {two Gauss\\nodes};
\node[process, text width=1.30cm] (picard) at (6.10,6.62)
    {operator-matched\\Picard correction};
\node[process, text width=0.88cm] (gl2) at (7.48,6.62)
    {GL2\\temporal\\contraction};

\draw[flow] (p2.east) -- (gauss.west);
\draw[flow] (gauss.east) -- (picard.west);
\draw[flow] (picard.east) -- (gl2.west);

\node at (3.45,6.08) {$\boldsymbol{W}^{n}(\boldsymbol{\xi})$};
\node at (4.72,6.08) {$\widetilde{\boldsymbol{W}}^{[0]}_\alpha(\boldsymbol{\xi})$};
\node at (6.10,6.08) {$\widetilde{\boldsymbol{W}}^{[1]}_\alpha(\boldsymbol{\xi})$};
\node at (7.50,6.08) {$\sum_\alpha w_\alpha$};

\node[anchor=west, font=\rmfamily\Large] at (3.12,5.50) {$\times$};
\node[font=\rmfamily\tiny, text width=4.35cm, align=center] at (5.67,5.51)
    {no global $\widetilde{\boldsymbol{W}}_\alpha$ fields $\cdot$ no temporal-state halo exchange};

\draw[aderblue, line width=0.62pt, fill=aderblue!2]
    (8.72,5.18) rectangle (11.88,8.02);
\draw[aderblue, line width=0.50pt] (8.72,7.43) -- (11.88,7.43);
\node[font=\rmfamily\footnotesize] at (10.30,7.80)
    {Materialized corrector data};
\node[font=\rmfamily\scriptsize\itshape] at (10.30,7.56)
    {global device memory};

\node[state, font=\rmfamily\tiny, minimum width=2.86cm, minimum height=0.43cm] at (10.30,7.08)
    {Time-averaged state \quad $\overline{\boldsymbol{W}}^{n}$};
\node[state, font=\rmfamily\tiny, minimum width=2.86cm, minimum height=0.43cm] at (10.30,6.52)
    {Compact physical flux \quad $\overline{\boldsymbol{F}}^{n}_{j}$};
\node[state, font=\rmfamily\tiny, minimum width=2.86cm, minimum height=0.53cm] at (10.30,5.89)
    {Signal radius and sensor \quad $\overline{\lambda}^{n}_{j},\,\nu^{n}$};

\draw[aderteal, line width=0.62pt, fill=aderteal!2]
    (12.48,5.18) rectangle (15.66,8.02);
\draw[aderteal, line width=0.50pt] (12.48,7.57) -- (15.66,7.57);
\node[font=\rmfamily\footnotesize] at (14.07,7.79) {Fused global corrector};

\node[process, minimum width=2.86cm, minimum height=0.39cm] at (14.07,7.20)
    {shared-memory tile};
\node[process, minimum width=2.86cm, minimum height=0.39cm] at (14.07,6.72)
    {radius-four central operator};
\node[process, minimum width=2.86cm, minimum height=0.39cm] at (14.07,6.24)
    {selective dissipation $+$ update};
\node[anchor=west, inner sep=0pt] at (12.64,5.72)
    {\resizebox{2.78cm}{!}{$
      \overline{\mathcal{L}}^{n}
      =-\sum_{j=1}^{3}\mathcal{D}_j\overline{\boldsymbol{F}}^{n}_{j}
      +\overline{\boldsymbol{S}}^{n}+\overline{\mathcal{A}}^{n}$}};
\node[anchor=west, inner sep=0pt] at (12.64,5.35)
    {\resizebox{2.63cm}{!}{$\boldsymbol{W}^{n+1}
      =\boldsymbol{W}^{n}+\Delta t\,\overline{\mathcal{L}}^{n}$}};

\node[state, minimum width=1.72cm, minimum height=1.38cm] (state-np1) at (17.03,6.63)
    {\footnotesize Accepted state\\[2pt]
     $\boldsymbol{W}^{n+1}$\\[2pt]
     \itshape single global commit};

\draw[flow] (state-n.east) -- (2.62,6.63);
\draw[flow] (8.26,6.63) -- (8.72,6.63);
\draw[flow] (11.88,6.63) -- (12.48,6.63);
\draw[flow] (15.66,6.63) -- (state-np1.west);

\draw[connector] (state-n.south) -- (1.05,4.86) -- (14.07,4.86);
\draw[flow] (14.07,4.86) -- (14.07,5.18);
\node[fill=white, inner sep=1pt] at (7.43,4.87)
    {$\boldsymbol{W}^{n}$ retained for final update};
\node[font=\rmfamily\footnotesize] at (9.12,4.48)
    {Only one compact global materialization between predictor and corrector};

\draw[black, line width=0.52pt] (0.05,4.24) -- (17.95,4.24);

\node[anchor=west, font=\rmfamily\bfseries\normalsize] at (0.05,3.98)
    {(b) Communication-overlapped execution timeline};
\node[anchor=east, font=\rmfamily\tiny\itshape] at (17.04,3.99)
    {Schematic timeline; bar widths are not to scale.};

\node[anchor=west, align=left, font=\rmfamily\tiny] at (0.13,3.47) {GPU compute\\stream};
\node[anchor=west, align=left, font=\rmfamily\tiny] at (0.13,2.62)
    {MPI halo transfer /\\communication stream};

\fill[black!4] (1.92,1.36) rectangle (4.88,3.63);
\fill[black!4] (8.50,1.36) rectangle (13.23,3.63);

\node[compute, font=\rmfamily\scriptsize, minimum width=3.72cm, minimum height=0.46cm] (ci1) at (3.78,3.42)
    {Predictor: interior};
\node[compute, font=\rmfamily\scriptsize, minimum width=2.64cm, minimum height=0.46cm] (cb1) at (6.99,3.42)
    {Predictor: boundary};
\node[compute, font=\rmfamily\scriptsize, minimum width=5.65cm, minimum height=0.46cm] (ci2) at (11.37,3.42)
    {Corrector: interior};
\node[compute, font=\rmfamily\scriptsize, minimum width=2.55cm, minimum height=0.46cm] (cb2) at (15.52,3.42)
    {Corrector: boundary};

\node[comm, font=\rmfamily\tiny, minimum width=2.82cm, minimum height=0.64cm] (h0) at (3.34,2.58)
    {$H_{0}$\\$\boldsymbol{W}^{n}$ halo exchange};
\node[comm, font=\rmfamily\tiny, minimum width=1.88cm, minimum height=0.64cm] (h1) at (9.45,2.58)
    {$H_{1}$\\$\overline{\boldsymbol{W}}^{n}$ and $\nu^{n}$\\halo exchange};
\node[comm, font=\rmfamily\tiny, minimum width=2.28cm, minimum height=0.64cm] (h2) at (11.59,2.58)
    {$H_{2}$\\directional compact-flux\\halo exchange};

\draw[connector] (h0.east) -- (5.68,2.58);
\draw[flow] (5.68,1.37) -- (5.68,3.18);
\node[align=center, font=\rmfamily\tiny] at (5.68,1.07) {dependency\\boundary 1};

\draw[connector] (h2.east) -- (13.65,2.58) -- (13.65,2.31) -- (14.23,2.31);
\draw[fill=white, line width=0.42pt] (14.35,2.31) circle (0.09);
\draw[flow] (14.35,2.40) -- (14.35,3.18);
\draw[connector, dashed] (14.35,2.22) -- (14.35,1.38);
\node[align=left, anchor=west, font=\rmfamily\tiny] at (14.55,2.02) {corrector-data\\completion};
\node[align=center, font=\rmfamily\tiny] at (14.35,1.07) {dependency\\boundary 2};

\node[font=\rmfamily\tiny\itshape, align=center] at (3.34,1.96)
    {communication hidden by\\interior work};
\node[font=\rmfamily\tiny\itshape, align=center] at (10.75,1.96)
    {communication hidden by\\interior work};

\draw[flow] (16.80,3.42) -- (17.48,3.42);
\node[anchor=west] at (17.50,3.42) {$\boldsymbol{W}^{n+1}$};
\draw[flow] (1.40,1.36) -- (17.72,1.36);
\node[anchor=east, font=\rmfamily\tiny\itshape] at (17.78,1.15) {time};

\draw[state] (0.40,0.22) rectangle (0.90,0.58);
\node[anchor=west, font=\rmfamily\tiny] at (1.06,0.40) {persistent device field};
\draw[compute] (3.46,0.22) rectangle (3.96,0.58);
\node[anchor=west, font=\rmfamily\tiny] at (4.12,0.40) {GPU computation};
\draw[comm] (6.54,0.22) rectangle (7.04,0.58);
\node[anchor=west, font=\rmfamily\tiny] at (7.20,0.40) {device-buffer MPI communication};
\draw[adergray, line width=0.40pt, dashed] (10.78,0.22) rectangle (11.34,0.58);
\node[anchor=west, font=\rmfamily\tiny] at (11.50,0.40) {kernel-local transient data};
\draw[flow] (15.40,0.40) -- (16.02,0.40);
\node[anchor=west, font=\rmfamily\tiny] at (16.23,0.40) {data dependency};
\end{tikzpicture}%
    }
    \caption{GPU-resident realization of one local ADER step. (a) The fused
    local predictor contracts kernel-local space--time data into compact
    globally materialized corrector fields, which are consumed by the fused
    global corrector. The signal-radius entry represents the three
    directional values $\overline{\lambda}_j^n$. (b) Interior work overlaps
    asynchronous halo exchange,
    whereas boundary kernels are issued only after their required data become
    available. Timeline widths are schematic and are not proportional to
    measured execution times.}
    \label{fig:gpu_resident_execution}
\end{figure}

The physical-flux interface exploits the structure of the compressible
Navier--Stokes equations. The three mass-flux components are recovered from
the momentum entries of $\overline{\boldsymbol W}^{n}$. Symmetry of the
total momentum-flux tensor leaves six independent entries, to which three
energy-flux components are added. These nine entries and the three averaged
directional signal radii occupy one five-component and one seven-component
work field; the frozen dissipation sensor occupies a separate scalar field.
The corrector loads the compact fields
into a shared-memory tile, reuses each value across the radius-four central
stencil, applies the selective dissipation, and writes
$\boldsymbol W^{n+1}$ in the same kernel. Direction-specific residual
arrays and a full three-direction physical-flux tensor are not materialized.

Table~\ref{tab:gpu_storage} relates this execution structure to persistent
storage. Local ADER requires the accepted state, two five-component work
fields, one seven-component work field, and one scalar sensor, totaling
$23N_{\mathrm a}$ doubles. RK3-F
additionally allocates one five-component stage-state array, giving
$28N_{\mathrm a}$ doubles. The corresponding full-domain capacities are
184 and 224~B per allocated node. Reusable exchange buffers scale with the
subdomain surface area. These capacities describe resident storage.
Section~\ref{sec:single_gpu_performance} measures the corresponding
complete-step cost and actual HBM traffic, including their repeated use
within the temporal construction.

\begin{table}[pos=!htbp]
    \caption{Persistent full-domain fields and reusable halo storage per MPI
    rank. The two time integrators use the same spatial discretization and
    double-precision representation.}
    \label{tab:gpu_storage}
    \centering
    \small
    \setlength{\tabcolsep}{3.2pt}
    \renewcommand{\arraystretch}{1.18}
    \begin{tabular*}{\linewidth}{@{\extracolsep{\fill}}p{0.19\linewidth}c
                    p{0.245\linewidth}p{0.245\linewidth}
                    >{\raggedleft\arraybackslash}p{0.18\linewidth}@{}}
        \toprule
        Storage object & Count & Local ADER content & RK3-F content
        & Allocated capacity \\
        \midrule
        \multicolumn{5}{@{}l}{\textit{Volume-scaled device fields}} \\
        Accepted state
        & 5
        & $\boldsymbol{W}^{n}\!\rightarrow\!\boldsymbol{W}^{n+1}$
        & $\boldsymbol{W}^{n}\!\rightarrow\!\boldsymbol{W}^{n+1}$
        & $5N_{\mathrm a}$ doubles \\
        Work field 1
        & 5
        & Averaged state $\overline{\boldsymbol{W}}^{n}$
        & Intermediate stage state
        & $5N_{\mathrm a}$ doubles \\
        Compact field A
        & 5
        & Averaged flux entries 1--5
        & Flux entries 1--5 at each stage
        & $5N_{\mathrm a}$ doubles \\
        Compact field B
        & 7
        & Averaged flux entries 6--9 and
          $\overline{\lambda}^{n}_{1:3}$
        & Flux entries 6--9 and $\lambda_{1:3}$ at each stage
        & $7N_{\mathrm a}$ doubles \\
        Dissipation sensor
        & 1
        & Frozen $\nu^{n}$
        & Stagewise $\nu$
        & $N_{\mathrm a}$ doubles \\
        Additional RK stage
        & 5
        & Not allocated
        & Intermediate stage state
        & $5N_{\mathrm a}$ doubles \\
        \addlinespace[2pt]
        \textbf{Full-domain total}
        & --
        & $23N_{\mathrm a}$ doubles
        & $28N_{\mathrm a}$ doubles
        & \textbf{184/224 B node$^{-1}$} \\
        \midrule
        \multicolumn{5}{@{}l}{\textit{Surface-scaled exchange storage}} \\
        Device send/receive
        & $4/$axis
        & Reused by all halo exchanges
        & Reused by all halo exchanges
        & $4(5g+1)A_{\Sigma}$ doubles \\
        Pinned-host staging
        & $4/$axis
        & Cross-node staging path
        & Cross-node staging path
        & $4(5g+1)A_{\Sigma}$ doubles \\
        \bottomrule
    \end{tabular*}
    \par\vspace{3pt}
    \begin{minipage}{0.98\linewidth}
        \footnotesize
        For volume-scaled fields, Count is the number of components; for
        exchange storage, it is the buffer multiplicity per coordinate axis.
        Here $N_{\mathrm a}=N_x^{\mathrm a}N_y^{\mathrm a}N_z^{\mathrm a}$,
        $A_{\Sigma}=N_yN_z+N_xN_z+N_xN_y$, and $g=4$.
        The four exchange buffers per axis are the low/high send and receive
        buffers. Compact field B contains four physical-flux components and
        three directional signal radii; directional flux exchanges send only
        the components required in each direction. Polynomial coefficients,
        temporal-node states, and Picard
        intermediates remain kernel-local and therefore do not contribute
        solution-sized global arrays. The optional positivity mask is disabled
        in the baseline configuration and is excluded.
    \end{minipage}
\end{table}

For the periodic block decompositions considered here, computation follows
the data dependencies in Figure~\ref{fig:gpu_resident_execution}(b).
Beginning-of-step state exchange overlaps the predictor on nodes whose
stencils are wholly local, after which the boundary predictor consumes the
completed halos. Exchange of the averaged state and sensor overlaps the
interior corrector, with direction-specific completion of the compact flux
data required by the boundary corrector. Thus, interior work can proceed
while boundary data are in transit. The scaling experiment in
Section~\ref{sec:multi_gpu_scalability} evaluates the complete advancement
cost of this distributed schedule, including communication.

\subsection{Controlled single-GPU comparison and performance analysis}
\label{sec:single_gpu_performance}
\label{sec:benchmark_protocol}

The controlled comparison evaluates the net cost of replacing three
full-residual RK stages by the local predictor--corrector construction.
Local ADER and RK3-F share the spatial operator, selective-dissipation
formulation, molecular transport, FP64 representation, boundary completion,
and time-step estimator. Their performance difference therefore measures
the complete temporal construction and its GPU execution within a common
spatial-discretization framework. \workprecisionrevision{The measurements first quantify
complete-step cost on a viscous benchmark, then determine the elapsed
time required to satisfy prescribed exact-solution error thresholds.
Kernel work and data movement finally explain the observed cost reduction.}

\paragraph{Common benchmark and timing protocol.}
\workprecisionrevision{The NS-TGV throughput comparisons in this subsection and
Section~\ref{sec:external_solver_comparison} use a common physical model,
workload, hardware, and timing protocol. The benchmark is the unforced
compressible Navier--Stokes Taylor--Green vortex (NS-TGV).} The reference quantities are
$\rho_0=U_0=L_0=R=1$, with

\[
M_0=0.1,\qquad Re_0=1600,\qquad \gamma=1.4,\qquad
\Omega=[0,2\pi)^3.
\]

The gas is calorically perfect, $\mu=1/1600$ is constant, and Stokes'
hypothesis is used. Thermal conductivity is zero, $\kappa=0$, while viscous
stresses and the viscous energy flux $u_i\tau_{ij}$ remain active.
This common transport model accommodates the native single-fluid viscous
formulations of all participating solvers. The initial field is

\[
\begin{aligned}
\rho&=1, & u&=\sin x\cos y\cos z,\\
v&=-\cos x\sin y\cos z, & w&=0,\\
p&=\frac{1}{\gamma M_0^2}
 +\frac{(\cos2x+\cos2y)(2+\cos2z)}{16},
&\rho E&=\frac{p}{\gamma-1}+\frac{\rho}{2}(u^2+v^2+w^2).
\end{aligned}
\]

Physical time is reported in the common units $t^*=tU_0/L_0$. Within this
benchmark, Local ADER and RK3-F both evaluate the combined convective
and viscous rate in Eq.~\eqref{eq:cfl_time_step_rewrite} once per complete
step, with $\mathrm{CFL}=0.5$ and the same device and MPI reductions.
With zero heat conduction, the viscous rate is $\sigma_{\mathrm v,\boldsymbol m}^{n}
=8\mu\sum_j h_{j,\boldsymbol m}^{-2}/(3\rho_{\boldsymbol m}^{n})$. The native
numerical and runtime configurations of the external solvers are
specified in Section~\ref{sec:external_solver_comparison}.

Native timing refers to execution without profiling instrumentation.
The five-implementation timing suite was run sequentially on the same
physical NVIDIA V100-SXM2-16GB GPU under an exclusive node allocation.
Each implementation in this timing suite uses $N_{\rm int}=302^3=27\,543\,608$ independent interior locations,
excluding ghost layers and repeated periodic end points.
Table~\ref{tab:ns_tgv_capacity} records both the allocation trials used
to select this common grid and the repeated timing schedule. HTR is
the limiting configuration: it completes $302^3$ and fails device
allocation at $303^3$, whereas all other implementations complete
larger grids. These trials determine the largest common cubic size for
the stated configurations without requiring the exact limits of the
higher-capacity implementations.

\begin{table}[pos=!htbp]
    \caption{Device-memory capacity trials and the repeated timing schedule.
    Successful capacity trials completed six full steps. An entry in the
    OOM column denotes a failed device allocation, not a measured throughput.
    \textcolor[rgb]{0,0,0}{\workprecisionrevision{All NS-TGV timing runs} use $302^3$ locations and enough
    batches to cover a common 30-s advancement window.}}
    \label{tab:ns_tgv_capacity}
    \centering
    \small
    \setlength{\tabcolsep}{4pt}
    \renewcommand{\arraystretch}{1.13}
    \begin{tabular*}{\linewidth}{@{\extracolsep{\fill}}lrrrr@{}}
    \toprule
    Implementation & Successful $N$ & OOM $N$ & Warm-up steps & Steps/batch \\
    \midrule
    Local ADER & 432 & 440 & \textcolor[rgb]{0,0,0}{47} & \textcolor[rgb]{0,0,0}{19} \\
    RK3-F & 408 & 416 & \textcolor[rgb]{0,0,0}{24} & \textcolor[rgb]{0,0,0}{10} \\
    STREAmS-2 & 392 & 400 & \textcolor[rgb]{0,0,0}{16} & \textcolor[rgb]{0,0,0}{7} \\
    MFC & \textcolor[rgb]{0,0,0}{479} & \textcolor[rgb]{0,0,0}{480} & \textcolor[rgb]{0,0,0}{16} & \textcolor[rgb]{0,0,0}{7} \\
    HTR & \textcolor[rgb]{0,0,0}{302} & \textcolor[rgb]{0,0,0}{303} & \textcolor[rgb]{0,0,0}{4} & \textcolor[rgb]{0,0,0}{2} \\
    \bottomrule
    \end{tabular*}
    \par\vspace{3pt}
    \begin{minipage}{0.98\linewidth}
    \footnotesize
    \textcolor[rgb]{0,0,0}{HTR uses a 15000-MiB framebuffer pool and one tile
    in these single-GPU trials.} Pool reservation is distinct from occupied
    solution storage.
    \end{minipage}
\end{table}

Each timing window contains complete physical steps: state-dependent CFL
evaluation and reductions, periodic and halo completion, all RK stages or
the complete ADER predictor and corrector, viscous stresses and work, and
the final state update. Initialization, runtime setup, restart reads,
optional diagnostics, and field output are excluded. Device or task-runtime
completion fences delimit synchronized batches. For elapsed wall time
$T_{\rm m}$ and $N_{\rm s}$ complete steps, the throughput and cost per
node per step are

\[
P_{\rm step}=\frac{N_{\rm int}N_{\rm s}}{10^6T_{\rm m}}
\quad[\mathrm{Mnode\ updates\,s^{-1}}],\qquad
g_{\rm step}=\frac{10^9T_{\rm m}}{N_{\rm int}N_{\rm s}}
\quad[\mathrm{ns\,node^{-1}\,step^{-1}}].
\]

All RK stages together count as one update, as do the ADER predictor and
corrector together, and $P_{\rm step}g_{\rm step}=10^3$. Physical-time advancement is
measured separately as $A=\Delta t^*_{\rm win}/T_{\rm win}$ with a common $T_{\rm win}=30$~s window.
For each independent launch, elapsed advancement wall time and physical-time
increments are accumulated from the first recorded batch boundary.
Linear interpolation between the two measured boundaries bracketing
30~s gives $\Delta t^*_{\rm win}$; no extrapolation is used. This quantity
complements complete-step throughput when time-step estimates differ.
Both metrics are evaluated over the measured initial smooth-flow window.

Four independent launches per implementation were organized into four
sweeps over the five implementations, with the order reversed in
alternating sweeps. Each launch began from the prescribed initial field.
Warm-up and synchronized batch lengths were calibrated to at least 5.5
and 2.25~s, respectively (Table~\ref{tab:ns_tgv_capacity}), and each
launch covered at least 30~s after the first recorded batch boundary.
Throughput is computed from the median batch step cost within each
launch; reported central values and interquartile ranges are then
calculated across the four launch estimates. All 20 reported launches
completed normally, and the relative range of median step costs was
below 0.14\% for every implementation. Median active SM clocks ranged
from 1518.5 to 1530~MHz, and the memory clock was 877~MHz.

\paragraph{Controlled complete-step results.}
For the matched pair, Local ADER reaches 230.87~Mnode updates/s,
compared with 118.76~Mnode updates/s for RK3-F. The corresponding costs
are 4.332 and 8.420~ns per node per complete step, giving a throughput
ratio of 1.944. This measurement includes the full local prediction
cost and establishes a net benefit of the one-step construction in the
tested GPU implementation.

Both local integrators take $\Delta t^*\simeq5.6043\times10^{-4}$. ADER advances at 0.004695
nondimensional time units per wall-clock second and RK3-F at 0.002416,
giving a physical-advancement ratio of approximately 1.943. The
throughput gain is therefore retained in physical advancement over these
measured smooth-flow windows. \workprecisionrevision{The next comparison examines whether
this reduction in step cost translates into lower cost at prescribed
solution accuracy when each integrator selects its own time step.}

\begin{workprecisionrevisiontext}
\paragraph{Time to solution at matched error thresholds.}
A reduction in complete-step cost is useful for an accurate simulation
when it also reduces the time required to meet a prescribed error
threshold. We therefore compare the two integrators using the sinusoidal
density wave and isentropic vortex with exact solutions defined in
Section~\ref{sec:convective_accuracy}. Both tests solve the Euler equations
with the same spatial operator and selective dissipation. The density
wave is advanced to $t=1$ on $128^3$ and $256^3$ grids; the vortex is
advanced through one advection period, $t=20\sqrt{\gamma}
\simeq23.664319$, on $256\times256\times10$ and
$384\times384\times10$ grids. The time step is recomputed using
Eq.~\eqref{eq:cfl_time_step_rewrite}, with zero molecular-transport rate,
and the last step is shortened to reach the common terminal time exactly.

The density criterion uses the absolute RMS error $E_2$ defined in
Eq.~\eqref{eq:smooth-advection-density-error}. To assess the complete
conservative state as well, the vortex tests also use the aggregate
relative error
\[
E_{W,\mathrm{rel}}=
\left[
\frac{\displaystyle\sum_{\boldsymbol m}
\left\|\boldsymbol W_{\boldsymbol m}^{\mathrm{num}}
       -\boldsymbol W_{\boldsymbol m}^{\mathrm e}\right\|_2^2}
{\displaystyle\sum_{\boldsymbol m}
\left\|\boldsymbol W_{\boldsymbol m}^{\mathrm e}\right\|_2^2}
\right]^{1/2}.
\]
Here the superscripts $\mathrm{num}$ and $\mathrm e$ denote numerical
and exact terminal states, and each vector norm includes all five
nondimensional conservative components. For each grid and error criterion,
both methods must satisfy the same threshold in
Table~\ref{tab:matched_error_cost}. Each integrator independently brackets
a passing and a failing CFL value and applies six bisections to select
the largest tested passing value in that bracket. Thus the comparison
accounts for differences in the time steps admitted by the specified
accuracy requirement.

Each paired comparison uses one V100-SXM2-16GB GPU with an exclusive
node allocation and FP64 arithmetic. After ten warm-up steps at
$\mathrm{CFL}=0.125$, the driver restores the common initial state and
times the complete trajectory. Synchronized wall time includes all CFL
evaluations and reductions, halo operations, and ADER or RK stages;
initialization, state resetting, and exact-error evaluation are excluded.
The selected CFL values are each measured in four independent runs, with
the method order reversed in alternating repetitions. The relative range
of repeated wall times is at most 1.32\% of the median across the selected
configurations. All 48 selected
trajectories reach their prescribed terminal times, satisfy their error
thresholds, maintain positive density and pressure, and have normalized
conservation errors below $10^{-10}$.

\begin{table}[pos=!htbp]
    \color{manuscriptworkprecision}
    \caption{\workprecisionrevision{Complete-trajectory cost at common
    exact-solution error thresholds on one V100 GPU in FP64. CFL values
    are selected independently for the two methods using
    Eq.~\eqref{eq:cfl_time_step_rewrite}. Times are medians of four
    independent runs; the speedup is $T_{\mathrm{RK3-F}}/T_{\mathrm{ADER}}$.}}
    \label{tab:matched_error_cost}
    \centering
    \small
    \setlength{\tabcolsep}{3pt}
    \renewcommand{\arraystretch}{1.12}
    \begin{tabular*}{\linewidth}{@{\extracolsep{\fill}}llcrrrrr@{}}
    \toprule
    Case & Grid & Error threshold & \multicolumn{2}{c}{CFL}
    & \multicolumn{2}{c}{Time [s]} & Speedup \\
    \cmidrule(lr){4-5}\cmidrule(lr){6-7}
    & & & ADER & RK3-F & ADER & RK3-F & \\
    \midrule
    Density wave & $128^3$ & $E_2\leq10^{-9}$ & 0.500000 & 0.640625 & 0.944 & 1.739 & 1.842 \\
    Density wave & $256^3$ & $E_2\leq10^{-10}$ & 0.460938 & 0.593750 & 15.810 & 30.859 & 1.952 \\
    Vortex & $256^2\times10$ & $E_2\leq10^{-6}$ & 0.531250 & 0.710938 & 6.297 & 11.897 & 1.889 \\
    Vortex & $384^2\times10$ & $E_2\leq10^{-7}$ & 0.324219 & 0.500000 & 32.350 & 54.465 & 1.684 \\
    Vortex & $256^2\times10$ & $E_{W,\mathrm{rel}}\leq10^{-6}$ & 0.398438 & 0.609375 & 8.394 & 13.892 & 1.655 \\
    Vortex & $384^2\times10$ & $E_{W,\mathrm{rel}}\leq10^{-7}$ & 0.218750 & 0.433594 & 48.166 & 62.983 & 1.308 \\
    \bottomrule
    \end{tabular*}
\end{table}

Table~\ref{tab:matched_error_cost} shows that the complete-step saving
translates into lower time to solution at every tested error threshold.
The density-wave speedups are 1.842 and 1.952. For vortex advection,
they are 1.684--1.889 under the density criterion and 1.308--1.655 under
the aggregate conservative-state criterion. For the stricter joint-state
threshold on the $384\times384\times10$ grid, the measured errors are
$9.86935\times10^{-8}$ for ADER and $9.86900\times10^{-8}$ for RK3-F.
RK3-F takes 4280 steps compared with 8482 for ADER, corresponding to a
mean time step larger by a factor of 1.982; nevertheless, ADER reduces
the complete-trajectory wall time by 23.5\%. These exact-solution
comparisons demonstrate that the lower step cost remains beneficial
after each integrator selects its own time step to meet the common
accuracy requirement.
\end{workprecisionrevisiontext}

\paragraph{Hardware work and execution mechanism.}
To explain the complete-step saving, we analyze a separate matched
kernel workload on another V100-SXM2-16GB GPU. This profiling group
uses the $268^3$ NS-TGV configuration and the same predictor,
corrector, and RK kernels, with a directional-maximum time-step
selector. This selector differs from the time-step estimator of
Eq.~\eqref{eq:cfl_time_step_rewrite} used for the native controlled
comparison. The kernel timings and counters identify the computation
and data movement of the temporal construction; the complete-step
throughput gain is supplied by the native comparison above.

NVIDIA Nsight Systems provides CUDA execution timelines that identify
the operation groups and kernel launches belonging to each complete
step. Nsight Compute provides hardware counters for the corresponding
floating-point work and data transfers. Each profiling mode uses two
independent launches per integrator with reversed method ordering;
each sampled window contains steps 65--69 after 64 warm-up steps.
Native checks use the same windows, and their wall times differ from
the traced measurements by less than 0.17\%. Application replay repeats
the preceding computation, with profiler cache flushing and clock
locking disabled. The collected kernel sequences contain no overlapping
GPU execution in these single-rank windows.

For one kernel, let $F$ denote the executed FP64 operation count,
$Q$ the bytes read from and written to high-bandwidth memory (HBM),
and $t_k$ its device execution time. The arithmetic intensity,
floating-point rate, and HBM bandwidth are

\[
I_{\rm HBM}=\frac{F}{Q},\qquad
P_{\rm FP64}=\frac{F}{t_k},\qquad
B_{\rm HBM}=\frac{Q}{t_k}.
\]

The FP64 count includes executed additions and multiplications, with
each fused multiply-add contributing two operations. Integer address
arithmetic and other instruction classes are not represented by this
count. Following the Roofline model~\citep{WilliamsWatermanPatterson2009},
the reference performance envelope is

\[
P_{\rm roof}(I_{\rm HBM})=
\min\!\left(P_{\rm ref},\,B_{\rm ref}I_{\rm HBM}\right),
\]

where $P_{\rm ref}$ and $B_{\rm ref}$ are the FP64 arithmetic and HBM bandwidth
ceilings for the specified device clocks. Arithmetic and bandwidth
utilization are normalized by their respective ceilings. The reference
values are 7.83~TFLOP/s in FP64 and 898~GB/s for HBM, evaluated from
the V100's nominal 1530/877-MHz clocks.

\begin{figure}[pos=!htbp]
    \centering
    \includegraphics[width=\linewidth]{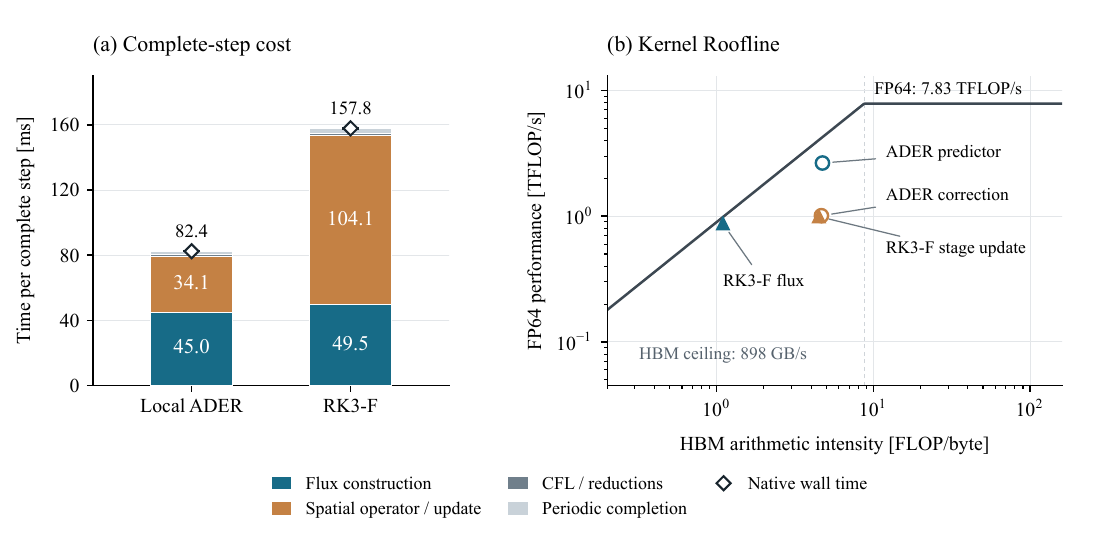}
    \caption{Hardware analysis of the matched $268^3$ NS-TGV computation.
    (a) GPU kernel time per complete step, grouped by operation; diamonds
    mark the corresponding native wall time. All three RK stages are
    accumulated. (b) FP64 HBM Roofline for the dominant kernel families.
    Open circles denote ADER and filled triangles denote RK3-F. Blue denotes
    flux construction: local prediction and integration in ADER, and
    instantaneous-flux evaluation in RK3-F (including sensors in panel a).
    Brown denotes spatial differentiation, dissipation, and state update:
    one ADER conservative correction or three RK stage updates.
    Intensities and rates are formed from summed counters and
    device times. Ranges span the two independent launches and are small
    on these axes. The ceilings use the nominal 1530/877-MHz device clocks.}
    \label{fig:ns_tgv_hardware_mechanism}
\end{figure}

Figure~\ref{fig:ns_tgv_hardware_mechanism}(a) first resolves the saving
by operation group. Local prediction and flux integration require
45.02~ms per ADER step, compared with 49.49~ms for the three
instantaneous-flux and sensor evaluations in RK3-F. Each RK stage uses
its instantaneous state without a local space--time predictor. The
ADER construction costs about 2.73 times one RK flux-evaluation stage.
The shared kernel for spatial differentiation, dissipation, and state
update is applied once for the ADER conservative correction, requiring
34.09~ms, and three times for the RK stage updates, requiring
104.09~ms in total. This operation group accounts for 92.6\% of the
reduction in traced GPU kernel time. The measured kernel-launch count
decreases from 38 to 19 per complete step, including periodic completion
and CFL reductions.

The counter totals then quantify the change in work and traffic.
For $K$ complete steps, the per-node work and traffic are formed
from $\sum F/(KN_{\rm int})$ and $\sum Q/(KN_{\rm int})$, respectively, with all executed kernels
included in the sums. In these normalizations, $N_{\rm int}$ denotes
the interior-node count of the profiling grid. The corresponding
aggregate intensity is $\sum F/\sum Q$. These totals include periodic completion and all three
RK stages. They quantify transferred data per update, whereas
Table~\ref{tab:gpu_storage} quantifies allocated resident capacity.

\begin{table}[pos=!htbp]
    \caption{Measured hardware work per complete step for the $268^3$
    viscous NS-TGV case. Operations and HBM bytes are summed over all
    kernels and normalized by the number of interior nodes and complete
    steps. Both independent launches contribute five steps after warm-up.
    The HBM column includes reads and writes; $I_{\rm step}$ is the ratio
    of the summed operation and byte counts.}
    \label{tab:ns_tgv_hardware_work}
    \centering
    \small
    \setlength{\tabcolsep}{4pt}
    \renewcommand{\arraystretch}{1.12}
    \begin{tabular*}{\linewidth}{@{\extracolsep{\fill}}lrrrr@{}}
    \toprule
    Integrator & FP64 work & HBM traffic & $I_{\rm step}$ & Launches \\
      & [FLOP/node/step] & [B/node/step] & [FLOP/B] & [/step] \\
    \midrule
    Local ADER & 8\,226.3 & 1\,782.6 & 4.615 & 19 \\
    RK3-F & 9\,766.1 & 3\,074.7 & 3.176 & 38 \\
    \bottomrule
    \end{tabular*}
\end{table}

Table~\ref{tab:ns_tgv_hardware_work} shows that ADER executes 15.8\%
fewer counted FP64 operations per complete step and transfers 42.0\%
fewer HBM bytes than RK3-F. The aggregate arithmetic intensity changes
from 3.176 to 4.615~FLOP/B. These quantities measure the complete temporal
construction, including its local predictor, rather than inferring
traffic from the number of resident arrays.

Finally, Figure~\ref{fig:ns_tgv_hardware_mechanism}(b) relates this work
to the characteristics of the dominant kernels. The ADER predictor
reaches 4.73~FLOP/B and 2.64~TFLOP/s, while the RK instantaneous-flux
kernel reaches 1.09~FLOP/B and 0.87~TFLOP/s. Their HBM bandwidth
utilizations are 62.3\% and 88.5\%, respectively. The shared kernel for
spatial differentiation and state update attains 24.1\% and 24.7\% of
the nominal HBM bandwidth in ADER and RK3-F, respectively. Its
per-invocation performance is comparable, while ADER requires one
global update instead of three RK stage updates. Together, the cost
decomposition and counter totals explain the execution mechanism of
the complete temporal construction and its GPU data flow.
\subsection{Application-level comparison with established GPU solvers}
\label{sec:external_solver_comparison}

The controlled comparison in Section~\ref{sec:single_gpu_performance}
quantifies the benefit of the local-ADER temporal construction, and the
kernel profiles identify its principal execution costs. The
present comparison places that implementation alongside STREAmS-2, MFC,
and HTR for application-level advancement of the same NS-TGV problem
\citep{BernardiniEtAl2023,RadhakrishnanEtAl2024,DiRenzoEtAl2020}.
It uses the common physical model, grid, FP64 precision, hardware, and
timing protocol of Section~\ref{sec:single_gpu_performance}. Each solver
retains its native spatial discretization, time integrator, data layout,
and GPU backend, as recorded in Table~\ref{tab:benchmark_configurations}.

\begin{table}[pos=!htbp]
    \caption{Native configurations used for the common viscous NS-TGV
    comparison. Constant viscosity and zero heat conduction apply to every row.}
    \label{tab:benchmark_configurations}
    \centering
    \footnotesize
    \setlength{\tabcolsep}{3pt}
    \renewcommand{\arraystretch}{1.15}
    \begin{tabular*}{\linewidth}{@{\extracolsep{\fill}}
    >{\raggedright\arraybackslash}p{0.16\linewidth}
    >{\raggedright\arraybackslash}p{0.36\linewidth}
    >{\raggedright\arraybackslash}p{0.42\linewidth}@{}}
    \toprule
    Implementation & Spatial and temporal path & GPU and runtime choices \\
    \midrule
    Local ADER / RK3-F
    & Shared central finite differences and selective dissipation;
      local ADER / three-stage full-residual RK
    & C++/CUDA; existing fused kernels; the same CFL estimator and
      transport operator; additional RK state allocated only for RK3-F. \\
    STREAmS-2
    & Sixth-order energy-preserving convection, sixth-order central
      conservative viscosity, native RK
    & CUDA Fortran; asynchronous production path; uniform periodic grid;
      shock sensor threshold 1.01; matching NVHPC MPI runtime. \\
    MFC
    & Native single-fluid IGR with fifth-order reconstruction,
      Lax--Friedrichs flux, RK3
    & Case-specialized OpenACC GPU execution; three fixed IGR iterations
      and warm-start iterations; no unified-memory spilling. \\
    HTR
    & Native skew-symmetric convection, viscous operator,
      and RK; TaylorGreen3DVortex initialization
    & Regent/Legion GPU tasks; tracing enabled; one tile in the final
      single-GPU runs; framebuffer pool 15000~MiB. \\
    \bottomrule
    \end{tabular*}
\end{table}

The native production advancement paths are retained. Adapters provide
analytic initialization where required, parameter conversion, allocation
bookkeeping, and synchronized timers. The common case excludes optional
forcing, chemistry, multiphase interfaces, and field output. For MFC, an
initialization adapter supplies the conservative state in binary format by
evaluating the common analytic field at the native finite-volume cell
centers before timing. STREAmS-2 uses its supported CUDA Fortran toolchain
and matching MPI libraries, while HTR retains task-runtime tracing and GPU
memory pools. Setup work is excluded from the advancement measurements.

Parameters are matched through the equations of state and transport
coefficients. The initial volume-averaged kinetic energy, enstrophy, and
total energy are $1/8$, $3/8$, and $178.6964285714$, respectively.
Finite-difference nodes and finite-volume cells retain their native
initialization conventions for the same periodic field. MFC's inclusive
upper indices $m=n=p=301$ represent $302^3$ cells, and its dimensional
time is converted with $U_0=0.1\sqrt{1.4\times101325}$ and $L_0=1$.

\begin{figure}[pos=!htbp]
    \centering
    \includegraphics[width=\linewidth]{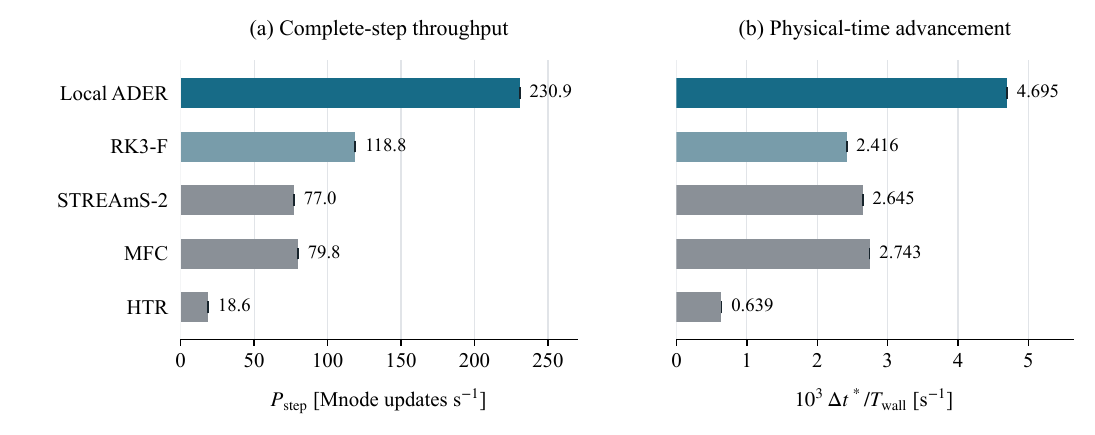}
    \caption{\textcolor[rgb]{0,0,0}{Single-GPU advancement of the same $302^3$ viscous
    NS-TGV problem. (a) Complete-step throughput. (b) Physical-time
    advancement per wall-clock second over a common 30-s window, obtained
    from the measured time stamps. Bars show medians over four independent
    launches; whiskers show interquartile ranges. Local ADER and RK3-F use
    Eq.~\eqref{eq:cfl_time_step_rewrite}; the external solvers retain their
    native CFL estimators. The measurements concern the initial smooth-flow
    window and do not constitute an equal-error comparison of complete
    turbulent decay.}}
    \label{fig:ns_tgv_single_gpu}
\end{figure}

All solvers evaluate their time-step estimator at every complete step with
an input CFL of 0.5. Local ADER and RK3-F use
Eq.~\eqref{eq:cfl_time_step_rewrite}, while STREAmS-2, MFC, and HTR retain
their native estimators. Table~\ref{tab:ns_tgv_results} and
Figure~\ref{fig:ns_tgv_single_gpu} collect the five implementations under
the common measurement protocol; the local-ADER and RK3-F entries are the
controlled results discussed in Section~\ref{sec:single_gpu_performance}.
The tabulated $t^*_{\rm end}$ is the terminal physical time of the
representative first launch. Figure~\ref{fig:ns_tgv_single_gpu}(b) instead
uses the measured physical-time increment over the common 30-s window,
as defined in Section~\ref{sec:single_gpu_performance}; the terminal times
are not those window increments.

\begin{table}[pos=!htbp]
    \color[rgb]{0,0,0}
    \caption{Measured single-GPU NS-TGV performance at $302^3$ in FP64.
    Brackets contain the interquartile range across four independent launch
    estimates. The last two columns give the final measured nondimensional
    step size and terminal physical time of the representative first launch.}
    \label{tab:ns_tgv_results}
    \centering
    \small
    \setlength{\tabcolsep}{3pt}
    \renewcommand{\arraystretch}{1.13}
    \begin{tabular*}{\linewidth}{@{\extracolsep{\fill}}lrrrr@{}}
    \toprule
    Implementation & $P_{\rm step}$ [Mnode/s] & $g_{\rm step}$ [ns] & $10^3\Delta t^*$ & $t^*_{\rm end}$ \\
    \midrule
    Local ADER & 230.87 [230.79, 230.88] & 4.332 & 0.56043 & 0.20736 \\
    RK3-F & 118.76 [118.69, 118.83] & 8.420 & 0.56043 & 0.10312 \\
    STREAmS-2 & 77.03 [77.02, 77.06] & 12.982 & 0.94568 & 0.11443 \\
    MFC & 79.79 [79.75, 79.84] & 12.532 & 0.94570 & 0.12105 \\
    HTR & 18.61 [18.60, 18.61] & 53.737 & 0.94569 & 0.02837 \\
    \bottomrule
    \end{tabular*}
\end{table}

The external complete-step throughputs are 77.03, 79.79, and
18.61~Mnode updates/s for STREAmS-2, MFC, and HTR, respectively.
Relative to these configurations, local ADER has throughput ratios of
2.997, 2.893, and 12.406. The three external configurations take
$\Delta t^*\simeq9.457\times10^{-4}$, compared with
$\Delta t^*\simeq5.6043\times10^{-4}$ for the two local integrators.
Their larger physical increments reduce the corresponding ADER
physical-advancement ratios to 1.775, 1.712, and 7.348 over the common
30-s window. The two panels of Figure~\ref{fig:ns_tgv_single_gpu} therefore
separate complete-step cost from physical-time advancement.

These measurements characterize the stated native configurations and
initial smooth-flow windows. Their numerical and runtime workloads differ
as specified in Table~\ref{tab:benchmark_configurations}, so the external
ratios cannot isolate the effect of ADER versus RK; that attribution rests
on the controlled RK3-F comparison in
Section~\ref{sec:single_gpu_performance}.
\subsection{Multi-GPU scalability}
\label{sec:multi_gpu_scalability}

The single-GPU comparisons establish the local execution cost and its
relation to the numerical workload. Distributed advancement adds halo
exchange and inter-rank synchronization to these operations. The following
experiments examine whether the local-ADER advancement rate is maintained
when these dependencies enter the complete-step cost. They use the same viscous NS-TGV equations,
zero heat conduction, and FP64 representation. One MPI rank is assigned
to each V100 GPU, with two host CPUs per rank, using 4--32 GPUs across
four eight-GPU nodes. Node-local exchanges use CUDA IPC, while cross-node
exchanges use pinned-host staging and the overlap schedule described in
Section~\ref{sec:gpu_resident_execution}, with OpenMPI 4.1.8 and UCX over
100-Gbit/s InfiniBand.

All configurations within the scaling sweep use the same executable and
evaluate a state-dependent time step at every step. For these distributed
measurements, the selector uses the summed inverse-time rate
\[
\sigma_{\boldsymbol m,\Sigma}^{n}
=\sum_{j=1}^{3}\frac{|u_{j,\boldsymbol m}^{n}|+c_{\boldsymbol m}^{n}}
                         {h_{j,\boldsymbol m}}
 +\frac{8\mu}{3\rho_{\boldsymbol m}^{n}}
  \sum_{j=1}^{3}\frac{1}{h_{j,\boldsymbol m}^{2}},
\]
with $\mathrm{CFL}=0.5$. \textcolor[rgb]{0,0,0}{This selector sums the
directional acoustic rates, whereas the single-GPU comparison uses their
Euclidean norm in Eq.~\eqref{eq:cfl_time_step_rewrite}.} The scaling ratios are formed entirely
within the distributed configuration and measure wall time per complete
step, including time-step selection and communication. No communication
profiler is active during these measurements.

Strong scaling fixes $768\times768\times384$ interior locations,
or $226\,492\,416$ nodes, in
$[0,4\pi)\times[0,4\pi)\times[0,2\pi)$. Weak scaling retains $384^3$
interior locations per GPU and enlarges the domain by complete periodic
vortex copies, preserving grid spacing, unit-vortex length, Mach number,
and Reynolds number. This local grid is selected within ADER's capacity
because the distributed experiment concerns that implementation alone.
The Cartesian partitions for 4, 8, 16, and 32 GPUs are
$2\times2\times1$, $2\times2\times2$, $4\times2\times2$, and
$8\times2\times2$, respectively. The corresponding weak-scaling grids
are $768\times768\times384$, $768^3$, $1536\times768\times768$, and
$3072\times768\times768$.

Two independent sweeps use forward and reverse configuration order. Each
launch records five synchronized batches, taking the maximum elapsed
rank time for each batch. Warm-up counts of 23--177 steps and batch lengths
of 9--71 steps provide at least five seconds of warm-up and two seconds
per batch. The central step cost is the median of the two launch medians.
With $T_p$ denoting the complete-step wall time on $p$ GPUs, the
four-GPU-referenced strong-scaling speedup and efficiency are
\[
S(p)=\frac{T_4}{T_p},\qquad
E_{\rm s}(p)=\frac{S(p)}{p/4},
\]
and the weak-scaling efficiency is $E_{\rm w}(p)=T_4/T_p$ at fixed work
per GPU. These efficiencies quantify parallel scaling of the complete
advancement operation.

\begin{figure}[pos=!htbp]
    \centering
    \includegraphics[width=\linewidth]{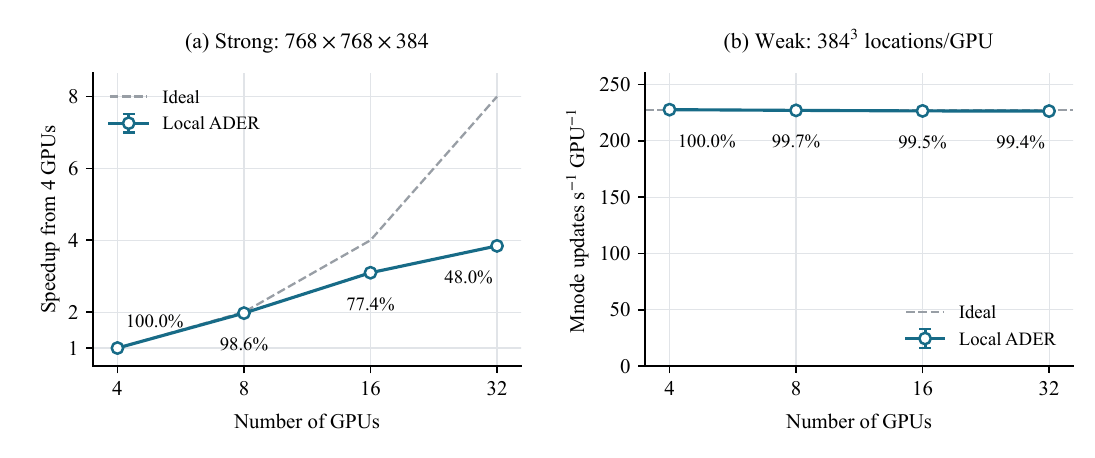}
    \caption{Local-ADER scalability on 4--32 V100 GPUs. (a) Strong scaling
    of $768\times768\times384$ locations, referenced to four GPUs.
    (b) Weak scaling with $384^3$ locations per GPU. Labels give parallel
    efficiencies. Symbols show the median over two independent launches;
    whiskers span their launch medians and are mostly smaller than the
    symbols. Dashed lines indicate ideal scaling.}
    \label{fig:ns_tgv_scaling}
\end{figure}

\begin{table}[pos=!htbp]
    \caption{Local-ADER strong and weak scaling for the zero-conduction
    NS-TGV benchmark. Rates count complete physical steps and include
    communication. Both scaling modes share the same four-GPU baseline.}
    \label{tab:ns_tgv_scaling}
    \centering
    \small
    \setlength{\tabcolsep}{3pt}
    \renewcommand{\arraystretch}{1.12}
    \begin{tabular*}{\linewidth}{@{\extracolsep{\fill}}lrrrrr@{}}
    \toprule
    Mode & GPUs & Step time [s] & $P_{\rm step}$ [Mnode/s] & $P_{\rm step}$/GPU & Efficiency [\%] \\
    \midrule
    Strong & 4 & 0.248745 & 910.5 & 227.6 & 100.00 \\
           & 8 & 0.126146 & 1795.5 & 224.4 & 98.59 \\
           & 16 & 0.080344 & 2819.0 & 176.2 & 77.40 \\
           & 32 & 0.064721 & 3499.5 & 109.4 & 48.04 \\
    \midrule
    Weak & 4 & 0.248745 & 910.5 & 227.6 & 100.00 \\
         & 8 & 0.249455 & 1815.9 & 227.0 & 99.72 \\
         & 16 & 0.249978 & 3624.2 & 226.5 & 99.51 \\
         & 32 & 0.250140 & 7243.7 & 226.4 & 99.44 \\
    \bottomrule
    \end{tabular*}
\end{table}

Figure~\ref{fig:ns_tgv_scaling} and Table~\ref{tab:ns_tgv_scaling} show
the different effects of reducing and maintaining local work. For the
fixed global grid, speedups relative to four GPUs are 1.972, 3.096, and
3.843 on 8, 16, and 32 GPUs. Strong-scaling efficiency decreases to
48.04\% at 32 GPUs as the subdomains become smaller. This trend is
consistent with the increasing importance of distributed dependencies
relative to local computation and a smaller opportunity for overlap;
the aggregate timings do not separately resolve those contributions.

At fixed $384^3$ work per GPU, the step time changes from 0.248745~s
on four GPUs to 0.250140~s on 32 GPUs. The largest configuration contains
$1\,811\,939\,328$ interior locations and reaches
7243.7~Mnode updates/s, or 226.4~Mnode updates/s per GPU, retaining
99.44\% weak-scaling efficiency. The weak-scaling result establishes that
the distributed execution preserves the per-GPU advancement rate for
large resident subdomains, while the strong-scaling result quantifies the
efficiency loss when the same global problem is divided into smaller
local workloads.

\section{Numerical results}
\label{sec:numerical_results}

\subsection{Accuracy of the local ADER convective discretization}
\label{sec:convective_accuracy}

This subsection evaluates the accuracy of the complete local ADER convective
discretization using two smooth periodic advection problems with known exact
solutions. \accuracyrevision{The tests assess coupled space--time convergence at fixed
CFL numbers. Varying the CFL number distinguishes background spatial error
from the temporal and mixed contributions of the local predictor--corrector
construction, without introducing boundary or molecular-transport errors.}

Two advection problems are considered. The first is a three-dimensional
sinusoidal density wave. It provides a simple setting in which the spatial and
temporal contributions to the error can be distinguished over a relatively
wide range of mesh resolutions. The compressible Euler equations are solved in
the periodic domain $[0,2\pi]^3$ with $\gamma=1.4$ and initial data
\begin{equation}
  (\rho,u,v,w,p)(x,y,z,0)
  =\left(1+0.1\sin(x+y+z),\,0.7,\,-0.2,\,0.3,\,1\right).
  \label{eq:sinwave3d-initial-condition}
\end{equation}
Since the velocity and pressure are uniform, the density profile is translated
without changing shape. At time $t$, the exact density is
\[
  \rho_{\mathrm{e}}(x,y,z,t)
  =1+0.1\sin\!\left[x+y+z-(u+v+w)t\right].
\]
The solution is advanced to $t=1$ on uniform $N^3$ meshes with
$N=10,11,\ldots,100$.

The second problem is the two-dimensional isentropic-vortex advection test,
which has been used extensively to assess the convergence of high-order
discretizations for smooth Euler flows~\citep{BalsaraShu2000,
SciacovelliEtAl2021,TitarevToro2004}. A vortex is superimposed on a uniform
perfect-gas flow. In nondimensional variables, the perturbations are
\[
  \delta u=-y\Omega,\qquad
  \delta v=x\Omega,\qquad
  \delta T=-\frac{\gamma-1}{2}\Omega^2,
  \qquad
  \Omega=\beta\exp\!\left[-\frac{x^2+y^2}{2}\right].
\]
The standard vortex strength corresponding to $\epsilon=5$ is used, together
with $R=\sigma=1$, $M_\infty=\sqrt{2/\gamma}$, and an advection angle of
$45^\circ$. The initial primitive variables follow from the isentropic
relations,
\[
  \rho=(1+\delta T)^{1/(\gamma-1)},\qquad
  u=\frac{1}{\sqrt{\gamma}}+\delta u,\qquad
  v=\frac{1}{\sqrt{\gamma}}+\delta v,\qquad
  p=\frac{1}{\gamma}(1+\delta T)^{\gamma/(\gamma-1)}.
\]
Periodic boundary conditions are imposed on $[-10,10]^2$. This larger domain
reduces the influence of the weak artificial shear layers associated with the
nonzero velocity perturbation near a periodic boundary
\citep{SciacovelliEtAl2021}. The problem is evaluated with the
three-dimensional solver by using $N\times N\times10$ nodes and a solution
that is uniform in the third direction. Meshes from $N=100$ to $500$ in
increments of five are used. The error is measured after one advection cycle,
when the exact vortex has returned to its initial position.

For both problems, the reported quantity is the discrete density error
\begin{equation}
  E_2=\left[\frac{1}{N_{\mathrm{node}}}
  \sum_{\boldsymbol{i}}
  \left(\rho_{\boldsymbol{i}}^{\mathrm{num}}
       -\rho_{\boldsymbol{i}}^{\mathrm{e}}\right)^2
  \right]^{1/2}.
  \label{eq:smooth-advection-density-error}
\end{equation}
The CFL numbers $0.8$, $0.4$, $0.2$, $0.1$, $0.05$, and $0.025$ are examined.
For each run, the time step is recomputed before every accepted step from the
current multidimensional convective spectral radius; only the last step is
shortened to reach the prescribed final time exactly.

\begin{accuracyrevisiontext}
The separation of spatial and temporal error used in the isentropic-vortex
study of Sciacovelli et al.~\citep{SciacovelliEtAl2021} is supplemented here by
the compact-predictor contribution derived in
Remark~\ref{rem:complete_euler_consistency_rewrite}. In the smooth regime where
background dissipation dominates the spatial error, the resulting estimate is
\[
  E_2\lesssim C_t\Delta t_{\max}^{3}
      +C_{\mathrm{mix}}\Delta t_{\max}^{2}h+C_xh^7,
\]
where $\Delta t_{\max}$ is the maximum accepted step over the fixed
integration interval and the constants depend on the smooth solution.
At fixed CFL, $\Delta t_{\max}$ is proportional to $h$, so both the cubic
temporal contribution and the mixed contribution scale as $h^3$.
Reducing CFL suppresses these terms in proportion to its third and second
powers, respectively, extending the mesh range over which the $h^7$
background spatial behavior can be observed. On sufficiently fine meshes
at any fixed nonzero CFL, the leading coupled error can therefore return
to a third-order slope.
\end{accuracyrevisiontext}

\begin{figure}[pos=!htbp]
  \centering
  \includegraphics[width=\linewidth]{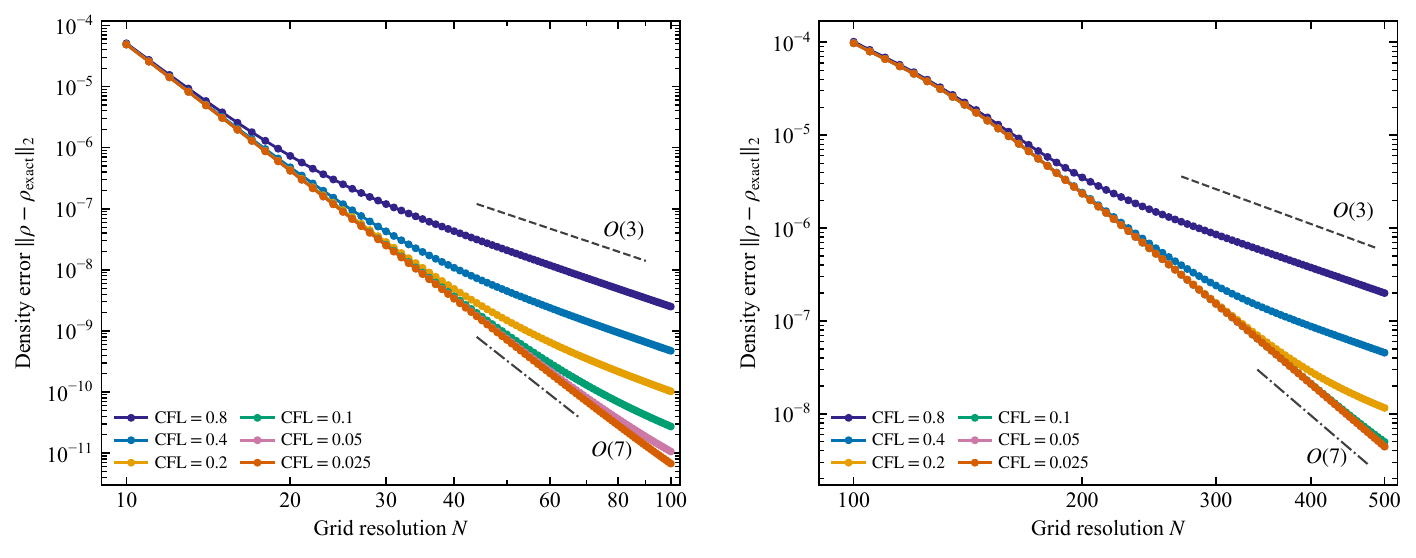}
  \caption{Density $L_2$ error for smooth periodic advection at six CFL
  numbers. The left panel shows the three-dimensional sinusoidal density wave
  on $N^3$ meshes at $t=1$; the right panel shows the isentropic vortex on
  $N\times N\times10$ meshes after one advection cycle. The dashed and
  dash-dotted segments indicate third- and seventh-order slopes,
  respectively.}
  \label{fig:smooth-advection-convergence}
\end{figure}

Figure~\ref{fig:smooth-advection-convergence} reports the two convergence
studies. For the sinusoidal wave, the low-CFL curves follow the seventh-order
reference slope over the resolved range. A log--log fit over $20\leq N\leq40$
at CFL $=0.025$ gives an observed order of 6.96. At CFL $=0.8$, the curve bends
toward the third-order reference slope as the mesh is refined, with an order of
3.04 over $60\leq N\leq100$. The intermediate CFL values progressively move
this transition toward finer meshes. The same separation of the two error
contributions is visible for the isentropic vortex. Over $300\leq N\leq500$,
the CFL $=0.8$ curve gives an observed order of 2.84, whereas the curves for
CFL $\leq0.1$ are nearly coincident and give orders between 6.73 and 6.93.
The CFL $=0.4$ and $0.2$ results lie between these temporal- and
spatial-error-dominated regimes. \accuracyrevision{These observations are consistent with
the complete Euler error balance in
Remark~\ref{rem:complete_euler_consistency_rewrite}. The measured slopes concern
coupled mesh and time-step refinement; they do not isolate temporal order
on a fixed spatial discretization.}

\subsection{Compressible Taylor--Green vortex}
\label{sec:tgv}

The compressible Taylor--Green vortex is used to assess the behavior of the
complete Navier--Stokes discretization after the smooth-advection tests. Its
initial large-scale vortical field develops thin shear layers and a broad
range of smaller scales before decaying. The calculation therefore tests, in a
single three-dimensional configuration, the nonlinear sensor, the
sensor-controlled artificial dissipation, the viscous operator, and the
local ADER predictor--corrector update.

The configuration adopts the canonical compressible Taylor--Green vortex
benchmark used in high-order-method assessments~\citep{DeBonis2013,
PengYang2018}. The nondimensional domain is
$\mathcal{V}=[0,2\pi]^3$ with periodic boundary conditions in all directions.
The initial velocity is
\begin{equation}
  \begin{aligned}
    u(x,y,z,0)&=\sin x\cos y\cos z,\\
    v(x,y,z,0)&=-\cos x\sin y\cos z,\\
    w(x,y,z,0)&=0.
  \end{aligned}
  \label{eq:tgv-initial-velocity}
\end{equation}
The initial pressure is
\begin{equation}
  p(x,y,z,0)=\frac{1}{\gamma M^2}
  +\frac{1}{16}\bigl(\cos 2x+\cos 2y\bigr)
   \bigl(2+\cos 2z\bigr),
  \label{eq:tgv-initial-pressure}
\end{equation}
with uniform initial density $\rho=\rho_{\mathrm{ref}}=1.292$. The remaining
thermodynamic variable is obtained from the perfect-gas equation of state. The
parameters are $\mathrm{Re}=1600$, $\mathrm{Pr}=0.71$, and $\gamma=1.4$.
Initial Mach numbers $M=0.1$, $0.5$, and $1.0$ are considered on uniform
nodal meshes with $64^3$, $128^3$, $256^3$, $512^3$, and $1024^3$ nodes.
All calculations use the complete local ADER scheme, including the
compression--Ducros--pressure sensor and both artificial-dissipation terms.
The time step is recomputed before every accepted step with
$\mathrm{CFL}=0.5$, and the final time is $t=16$.

The primary reported quantities are the volume-averaged kinetic energy and the
resolved solenoidal enstrophy,
\begin{equation}
  \begin{aligned}
    K(t)&=\frac{1}{\rho_{\mathrm{ref}}|\mathcal{V}|}
    \int_{\mathcal{V}}\frac{1}{2}\rho u_i u_i\,\mathrm{d}\mathcal{V},\\
    \Omega(t)&=\frac{1}{\rho_{\mathrm{ref}}\mathrm{Re}|\mathcal{V}|}
    \int_{\mathcal{V}}\omega_i\omega_i\,\mathrm{d}\mathcal{V},
    \qquad \boldsymbol{\omega}=\nabla\times\boldsymbol{u}.
  \end{aligned}
  \label{eq:tgv-statistics}
\end{equation}
The velocity derivatives used for $\Omega$ employ the same eighth-order
centered stencil as the inviscid central operator. The kinetic-energy spectra
are obtained from a three-dimensional Fourier transform followed by spherical
shell binning. The spectra are examined near the enstrophy maximum, at
$t=9.0$, $9.3$, and $11.6$ for $M=0.1$, $0.5$, and $1.0$, respectively.

\begin{figure}[pos=!htbp]
  \centering
  \includegraphics[width=\linewidth]{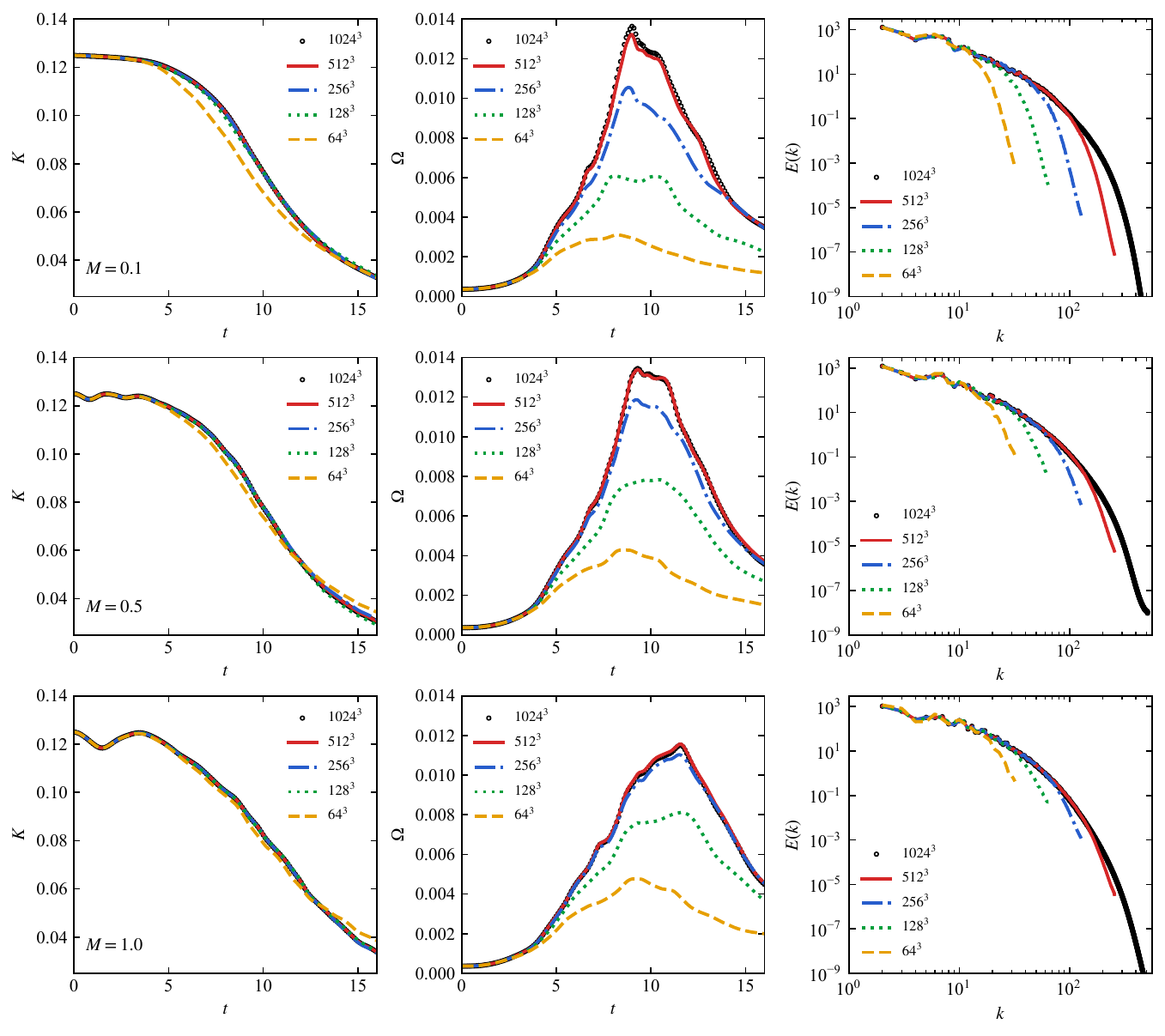}
  \caption{Grid-resolution dependence of the compressible Taylor--Green
  vortex computed with the local ADER method. The rows correspond to
  $M=0.1$, $0.5$, and $1.0$ from top to bottom; the columns show the
  volume-averaged kinetic energy, resolved solenoidal enstrophy, and
  kinetic-energy spectrum. Results span $64^3$--$1024^3$ for all three
  Mach numbers. The spectra are evaluated at $t=9.0$, $9.3$, and $11.6$,
  respectively.}
  \label{fig:tgv-statistics}
\end{figure}

Figure~\ref{fig:tgv-statistics} shows the grid-refinement results. The
kinetic-energy histories are comparatively insensitive to resolution because
they are dominated by the large scales. In contrast, the coarse meshes
substantially underestimate the enstrophy maximum; refinement recovers
progressively smaller vortical structures, and the two finest meshes give
closely agreeing peaks. On the $1024^3$ meshes, the sampled maxima are
$1.3630\times10^{-2}$ at $t=9.0$ for $M=0.1$,
$1.3443\times10^{-2}$ at $t=9.3$ for $M=0.5$, and
$1.1507\times10^{-2}$ at $t=11.6$ for $M=1.0$. The relative differences
between the $512^3$ and $1024^3$ enstrophy maxima, computed as the absolute
difference divided by the corresponding $1024^3$ value, are approximately
$2.96\%$, $0.34\%$, and $0.51\%$, respectively. Using the same normalization,
the kinetic-energy differences at $t=16$ are approximately $0.04\%$, $0.29\%$,
and $0.07\%$ for the three Mach numbers.

The spectra provide the more discriminating resolution evidence. Refinement
extends the high-wavenumber cutoff while leaving the energy-containing range
largely unchanged. The $64^3$ and $128^3$ calculations therefore reproduce
the large-scale decay but truncate a substantial part of the small-scale
content; the $256^3$ results recover the main trend but retain an earlier
cutoff. The $512^3$ and $1024^3$ spectra remain close over a substantially
broader range, although differences persist in the high-wavenumber tail.
No terminal upturn is visible near the cutoff in the plotted cases,
indicating no apparent accumulation of energy at the smallest resolved scales.

These results show that agreement in kinetic energy alone does not establish
resolution of the small-scale flow.

The spatial organization of the vortical structures provides a complementary
view of the same resolution effect. The second invariant of the
velocity-gradient tensor is defined as
\[
  Q=\frac{1}{2}\left(\|\boldsymbol{R}\|^2-\|\boldsymbol{S}\|^2\right),
\]
where
\[
  \boldsymbol{S}=\frac{1}{2}\left(\nabla\boldsymbol{u}
  +\nabla\boldsymbol{u}^{\mathsf T}\right),
  \qquad
  \boldsymbol{R}=\frac{1}{2}\left(\nabla\boldsymbol{u}
  -\nabla\boldsymbol{u}^{\mathsf T}\right).
\]
\begin{figure}[pos=!htbp]
  \centering
  \includegraphics[width=0.92\linewidth]{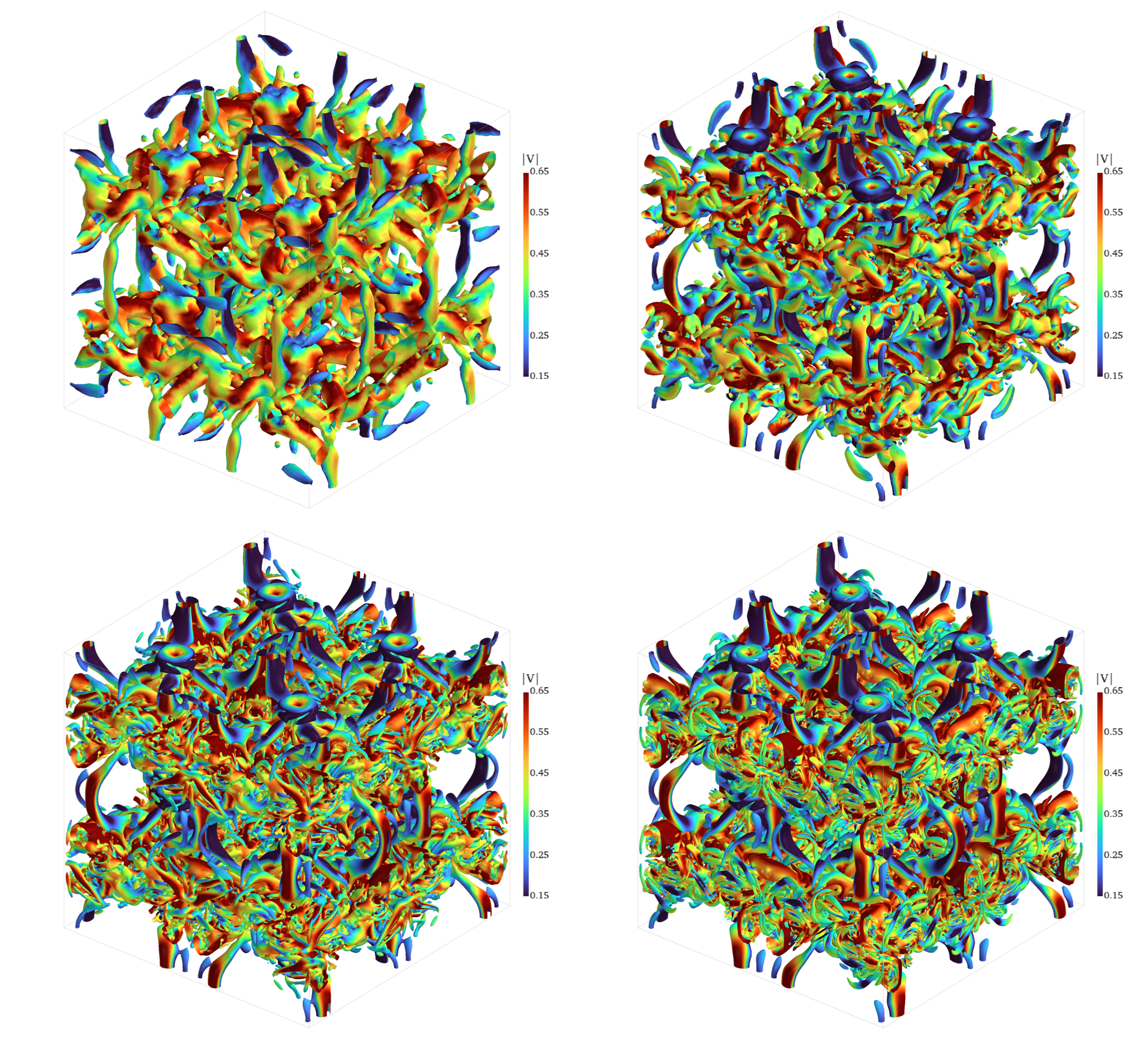}
  \caption{Instantaneous $Q=0.5$ isosurfaces for the $M=0.1$
  Taylor--Green vortex at $t=10$, computed with the local ADER method.
  The top-left, top-right, bottom-left, and bottom-right panels correspond
  to $64^3$, $128^3$, $256^3$, and $512^3$ meshes, respectively, and
  use a common color scale for the velocity magnitude.}
  \label{fig:tgv-qcriterion}
\end{figure}

The derivatives are evaluated with the same eighth-order centered stencil
used by the production spatial operator. Figure~\ref{fig:tgv-qcriterion}
shows the $Q=0.5$ isosurfaces for the $M=0.1$ case at $t=10$ on four mesh
resolutions. Refinement preserves the large-scale organization while
revealing progressively thinner and more numerous vortex filaments. This
visual comparison is not used as a standalone convergence criterion; it
complements the enstrophy and spectral measures by showing where the
additional resolved small-scale content is located.

\subsection{Decaying compressible isotropic turbulence}
\label{sec:compressible-isotropic}

The decaying compressible isotropic-turbulence problem complements the
organized structures of the Taylor--Green vortex with a broadband random
field. It tests the local ADER method under temperature-dependent molecular
transport and compressibility-dependent velocity gradients, and distinguishes
the resolution requirements of integral statistics from those of
gradient-based intermittency.

The configuration follows Wang et al.~\citep{WangWangPan2024}. The
computational domain is the periodic cube $\mathcal{V}=[-\pi,\pi]^3$.
A Gaussian random velocity field is projected in Fourier space onto the
plane normal to each nonzero wave vector. Its initial shell-integrated
energy spectrum is prescribed as
\begin{equation}
  E(k)=A_0 k^4\exp\left(-\frac{2k^2}{k_0^2}\right),
  \qquad A_0=1.3\times10^{-4},\quad k_0=8.
  \label{eq:cit-initial-spectrum}
\end{equation}
Unresolved modes are removed and the solenoidal field is rescaled to the
prescribed initial kinetic energy. The same velocity realization is used
for both Mach numbers and all mesh resolutions; the finer initial fields
are obtained by Fourier zero-padding of the resolved $64^3$ modes.
The initial density is uniform, $\rho_0=1$, and the uniform temperature
sets the initial turbulent Mach number,
\[
  M_{t,0}=\frac{\sqrt{\langle u_i u_i\rangle_0}}{a_0},
  \qquad a_0=\sqrt{\gamma T_0}.
\]
Here $\langle\cdot\rangle$ denotes a volume average and the gas constant
is normalized to unity. The two cases use $M_{t,0}=0.3$ and $1.2$,
with $\gamma=1.4$, $\mathrm{Pr}=0.7$, and reference Taylor-scale Reynolds
number $\mathrm{Re}_{\lambda,0}=72$. The viscosity follows
$\mu=\mu_0(T/T_0)^{0.76}$, with
$\mu_0=2.0057376824\times10^{-3}$.

Uniform nodal meshes of $64^3$, $128^3$, $256^3$, $512^3$, and $1024^3$
are used; the finest contains $1\,073\,741\,824$ spatial nodes.
All runs use the complete local ADER scheme with the
compression--Ducros--pressure sensor, the prescribed artificial dissipation,
temperature-dependent viscosity and heat conduction, and positivity recovery.
The time step is recomputed before every accepted step with
$\mathrm{CFL}=0.3$. The reference turnover time is
\[
  \tau_0=\frac{L_0}{u'_0}=0.5424170415,
  \qquad L_0=\frac{\sqrt{2\pi}}{k_0},
  \qquad u'_0=\sqrt{\frac{\langle u_i u_i\rangle_0}{3}}.
\]
Each calculation covers $0\leq t/\tau_0\leq5$, with 251 equally spaced
statistical samples including both endpoints.

The reported quantities are the normalized kinetic energy and density
fluctuation,
\begin{equation}
  \frac{K(t)}{K_0},\qquad
  K(t)=\frac{\langle\rho u_i u_i\rangle}{2\rho_0},\qquad
  \frac{\rho_{\mathrm{rms}}}{M_{t,0}^2},\qquad
  \rho_{\mathrm{rms}}=
  \left\langle(\rho-\langle\rho\rangle)^2\right\rangle^{1/2},
  \label{eq:cit-statistics}
\end{equation}
where $K_0=K(0)$, together with the longitudinal velocity-gradient skewness
\begin{equation}
  S_u=\frac{\left\langle(u_x^3+v_y^3+w_z^3)/3\right\rangle}
  {\left\langle(u_x^2+v_y^2+w_z^2)/3\right\rangle^{3/2}}.
  \label{eq:cit-skewness}
\end{equation}
The diagonal velocity derivatives use the same eighth-order centered
stencil as the production spatial operator. The initial and final
statistics are independently recomputed from the conservative fields.
All sampled states retain positive density and pressure, and the relative
mass and total-energy drifts remain at double-precision roundoff level
throughout the ten calculations.

\begin{figure}[pos=!htb]
  \centering
  \includegraphics[width=0.92\linewidth]{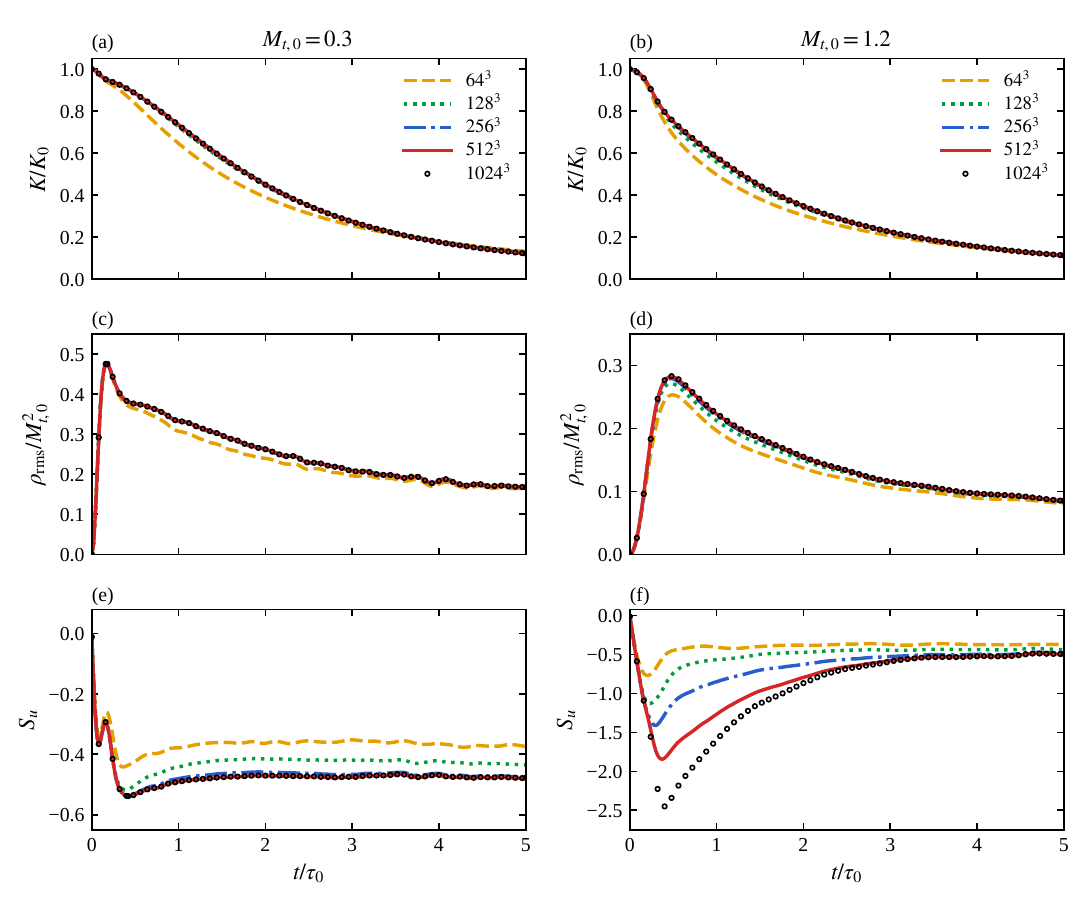}
  \caption{Grid-resolution dependence of decaying compressible isotropic
  turbulence computed with the local ADER method. The columns correspond
  to $M_{t,0}=0.3$ and $1.2$; the rows show $K/K_0$,
  $\rho_{\mathrm{rms}}/M_{t,0}^2$, and $S_u$.
  Colored lines denote $64^3$--$512^3$ results; black open circles denote
  selected $1024^3$ samples, including the sampled extrema.}
  \label{fig:cit-statistics}
\end{figure}

Figure~\ref{fig:cit-statistics} shows that the kinetic-energy histories
and density-fluctuation peaks approach their fine-grid values earlier
than the gradient skewness. For $M_{t,0}=0.3$, the final $K/K_0$ values
on the $512^3$ and $1024^3$ meshes are $0.120779$ and $0.120883$;
for $M_{t,0}=1.2$, they are $0.111915$ and $0.111851$.
The corresponding differences, normalized by the $1024^3$ values, are
$0.086\%$ and $0.057\%$. The finest-grid normalized density-fluctuation
peaks are $0.475272$ and $0.283036$, reached at
$t/\tau_0=0.18$ and $0.48$, respectively. Their differences from the
$512^3$ peaks are $0.0024\%$ and $0.329\%$ under the same normalization.

The skewness gives a more discriminating resolution measure. At
$M_{t,0}=0.3$, its sampled minimum changes only from $-0.538861$ to
$-0.538509$ between the two finest meshes. At $M_{t,0}=1.2$, however,
the minimum changes from $-1.84156$ at $t/\tau_0=0.36$ to $-2.44963$
at $t/\tau_0=0.40$, a difference of $24.8\%$ relative to the finer-grid
magnitude. Thus, the close agreement of the integral measures does not
imply grid convergence of the intermittent velocity gradients in the
supersonic case.

\begin{figure}[pos=!htb]
  \centering
  \includegraphics[width=\linewidth]{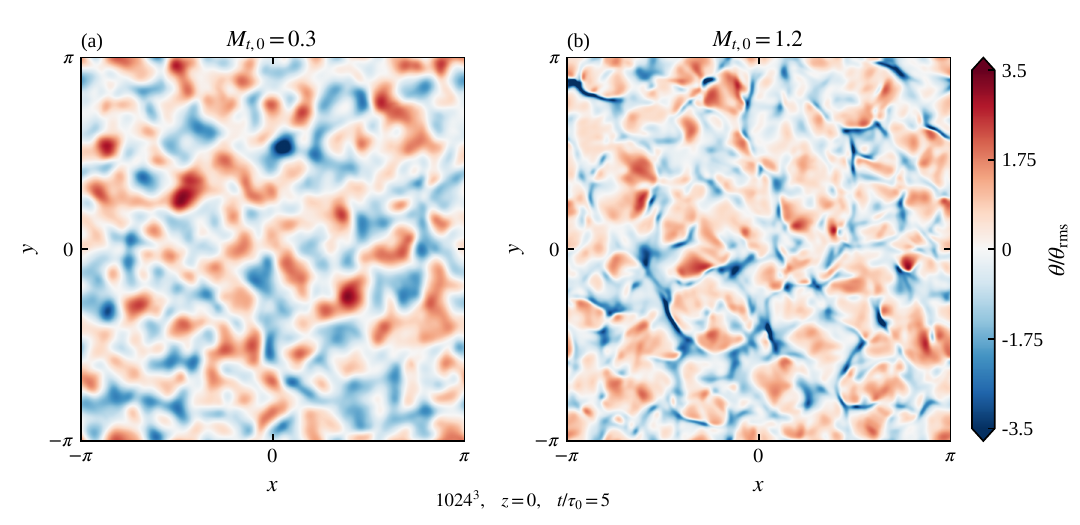}
  \caption{Normalized dilatation $\theta/\theta_{\mathrm{rms}}$ on the
  $z=0$ plane at $t/\tau_0=5$ for the two $1024^3$ cases.
  Blue denotes compression and red denotes expansion. Both panels use
  the range $[-3.5,3.5]$; colorbar extensions indicate values outside
  this range. The original slice values are shown without smoothing
  or interpolation.}
  \label{fig:cit-dilation}
\end{figure}

The instantaneous dilatation provides a spatial view of this compressible
response. Defining
\[
  \theta=\nabla\cdot\boldsymbol{u},
  \qquad \theta_{\mathrm{rms}}=\langle\theta^2\rangle^{1/2},
\]
Fig.~\ref{fig:cit-dilation} shows the physical $z=0$ plane at
$t/\tau_0=5$ on the $1024^3$ meshes. The high-Mach-number field contains
more localized intense compressive regions, whereas the low-Mach-number
field is dominated by broader, smoother structures. Each slice is
normalized by its own full-volume RMS, which is $0.366494$ for
$M_{t,0}=0.3$ and $1.00135$ for $M_{t,0}=1.2$.
The common signed color scale therefore compares the spatial organization
relative to each case's fluctuation level, while these RMS values quantify
the difference in absolute dilatation intensity.

\subsection{Supersonic turbulent channel flow}
\label{sec:supersonic-channel}

The supersonic turbulent channel extends the validation from periodic,
homogeneous turbulence to a wall-bounded flow with sustained molecular
transport. It assesses the local ADER update on a stretched mesh together
with the isothermal no-slip wall treatment, constant-mass-flux forcing,
and the accumulation of long-time turbulence statistics. The mean
thermodynamic stratification and anisotropic Reynolds stresses provide
complementary measures of the resulting compressible channel state.

The configuration follows the $\mathrm{Re}_b=3000$ case documented for the
HTR solver by Di Renzo et al.~\citep{DiRenzoEtAl2020}, which reproduces the
DNS benchmark of Coleman et al.~\citep{Coleman1995}. The bulk Reynolds and
Mach numbers are
\[
  \mathrm{Re}_b=\frac{\dot m_b''h}{\mu_w}=3000,
  \qquad
  M_b=\frac{U_b}{a_w}=1.5,
  \qquad
  U_b=\frac{\dot m_b''}{\rho_b},
\]
where $h$ is the channel half-height, $\dot m_b''$ is the bulk mass flux per
unit cross-sectional area, and the subscript $w$ denotes a wall value.
The calorically perfect gas has $\gamma=1.4$ and $\mathrm{Pr}=0.7$.
With $h$, the prescribed bulk velocity, and the reference bulk density
normalized to unity, the isothermal-wall temperature is $T_w=0.31746$
and the viscosity follows
$\mu=(1/\mathrm{Re}_b)(T/T_w)^{0.7}$.
A multiplicative forcing correction is applied once per accepted physical
step to maintain the prescribed bulk mass flux $\rho_bU_b=1$.

The domain dimensions are $4\pi h\times2h\times2\pi h$ in the streamwise,
wall-normal, and spanwise directions, respectively. Periodicity is imposed
in the two homogeneous directions, and both walls are isothermal and
no-slip. The $256\times128\times128$ mesh is uniform in the homogeneous
directions and follows a hyperbolic-tangent stretching in the wall-normal
direction. The stretching parameter is selected using the reference
$\mathrm{Re}_{\tau,\mathrm{ref}}=219$ to give a first-cell width of
$\Delta y^+_{\mathrm{ref}}\simeq0.8$. The complete local ADER scheme is used
with temperature-dependent viscosity and heat conduction, the
compression--Ducros--pressure sensor, and the prescribed artificial
dissipation. Wall-normal derivatives are evaluated through the coordinate
mapping, with a dedicated wall-adjacent closure for the isothermal no-slip
conditions. The time step is recomputed before every accepted step with
$\mathrm{CFL}=0.5$.

\begin{figure}[pos=!htbp]
  \centering
  \includegraphics[width=\linewidth]{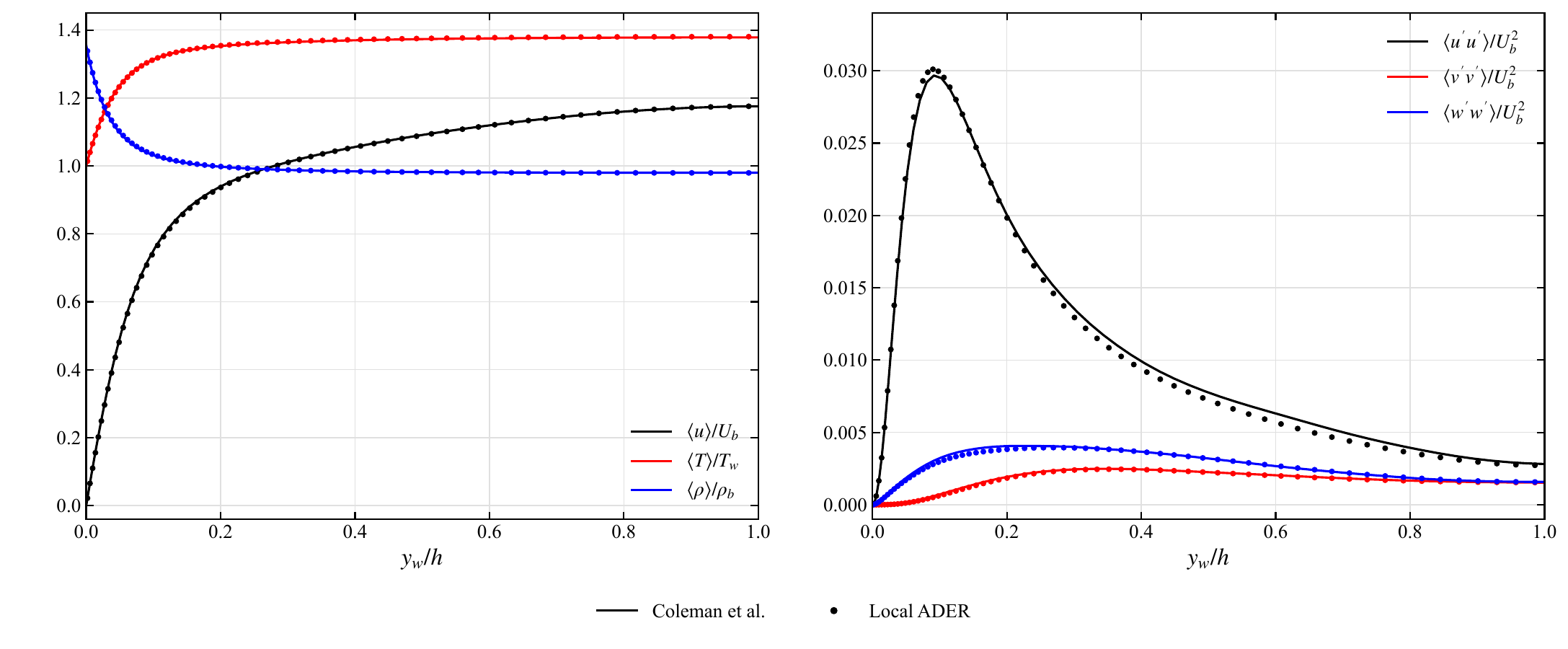}
  \caption{Reynolds-averaged statistics for the $\mathrm{Re}_b=3000$,
  $M_b=1.5$ supersonic turbulent channel computed with the local ADER
  method using a sampling duration of $\Delta t^*=3000$ after equilibration.
  The left panel shows mean
  velocity, temperature, and density; the right panel shows the diagonal
  Reynolds stresses. Solid lines denote the lower-half Coleman reference
  profiles distributed with the HTR benchmark~\citep{Coleman1995,DiRenzoEtAl2020};
  filled circles denote the present statistics averaged over both channel
  halves. The coordinate $y_w/h$ measures distance from the nearest wall.}
  \label{fig:coleman-channel}
\end{figure}

The initial pressure and temperature are uniform. A fourth-power
streamwise-velocity profile is prescribed in the wall-normal direction,
with counter-rotating streamwise vortices and a uniformly distributed
velocity perturbation superposed to initiate transition. After the bulk
temperature and pressure have reached statistically stationary levels,
the reported statistics are accumulated over a reference sampling duration
$\Delta t^*=3000$, where
\[
  t^*=t\,\frac{\mathrm{Re}_{\tau,\mathrm{ref}}^2}{\mathrm{Re}_b}
\]
is the reference wall-time coordinate for the nondimensional solver time.
The complete averaging interval contains $11298$ samples collected every
ten physical steps, including the final partial sampling interval. First- and
second-order moments are averaged over the homogeneous planes and weighted
by the elapsed physical time between samples. The raw moments are then
combined across the two mirrored channel halves, with the appropriate
sign reversal for wall-normal velocity, before forming the Reynolds
covariances. The normalization uses the mass-weighted bulk velocity
obtained by integrating the mean density and streamwise momentum across
the channel.

Figure~\ref{fig:coleman-channel} compares the mean velocity, temperature,
and density profiles and the three diagonal Reynolds stresses with the
Coleman reference data distributed with the HTR benchmark. The mean
profiles reproduce the thermodynamic stratification associated with
isothermal-wall cooling and viscous heating. Near the channel centerline,
$\langle u\rangle/U_b$, $\langle T\rangle/T_w$, and
$\langle\rho\rangle/\rho_b$ approach $1.17537$, $1.38030$, and $0.97932$,
respectively. Over the sampled half-channel, the relative $L_2$
differences of these profiles from the reference are $0.21\%$, $0.13\%$,
and $0.09\%$. The friction Reynolds number evaluated from the
time-averaged density and shear stress at the two walls is
$\mathrm{Re}_\tau=221.56$.

The Reynolds-stress profiles capture the pronounced near-wall peak of the
streamwise component, the broader maxima of the wall-normal and spanwise
components, and their relative magnitudes across the channel. The peak
values of $\langle u'u'\rangle/U_b^2$, $\langle v'v'\rangle/U_b^2$, and
$\langle w'w'\rangle/U_b^2$ are $0.030109$, $0.002457$, and $0.003949$,
respectively. Under the reference-data convention used in
Fig.~\ref{fig:coleman-channel}, these peaks agree with the benchmark to
within $3.0\%$. The corresponding relative $L_2$ differences of the stress
profiles are $3.49\%$, $2.19\%$, and $3.60\%$. Together, the mean-field
and Reynolds-stress comparisons demonstrate that the mapped molecular
transport, wall treatment, forcing, and local ADER update reproduce the
principal statistical features of this compressible wall-bounded flow.

\subsection{Mach-6 hypersonic transitional boundary layer}
\label{sec:ma6-transitional-boundary-layer}

The final case considers a spatially developing hypersonic boundary layer
over a flat plate. It complements the periodic and channel-flow tests by
combining a laminar similarity inflow, characteristic open boundaries,
localized wall forcing, and wall heat transfer with the amplification and
breakdown of instability waves. The configuration follows the first-mode
oblique-breakdown case of Franko and Lele~\citep{FrankoLele2013} and its
calorically perfect-gas realization in the HTR
benchmark~\citep{DiRenzoEtAl2020}.

The free-stream Mach number and Reynolds number based on the inlet
displacement thickness $\delta_0^*$ are
\[
  M_\infty=6,
  \qquad
  \mathrm{Re}_{\delta_0^*}
  =\frac{\rho_\infty U_\infty\delta_0^*}{\mu_\infty}=3000.
\]
The gas has $\gamma=1.4$ and $\mathrm{Pr}=0.71$. The isothermal wall is
held at $T_w/T_\infty=6.5$, and the viscosity follows
$\mu/\mu_\infty=(T/T_\infty)^{0.7}$. The domain dimensions are
\[
  L_x\times L_y\times L_z
  =1000\delta_0^*\times75\delta_0^*\times20\pi\delta_0^*.
\]
The leading edge lies outside the domain. At the inflow plane $x=x_0$,
the velocity and temperature are prescribed from the compressible laminar
similarity solution. The spanwise direction is periodic, and
characteristic open conditions are imposed at the upper and downstream
boundaries. The tangential wall velocities vanish, while the prescribed
wall-normal velocity provides the transition forcing.

Blowing and suction are applied over
$15\leq(x-x_0)/\delta_0^*\leq20$ as
\[
  v_w=f(x)g(z)\sum_{m=1}^{2}A_m
  \sin(\omega_m t-\beta_m z).
\]
Here $f(x)$ is a Gaussian centered at
$(x_s-x_0)/\delta_0^*=17.5$ with width
$\sigma/\delta_0^*=0.75$, and $g(z)$ introduces a weak spanwise
modulation to break the remaining symmetry. The two oblique modes have
equal amplitudes $A_m/U_\infty=0.05$, frequencies
$\omega_m\delta_0^*/a_\infty=0.9$, and spanwise wavenumbers
$\beta_m\delta_0^*=\pm0.3$. The wall-normal velocity is zero outside
the forcing strip.

The billion-node configuration uses a
$6336\times384\times420$ mesh, giving
$1\,021\,870\,080$ spatial nodes. The mesh is uniform in $x$ and $z$
and hyperbolically stretched toward the wall, with a target first spacing
$\Delta y^+\simeq0.32$ based on the estimated outlet viscous scale.
The mapped local ADER method uses the dynamic
spectral-radius estimate at $\mathrm{CFL}=0.6$.

The flow is evolved to $t^*=5000$, where
$t^*=tU_\infty/\delta_0^*$, corresponding to five domain flow-through
times. The statistics accumulators are then reset, and wall quantities
are averaged over the span and over $5000\leq t^*\leq8000$. This
sampling interval spans another three flow-through times. Samples are
collected every ten physical steps and weighted by the elapsed physical
time, including the final partial interval.

\begin{figure}[pos=!htb]
  \centering
  \includegraphics[width=\linewidth]{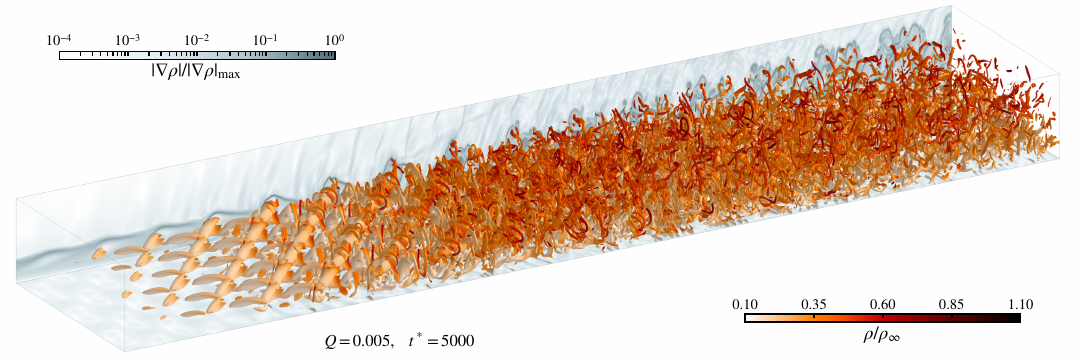}
  \caption{Instantaneous structure of the Mach-6 transitional boundary
  layer at $t^*=5000$.
  The $Q=0.005$ isosurface is colored by $\rho/\rho_\infty$;
  the wall-adjacent and side planes show
  $|\nabla\rho|/|\nabla\rho|_{\max}$, normalized over the displayed
  volume. The displayed region covers
  $400\leq(x-x_0)/\delta_0^*\leq1000$ and
  $0\leq y/\delta_0^*\leq25$.}
  \label{fig:ma6-transition-structure}
\end{figure}

Figure~\ref{fig:ma6-transition-structure} shows the instantaneous
downstream structure at $t^*=5000$. The $Q=0.005$ isosurface is colored
by $\rho/\rho_\infty$, and the wall-adjacent and side planes show the
normalized density-gradient magnitude. The inclined vortical structures
retain a relatively ordered pattern near the upstream end of the
displayed interval. Farther downstream, their spanwise interaction and
breakdown produce shorter, more irregular structures. This progression
connects the growth of the imposed oblique disturbance to the development
of a three-dimensional small-scale field.

\begin{figure}[pos=!htb]
  \centering
  \includegraphics[width=\linewidth]{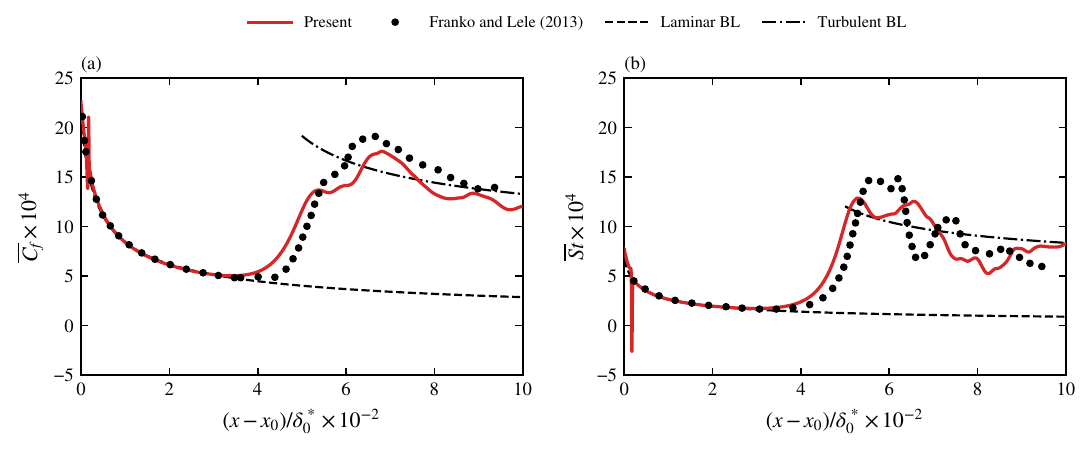}
  \caption{Time- and spanwise-averaged skin-friction coefficient (left)
  and Stanton number (right) for the Mach-6 transitional boundary layer.
  Red lines denote the present
  local ADER results, black symbols denote the DNS data of Franko and
  Lele~\citep{FrankoLele2013}, and dashed and dash-dotted lines denote
  the laminar and turbulent boundary-layer correlations, respectively.
  Statistics are accumulated over $5000\leq t^*\leq8000$.}
  \label{fig:ma6-wall-statistics}
\end{figure}

The wall response is measured by the time- and spanwise-averaged
skin-friction coefficient and Stanton number,
\[
  \overline{C_f}
  =\frac{2\overline{\tau_w}}{\rho_\infty U_\infty^2},
  \qquad
  \overline{St}
  =\frac{-\overline{q_w}}
  {\rho_\infty U_\infty c_p(T_{aw}-T_w)},
\]
where $q_w$ is the conductive heat flux in the positive wall-normal
direction and
$T_{aw}/T_\infty=1+\mathrm{Pr}^{1/3}(\gamma-1)M_\infty^2/2$.
Figure~\ref{fig:ma6-wall-statistics} compares the present results with
the DNS data of Franko and Lele~\citep{FrankoLele2013} and the laminar
and turbulent boundary-layer correlations used in the HTR benchmark.
The present curves contain the complete streamwise statistics without
spatial smoothing.

Upstream of transition, the friction and heat-transfer distributions
follow the laminar decay and closely track the reference data. Their
subsequent increase, the broad skin-friction maximum, and the two-stage
heat-transfer overshoot recover the main signatures of the prescribed
oblique breakdown. The peak levels are lower than the reference values,
and the detailed extrema occur at different streamwise locations.
Farther downstream, both distributions tend toward the turbulent
correlations. The localized negative Stanton number inside the forcing
strip is retained in the comparison.

Taken together, the instantaneous structures and wall statistics show
the preservation of the upstream laminar layer, amplification of the
forced oblique modes, and subsequent breakdown with increased friction
and heat transfer.

\section{Conclusions}
\label{sec:conclusions}

A conservative local ADER finite-difference method has been developed for
compressible-flow DNS on GPUs. A ten-point quadratic representation,
two-node Gauss--Legendre quadrature, and one Picard correction supply
time-averaged fluxes to a single global conservative update. Operator
matching couples the compact predictor to the wide central corrector.
\accuracyrevision{The complete Euler update is third-order consistent under
fixed-CFL refinement when its temporal--spatial mixed errors are included.
Smooth convective tests recover this third-order convergence and
near-seventh-order behavior in the regime dominated by background spatial error.}

The sinusoidal-wave and isentropic-vortex calculations recover the
expected spatial- and temporal-error regimes. The compressible
Taylor--Green studies through $1024^3$ nodes resolve finer
vortical scales, with the two finest meshes giving closely agreeing
kinetic-energy histories and enstrophy peaks. In decaying compressible
isotropic turbulence, mass and total energy are conserved to
double-precision roundoff with temperature-dependent molecular transport.
The kinetic-energy and density-fluctuation statistics converge more
rapidly than the velocity-gradient skewness. Its sensitivity at
$M_{t,0}=1.2$ shows that integral statistics alone do not establish
resolution of intermittent compressive structures.

The wall-bounded cases assess molecular transport and boundary treatment
on stretched meshes. For the supersonic channel, the relative $L_2$
differences from the reference are at most $0.21\%$ for the mean velocity,
temperature, and density profiles and $3.60\%$ for the diagonal Reynolds
stresses. The Mach-6 boundary-layer calculation reproduces oblique-mode
amplification, three-dimensional breakdown, and the associated increase
in skin friction and heat transfer. These results support the mapped
formulation for statistically stationary wall turbulence and spatially
developing hypersonic transition.

The GPU implementation contracts temporal-node data within the predictor
kernel, reuses resident work arrays, and overlaps halo communication with
interior computation. On the \textcolor[rgb]{0,0,0}{$302^3$} viscous NS-TGV benchmark with zero
heat conduction, local ADER and RK3-F attain
\textcolor[rgb]{0,0,0}{$230.87$ and $118.76$}~Mnode updates/s on one V100 GPU, giving a complete-step
speedup of \textcolor[rgb]{0,0,0}{$1.944$} with a common spatial operator,
dissipation, transport model, and CFL selector. \textcolor[rgb]{0,0,0}{The separate $268^3$ hardware profiles show} that local ADER reduces HBM traffic by $42.0\%$ and counted FP64
operations by $15.8\%$ per complete step relative to RK3-F.

In the distributed benchmark, the weak-scaling efficiency from 4 to 32 V100 GPUs is
$99.44\%$, and the configuration containing $1.812\times10^9$ nodes
attains $7243.7$~Mnode updates/s. For the fixed-size problem,
strong-scaling efficiency decreases to $48.04\%$ at 32 GPUs as the
local subdomains become smaller.

With the spatial discretization, dissipation, and transport model held
fixed, the controlled comparison with RK3-F demonstrates that the local
ADER construction improves complete-step advancement efficiency,
providing an efficient time-integration approach for large-scale
compressible-flow simulations on GPUs.

\section*{Acknowledgements}

This work was funded by the Funding of National Key Laboratory of
Computational Physics, the National Natural Science Foundation of China
(Grant No. 12302378 and 92371107) and the Natural Science Basic
Research Plan in Shaanxi Province of China (No. 2025SYS-SYSZD-070).

\bibliographystyle{cas-model2-names}
\bibliography{references}

@article{Coleman1995,
    author = {Coleman, Gary N. and Kim, John and Moser, Robert D.},
    title = {A Numerical Study of Turbulent Supersonic Isothermal-Wall Channel Flow},
    journal = {Journal of Fluid Mechanics},
    year = {1995},
    volume = {305},
    pages = {159--183},
    doi = {10.1017/S0022112095004587}
}

@article{MoinMahesh1998,
    author = {Moin, Parviz and Mahesh, Krishnan},
    title = {Direct Numerical Simulation: A Tool in Turbulence Research},
    journal = {Annual Review of Fluid Mechanics},
    year = {1998},
    volume = {30},
    number = {1},
    pages = {539--578},
    doi = {10.1146/annurev.fluid.30.1.539}
}

@article{JohnsenEtAl2010,
    author = {Johnsen, Eric and Larsson, Johan and Bhagatwala, Ankit V. and Cabot, William H. and Moin, Parviz and Olson, Britton J. and Rawat, Pradeep S. and Shankar, Santhosh K. and Sj{\"o}green, Bj{\"o}rn and Yee, H. C. and Zhong, Xiaolin and Lele, Sanjiva K.},
    title = {Assessment of High-Resolution Methods for Numerical Simulations of Compressible Turbulence with Shock Waves},
    journal = {Journal of Computational Physics},
    year = {2010},
    volume = {229},
    number = {4},
    pages = {1213--1237},
    doi = {10.1016/j.jcp.2009.10.028}
}

@article{Lele1992,
    author = {Lele, Sanjiva K.},
    title = {Compact Finite Difference Schemes with Spectral-Like Resolution},
    journal = {Journal of Computational Physics},
    year = {1992},
    volume = {103},
    number = {1},
    pages = {16--42},
    doi = {10.1016/0021-9991(92)90324-R}
}

@article{BhagatwalaLele2009,
    author = {Bhagatwala, Ankit V. and Lele, Sanjiva K.},
    title = {A Modified Artificial Viscosity Approach for Compressible Turbulence Simulations},
    journal = {Journal of Computational Physics},
    year = {2009},
    volume = {228},
    number = {14},
    pages = {4965--4969},
    doi = {10.1016/j.jcp.2009.04.009}
}

@article{SciacovelliEtAl2021,
    author = {Sciacovelli, Luca and Passiatore, Donatella and Cinnella, Paola and Pascazio, Giuseppe},
    title = {Assessment of a High-Order Shock-Capturing Central-Difference Scheme for Hypersonic Turbulent Flow Simulations},
    journal = {Computers \& Fluids},
    year = {2021},
    volume = {230},
    pages = {105134},
    doi = {10.1016/j.compfluid.2021.105134}
}

@article{KennedyCarpenterLewis2000,
    author = {Kennedy, Christopher A. and Carpenter, Mark H. and Lewis, R. Michael},
    title = {Low-Storage, Explicit {Runge--Kutta} Schemes for the Compressible {Navier--Stokes} Equations},
    journal = {Applied Numerical Mathematics},
    year = {2000},
    volume = {35},
    number = {3},
    pages = {177--219},
    doi = {10.1016/S0168-9274(99)00141-5}
}

@article{DiRenzoEtAl2020,
    author = {Di Renzo, Mario and Fu, Lin and Urzay, Javier},
    title = {{HTR} Solver: An Open-Source Exascale-Oriented Task-Based Multi-{GPU} High-Order Code for Hypersonic Aerothermodynamics},
    journal = {Computer Physics Communications},
    year = {2020},
    volume = {255},
    pages = {107262},
    doi = {10.1016/j.cpc.2020.107262}
}

@article{RadhakrishnanEtAl2024,
    author = {Radhakrishnan, Anand and Le Berre, Henry and Wilfong, Benjamin and Spratt, James S. and Rodriguez, Mauro, Jr. and Colonius, Tim and Bryngelson, Spencer H.},
    title = {Method for Portable, Scalable, and Performant {GPU}-Accelerated Simulation of Multiphase Compressible Flow},
    journal = {Computer Physics Communications},
    year = {2024},
    volume = {302},
    pages = {109238},
    doi = {10.1016/j.cpc.2024.109238}
}

@article{DangEtAl2022,
    author = {Dang, Guanlin and Liu, Shiwei and Guo, Tongbiao and Duan, Junyi and Li, Xinliang},
    title = {Direct Numerical Simulation of Compressible Turbulence Accelerated by Graphics Processing Unit: An Open-Source High Accuracy Accelerated Computational Fluid Dynamic Software},
    journal = {Physics of Fluids},
    year = {2022},
    volume = {34},
    number = {12},
    pages = {126106},
    doi = {10.1063/5.0127684}
}

@article{BernardiniEtAl2023,
    author = {Bernardini, Matteo and Modesti, Davide and Salvadore, Francesco and Sathyanarayana, Srikanth and Della Posta, Gabriele and Pirozzoli, Sergio},
    title = {{STREAmS}-2.0: Supersonic Turbulent Accelerated Navier--Stokes Solver Version 2.0},
    journal = {Computer Physics Communications},
    year = {2023},
    volume = {285},
    pages = {108644},
    doi = {10.1016/j.cpc.2022.108644}
}

@article{JiangShu1996,
    author = {Jiang, Guang-Shan and Shu, Chi-Wang},
    title = {Efficient Implementation of Weighted {ENO} Schemes},
    journal = {Journal of Computational Physics},
    year = {1996},
    volume = {126},
    number = {1},
    pages = {202--228},
    doi = {10.1006/jcph.1996.0130}
}

@article{FuHuAdams2016,
    author = {Fu, Lin and Hu, Xiangyu Y. and Adams, Nikolaus A.},
    title = {A Family of High-Order Targeted {ENO} Schemes for Compressible-Fluid Simulations},
    journal = {Journal of Computational Physics},
    year = {2016},
    volume = {305},
    pages = {333--359},
    doi = {10.1016/j.jcp.2015.10.037}
}

@article{CockburnShu1998,
    author = {Cockburn, Bernardo and Shu, Chi-Wang},
    title = {The {Runge--Kutta} Discontinuous {Galerkin} Method for Conservation Laws {V}: Multidimensional Systems},
    journal = {Journal of Computational Physics},
    year = {1998},
    volume = {141},
    number = {2},
    pages = {199--224},
    doi = {10.1006/jcph.1998.5892}
}

@article{LaxWendroff1960,
    author = {Lax, Peter D. and Wendroff, Burton},
    title = {Systems of Conservation Laws},
    journal = {Communications on Pure and Applied Mathematics},
    year = {1960},
    volume = {13},
    number = {2},
    pages = {217--237},
    doi = {10.1002/cpa.3160130205}
}

@article{QiuShu2003,
    author = {Qiu, Jianxian and Shu, Chi-Wang},
    title = {Finite Difference {WENO} Schemes with {Lax--Wendroff}-Type Time Discretizations},
    journal = {SIAM Journal on Scientific Computing},
    year = {2003},
    volume = {24},
    number = {6},
    pages = {2185--2198},
    doi = {10.1137/S1064827502412504}
}

@article{ShuOsher1988,
    author = {Shu, Chi-Wang and Osher, Stanley},
    title = {Efficient Implementation of Essentially Non-Oscillatory Shock-Capturing Schemes},
    journal = {Journal of Computational Physics},
    year = {1988},
    volume = {77},
    number = {2},
    pages = {439--471},
    doi = {10.1016/0021-9991(88)90177-5}
}

@misc{GaoEtAl2026,
    author = {Gao, Qihui and Ji, Xing and Du, Zhifang and Li, Shiyi and Chen, Yibing and Xu, Kun},
    title = {The Stability Priority of Spatial-Temporal Coupled Compact Element Methods over Decoupled Compact Element Methods},
    year = {2026},
    howpublished = {arXiv preprint arXiv:2601.17264},
    doi = {10.48550/arXiv.2601.17264},
    url = {https://arxiv.org/abs/2601.17264}
}

@article{XuanXu2018,
    author = {Xuan, Li-Jun and Xu, Kun},
    title = {An Efficient High-Order Finite Difference Gas-Kinetic Scheme for the {Euler} and {Navier--Stokes} Equations},
    journal = {Computers \& Fluids},
    year = {2018},
    volume = {166},
    pages = {243--252},
    doi = {10.1016/j.compfluid.2018.01.026}
}

@misc{ZhaoChenXu2026,
    author = {Zhao, Fengxiang and Chen, Yibing and Xu, Kun},
    title = {Finite Difference Implementation of a High-Order Space-Time Coupled Compact Gas-Kinetic Scheme},
    year = {2026},
    howpublished = {arXiv preprint arXiv:2606.17772},
    doi = {10.48550/arXiv.2606.17772},
    url = {https://arxiv.org/abs/2606.17772}
}

@article{LiChenJiang2020,
    author = {Li, Shiyi and Chen, Yibing and Jiang, Song},
    title = {An Efficient High-Order Gas-Kinetic Scheme ({I}): {Euler} Equations},
    journal = {Journal of Computational Physics},
    year = {2020},
    volume = {415},
    pages = {109488},
    doi = {10.1016/j.jcp.2020.109488}
}

@inproceedings{LampleCharton2020,
    author = {Lample, Guillaume and Charton, Fran{\c{c}}ois},
    title = {Deep Learning for Symbolic Mathematics},
    booktitle = {International Conference on Learning Representations},
    year = {2020},
    url = {https://openreview.net/forum?id=S1eZYeHFDS}
}

@article{ChristliebGucluSeal2015,
    author = {Christlieb, Andrew J. and G{\"u}{\c{c}}l{\"u}, Yaman and Seal, David C.},
    title = {The Picard Integral Formulation of Weighted Essentially Nonoscillatory Schemes},
    journal = {SIAM Journal on Numerical Analysis},
    year = {2015},
    volume = {53},
    number = {4},
    pages = {1833--1856},
    doi = {10.1137/140959936}
}

@article{LeeLee2021,
    author = {Lee, Youngjun and Lee, Dongwook},
    title = {A Single-Step Third-Order Temporal Discretization with {Jacobian}-Free and {Hessian}-Free Formulations for Finite Difference Methods},
    journal = {Journal of Computational Physics},
    year = {2021},
    volume = {427},
    pages = {110063},
    doi = {10.1016/j.jcp.2020.110063}
}

@article{DumbserEtAl2008,
    author = {Dumbser, Michael and Balsara, Dinshaw S. and Toro, Eleuterio F. and Munz, Claus-Dieter},
    title = {A Unified Framework for the Construction of One-Step Finite Volume and Discontinuous Galerkin Schemes on Unstructured Meshes},
    journal = {Journal of Computational Physics},
    year = {2008},
    volume = {227},
    number = {18},
    pages = {8209--8253},
    doi = {10.1016/j.jcp.2008.05.025}
}

@article{BalsaraEtAl2013,
    author = {Balsara, Dinshaw S. and Meyer, Chad and Dumbser, Michael and Du, Huijing and Xu, Zhiliang},
    title = {Efficient Implementation of {ADER} Schemes for {Euler} and Magnetohydrodynamical Flows on Structured Meshes---Speed Comparisons with {Runge--Kutta} Methods},
    journal = {Journal of Computational Physics},
    year = {2013},
    volume = {235},
    pages = {934--969},
    doi = {10.1016/j.jcp.2012.04.051}
}

@article{Norman2014,
    author = {Norman, Matthew R.},
    title = {A {WENO}-Limited, {ADER-DT}, Finite-Volume Scheme for Efficient, Robust, and Communication-Avoiding Multi-Dimensional Transport},
    journal = {Journal of Computational Physics},
    year = {2014},
    volume = {274},
    pages = {1--18},
    doi = {10.1016/j.jcp.2014.05.043}
}

@inproceedings{Jameson1981,
    author = {Jameson, Antony and Schmidt, Wolfgang and Turkel, Eli},
    title = {Numerical Solutions of the {Euler} Equations by Finite Volume Methods Using {Runge--Kutta} Time-Stepping Schemes},
    booktitle = {14th Fluid and Plasma Dynamics Conference},
    year = {1981},
    note = {AIAA Paper 81-1259},
    doi = {10.2514/6.1981-1259}
}

@article{Ducros1999,
    author = {Ducros, Fr\'ed\'eric and Ferrand, Val\'erie and Nicoud, Franck and Weber, Christophe and Darracq, Dominique and Gacherieu, Claude and Poinsot, Thierry},
    title = {Large-Eddy Simulation of the Shock/Turbulence Interaction},
    journal = {Journal of Computational Physics},
    volume = {152},
    number = {2},
    pages = {517--549},
    year = {1999},
    doi = {10.1006/jcph.1999.6238}
}

@article{BalsaraShu2000,
    author = {Balsara, Dinshaw S. and Shu, Chi-Wang},
    title = {Monotonicity Preserving Weighted Essentially Non-Oscillatory Schemes with Increasingly High Order of Accuracy},
    journal = {Journal of Computational Physics},
    year = {2000},
    volume = {160},
    number = {2},
    pages = {405--452},
    doi = {10.1006/jcph.2000.6443}
}

@article{TitarevToro2004,
    author = {Titarev, Vladimir A. and Toro, Eleuterio F.},
    title = {Finite-Volume {WENO} Schemes for Three-Dimensional Conservation Laws},
    journal = {Journal of Computational Physics},
    year = {2004},
    volume = {201},
    number = {1},
    pages = {238--260},
    doi = {10.1016/j.jcp.2004.05.015}
}

@inproceedings{DeBonis2013,
    author = {DeBonis, James R.},
    title = {Solutions of the {Taylor--Green} vortex problem using high-resolution explicit finite difference methods},
    booktitle = {51st AIAA Aerospace Sciences Meeting including the New Horizons Forum and Aerospace Exposition},
    year = {2013},
    note = {{AIAA} Paper 2013-0382},
    doi = {10.2514/6.2013-382}
}

@article{PengYang2018,
    author = {Peng, Naifu and Yang, Yue},
    title = {Effects of the {Mach} number on the evolution of vortex-surface fields in compressible {Taylor--Green} flows},
    journal = {Physical Review Fluids},
    volume = {3},
    number = {1},
    pages = {013401},
    year = {2018},
    doi = {10.1103/PhysRevFluids.3.013401}
}

@article{FrankoLele2013,
    author = {Franko, Kenneth J. and Lele, Sanjiva K.},
    title = {Breakdown mechanisms and heat transfer overshoot in hypersonic zero pressure gradient boundary layers},
    journal = {Journal of Fluid Mechanics},
    volume = {730},
    pages = {491--532},
    year = {2013},
    doi = {10.1017/jfm.2013.350}
}

@article{WangWangPan2024,
  author  = {Wang, Yibo and Wang, Yuhang and Pan, Liang},
  title   = {Compact high-order gas-kinetic scheme for direct numerical simulation of compressible turbulent flows},
  journal = {Physics of Fluids},
  volume  = {36},
  number  = {1},
  pages   = {016137},
  year    = {2024},
  doi     = {10.1063/5.0181987}
}

@misc{SathyanarayanaEtAl2023,
    author = {Sathyanarayana, Srikanth and Bernardini, Matteo and Modesti, Davide and Pirozzoli, Sergio and Salvadore, Francesco},
    title = {High-Speed Turbulent Flows towards the Exascale: {STREAmS}-2 Porting and Performance},
    year = {2023},
    howpublished = {arXiv preprint arXiv:2304.05494},
    url = {https://arxiv.org/abs/2304.05494}
}

@article{WilliamsWatermanPatterson2009,
  author = {Williams, Samuel and Waterman, Andrew and Patterson, David},
  title = {{Roofline}: An insightful visual performance model for multicore architectures},
  journal = {Communications of the ACM},
  year = {2009},
  volume = {52},
  number = {4},
  pages = {65--76},
  doi = {10.1145/1498765.1498785}
}

\end{document}